\documentclass[aps,prd,twocolumn,groupedaddress]{revtex4-2}
\usepackage[dvipsnames]{xcolor}
\usepackage[breaklinks,colorlinks,citecolor=blue,linkcolor=Green]{hyperref}
\usepackage{amssymb}
\usepackage{amsmath}
\usepackage{float}
\usepackage{graphicx}
\usepackage{gensymb}

\begin{document}


\title{Updated constraints on WIMP dark matter annihilation by radio observations of M31}


\author{Andrei E. Egorov}
\email[E-mail: ]{aegorov@runbox.com}
\affiliation{Nuclear Physics and Astrophysics Division, Lebedev Physical Institute, Leninskii prospect - 53, Moscow, Russia}

\date{\today}

\begin{abstract}
The present work derived the robust constraints on annihilating WIMP parameters utilizing new radio observations  of M31, as well as new studies of its dark matter distribution and other properties. The characteristics of emission due to DM annihilation were computed in the frame of 2D galactic model employing GALPROP code adapted specifically for M31. This enabled us to refine various inaccuracies of previous studies on the subject. DM constraints were obtained for two representative annihilation channels: $\chi\chi \rightarrow b\overline{b}$ and $\chi\chi \rightarrow \tau^+\tau^-$. A wide variety of radio data was utilized in the frequency range $\approx$(0.1--10) GHz. As the result the thermal WIMP lighter than fiducially $\approx$ 70 GeV for $b\overline{b}$ channel and $\approx$ 40 GeV for $\tau^+\tau^-$ was excluded. The corresponding mass threshold uncertainty ranges were estimated to be 20--210 GeV and 18--89 GeV. The obtained exclusions are competitive to those from Fermi-LAT observations of dwarfs and AMS-02 measurements of antiprotons. Our constraints do not exclude the explanation of the gamma-ray outer halo of M31 and the Galactic center excess by annihilating DM. The thermal WIMP with $m_x \approx 70$ GeV, which explains the outer halo, would make a significant contribution to the non-thermal radio flux in M31 nucleus, fitting well both the spectrum and morphology. And, finally, we questioned the possibility claimed in other studies to robustly constrain heavy thermal WIMP with $m_x > 100$ GeV by radio data on M31. 
\end{abstract}


\maketitle

\section{\label{sec:i}Introduction and motivation}

The physical nature of dark matter (DM) remains to be one of the biggest puzzles in modern physics and astronomy. Currently, we may outline three most popular candidates for the role of DM: weakly interacting massive particles (WIMPs), axionlike particles (ALPs) and sterile neutrinos (see e.g. \cite{2019arXiv191204727R} for a review). This work is dedicated to indirect searches of the first candidate -- WIMPs, which historically have been the most anticipated. The idea of indirect searches is based on the opportunity of pair annihilation of WIMPs, which are assumed to be Majorana fermions, particularly e.g. neutralinos. This annihilation produces various highly-energetic Standard Model particles, whose signatures can be potentially detected or constrained astrophysically. Historically this idea was introduced and explored for the first time probably in \cite{1980SvJNP..31..664Z}. For a review of the contemporary status of this field, one may see e.g. \cite{2021arXiv210902696S}. In our case we consider DM halo of the big neighbor galaxy M31 (or Andromeda galaxy, or NGC 224). Annihilating WIMPs in its halo would produce relativistic electrons and positrons ($e^\pm$), which in turn would generate synchrotron emission in the galactic magnetic field (MF) at radio frequencies. Hence, the radio observations of the galaxy may infer some constraints on annihilating DM or even hints of DM signal.

This work mainly aims to derive the conservative and robust constraints on WIMP parameters -- mass and annihilation cross section. The considered WIMP mass range is 10--1000 GeV. Indeed, theoretically WIMPs may exist beyond this range. However, WIMPs lighter than $\sim$ 10 GeV with the cross section around the thermal value are already excluded robustly by various strategies \cite{2021arXiv210902696S}, at least for canonical annihilation channels. And WIMPs heavier than $\sim$ 1 TeV produce too faint radio signals to be separable from other astrophysical emissions. Then inside the chosen mass range thermal WIMPs lighter than $\approx$(20--100) GeV were excluded by Fermi-LAT observations of dwarf Milky Way (MW) satellites \cite{2020PhRvD.102f1302A,2020JCAP...02..012H}. The quoted mass threshold is so uncertain due to uncertainties in the assumed DM densities in dwarfs and a variety of possible annihilation channels. Then the masses above $\approx$ 100 GeV are expected to be probed by future Cherenkov Telescope Array (CTA) observations of the Galactic halo \cite{2021JCAP...01..057A}. Thus, gamma-ray observations, although producing rather model-independent constraints, tend to have some sensitivity gap around $m_x = 100$ GeV. Hence, other independent probes should be utilized in order to complement searches in the gamma-ray band, also verifying them independently. And radio observations are one such probe. Unlike the prompt gamma radiation, the radio emission is produced by secondary $e^\pm$, which makes the respective constraints more model-dependent. However, the advantage of the radio band is much higher angular and spectral resolutions in comparison with the gamma-ray band, which provides very detailed imaging of targets. The polarization information can be utilized too, which enhances the sensitivity even further \cite{2022arXiv220404232M}. Thus, the ultimate goal of our work is to probe WIMP masses up to $\approx$ 100 GeV at the level of thermal cross section as the most physically motivated one. 

M31 is one of the best targets in the sky for DM searches in radio due to its proximity, which allows detailed imaging and studies of the galactic environment -- particularly mass and MF distributions. Dwarf MW satellites are not so attractive in radio as in the gamma-ray band due to the absence of any direct measurements of their MF. Objects beyond the Local Group are more distant and, hence, would have fainter annihilation signals. Our own Galaxy MW might be worse target than M31 due to brighter central region, which is illustrated by table \ref{tab:i} at a couple of relevant frequencies. The total radio emission is indeed composed by a wide variety of mechanisms, and it is quite difficult to separate the various emission components. Hence, in general, a lower total emission intensity would imply a lower limit on the intensity of emission due to DM annihilation and, therefore, stronger DM constraints. A majority of the emission from DM is concentrated in a small region around the galactic center with radius $R \sim 1$ kpc; since typically DM density falls off with distance fast, and the volumic annihilation rate is proportional to the square of DM density. We see from table \ref{tab:i} that M31 nucleus is fainter than that of MW by more than an order of magnitude! This suggests potentially high sensitivity to weak DM contribution and motivates to study M31 very thoroughly as a promising target.   
\begin{table}[h]
\caption{\label{tab:i}The estimated intensities of the nuclear regions ($R \leqslant 1$ kpc) of MW and M31 at relevant frequencies with respective references.}
\begin{ruledtabular}
\centering
\begin{tabular}{ccc}
Galaxy & Intensity at      & Intensity at  \\
       & 0.408 GHz, kJy/sr & 1.42 GHz, kJy/sr \\                     
\hline
MW  & 1400 {\cite{Haslam}} & 440 {\cite{2001A&A...376..861R}} \\
M31 & 78 {\cite{1985A&A...150L...1W}} & 32 {\cite{1985A&A...150L...1W}} \\
\end{tabular}
\end{ruledtabular}
\end{table}				

At the frequencies relevant for WIMP searches, the galactic radio emission has two main components -- thermal and non-thermal (illustrated in fig. \ref{fig:sed} below). The thermal emission is generated by ionized galactic gas and can be relatively easily subtracted due to a distinct spectral shape. The non-thermal emission represents the synchrotron from relativistic $e^\pm$ of cosmic rays (CR) and may contain the component from $e^\pm$ of DM. To distinguish these two populations, they are refereed as CR and DM $e^\pm$ respectively. It is difficult in general to separate these two components of synchrotron emission, since they have quite similar spectral shapes as shown below. For this reason, in order to derive conservative and robust DM constraints, I separated out only the thermal emission when possible and did not attempt to estimate the level of CR synchrotron. Such an attempt would indeed strengthen the constraints, but make them very model-dependent at the same time. 

In order to derive DM constraints at first we have to model theoretically the characteristics of anticipated emission due to DM, i.e. its intensity dependence on frequency and direction. This is done by utilization of GALPROP code and is described in sec. \ref{sec:dm}. The obtained DM emission maps and spectra are discussed in sec. \ref{sec:maps}. Then the relevant observational data were collected and compared with the theoretically calculated DM emission intensity. The latter is linearly proportional to the annihilation cross section. This fact allows to calculate the limits on cross section relatively easily for a fixed WIMP mass. The derivation of constraints is described in sec. \ref{sec:constr}. Then sec. \ref{sec:gamma} relates the derived constraints with those from gamma-ray band and antiprotons. And sec. \ref{sec:signal} discusses semi-qualitatively possible DM contribution into the central radio emission in M31.

In fact, this paper continues our previous work on the subject \cite{2013PhRvD..88b3504E}. Currently, an attempt of comprehensive study has been made, which includes the following key novelties and improvements:
\begin{enumerate}
	\item Multiple inaccuracies made in the previous works (they are discussed in sec. \ref{sec:cr}) were refined.
	\item New information on DM density distribution and galactic environment emerged over last 10 years were incorporated.
	\item For the first time, the transport equation for DM $e^\pm$ was solved in 2D including the diffusion of $e^\pm$, before only 1D (i.e. spherically-symmetric) approximation was implemented -- more in sec. \ref{sec:cr}.
	\item New radio observational data were included, particularly very valuable low-frequency images from LOFAR.
	\item Emission component separation was included: at some frequencies the utilized maps are cleaned of the discrete projected nuisance sources and diffuse thermal emission.
\end{enumerate}

Table \ref{tab:i-par} shows the utilized parameters of M31, which define its position with respect to (w.r.t.) the observer. Over the text of paper variable $R$ denotes the 3D radial distance, while $(r,z)$ are cylindrical coordinates attached to the galactic plane. They correspond to $(\rho',z')$ coordinates in our paper \cite{2013PhRvD..88b3504E}.
\begin{table}[h]
	\caption{\label{tab:i-par}The values of basic parameters of M31 used in this work.}
	\begin{ruledtabular}
		\centering
		\begin{tabular}{ccc}
			Parameter & Value & Reference  \\
			\hline
			Galactic longitude of center & 121.17$\degree$ & \cite{NED} \\
			Galactic latitude of center & --21.57$\degree$ & \cite{NED} \\
			Distance $d$  & 760 kpc &  \cite{2021ApJ...920...84L} \\
			Inclination & 74$\degree$ & \cite{2021ApJ...920...84L} \\
		\end{tabular}
	\end{ruledtabular}
\end{table}							

\section{\label{sec:cr}Critical review of the previous works on the subject}

Here is the list of papers dedicated to DM searches in M31 in radio best to the author's knowledge -- \cite{2013PhRvD..88b3504E,2018PhRvD..97j3021M,2019ApJ...872..177C,2019JCAP...08..019B,2021MNRAS.501.5692C}. Each of these works has certain weaknesses, which are briefly reviewed below.

\textbf{\cite{2013PhRvD..88b3504E}} (our work) was the first approximation to the task. In that study we neglected by the spatial diffusion of DM $e^\pm$ in the central region. Then eqs. (8) and (10) there, which define the energy loss rates for $e^\pm$, are unfortunately incorrect at some level, which was revealed in verification procedure conducted now. Thus, inverse Compton scattering (ICS) losses (8) must have the prefactor 1.0 instead of 0.76. Then the bremsstrahlung losses (10) there must have the form as (\ref{eq:dm-loss-b}) here below. Moreover, historically these errors propagated through many papers due to copying without checking. Thus, the following non-exhaustive list of papers has wrong expressions for some of the energy loss components -- \cite{2018PhRvD..97j3021M,2019JCAP...08..019B,2006A&A...455...21C,2020ApJ...900..126C,2019PDU....2600355C,2013ApJ...768..106S,2021Galax...9...11C,2017JCAP...09..027M}. These loss rates play an important role in the calculations. For this reason the verified expressions for them are provided below -- eqs. (\ref{eq:dm-loss-s})--(\ref{eq:dm-loss-c}). Another possible problem in the above papers is the expression for electron's synchrotron power. This can be seen, for example, in \cite[eq. (38)]{2006A&A...455...21C}: the physical meaning of the factor $(1+(\gamma \nu_p / \nu)^2)^{3/2}$ is unclear, its explanation can not be found. Returning back to the discussion of \cite{2013PhRvD..88b3504E}: we assumed there for the interstellar radiation field (ISRF) energy density to be constant and equal to that in MW center. This energy density defines the ICS loss rate (eq. (\ref{eq:dm-loss-i})). Indeed, the density is not uniform in the galactic environment and, as shown in sec. \ref{sec:dm-prop} below, larger in M31 nucleus by $\approx$ 2 times w.r.t. that in MW nucleus.

\textbf{\cite{2018PhRvD..97j3021M}} calculated multifrequency emissions from annihilating WIMPs, which would fit the gamma-ray excess around M31 center seen by Fermi-LAT \cite{2017ApJ...836..208A}. For this purpose the authors solved analytically the transport equation for DM $e^\pm$ with spatial diffusion in the spherically-symmetric approximation employing their RX-DMFIT tool \cite{2017JCAP...09..027M}. Although such universal tools are generally useful, 1D approximation presumably applies well to dwarf spheroidals and virialized galaxy clusters; but not to spiral galaxies, which inherently have (at least) 2D geometry. Also, each target has its own peculiar properties. The authors extrapolated (similarly to \cite{2013PhRvD..88b3504E}) MW central ISRF energy density to M31, which would overestimate the emission intensity from $e^\pm$ due to underestimated cooling rate.

\textbf{\cite{2019ApJ...872..177C}} claimed abnormally strong DM constraints from their observations of M31 at 5 GHz (just one frequency) by 26 m single-dish radio telescope. Thus, these authors stated the exclusion of WIMP masses up to $\approx$ 300 GeV for all annihilation channels! They claimed that by somehow their observational technique called "cross-scans" managed to avoid the majority of usual astrophysical emissions and left mainly the diffuse large-scale emission potentially attributable to DM. The described technique is not understandable, especially taking into account their angular resolution FWHM $\approx 10'$. They utilized the region, which captures essentially the whole galaxy, for the derivation of constraints. The measured diffuse signal looks to be around the noise level from their fig. 1. Likely, the authors severely underestimated the large-scale diffuse flux, getting its value much below other independent measurements. Then in the calculations they neglected completely by the spatial diffusion of DM $e^\pm$ in the whole galaxy, and by spatial variations of MF and ISRF density. These are clearly crude approximations, as can be seen e.g. from \cite[fig. 6]{2018PhRvD..97j3021M}, where the impact of diffusion coefficient value on the radio flux is shown for the central kiloparsec. This impact is significant even in the central region, where $e^\pm$ cool fast. In outer regions, $e^\pm$ cool much slower, hence migrate much further sinking into the halo -- and the diffusion is much more important. Thus, overall the claimed constraints can not be considered as justified.

\textbf{\cite{2019JCAP...08..019B}} basically solved the transport equation again in 1D and possibly (not really clear from the text) neglected by the diffusion too, producing even stronger constraints than \cite{2019ApJ...872..177C} with large ROIs up to 50$'$. Thus, the reasoning from the previous two paragraphs applies here directly, too. One advantage here in comparison with \cite{2019ApJ...872..177C} is the spatially-dependent MF (1D). However, MF model in \cite{2019JCAP...08..019B} (their eq. (4.2), taken from \cite{2010ApJ...723L..44R}) likely has the wrong meaning of "r" variable: originally in \cite{2010ApJ...723L..44R} it means 2D (radial-in-plane) distance, but the author of \cite{2019JCAP...08..019B} interpreted "r" as 3D distance, hence distorting the model significantly. Meanwhile, the same misinterpretation likely happened in \cite{2019ApJ...872..177C} too. Then the author claimed that the derived constraints rule out DM interpretation of Galactic center (GC) gamma-ray and antiproton CR excesses. However, the derived constraints are again doubtful, as it is shown below. 

\textbf{\cite{2021MNRAS.501.5692C}} utilized the radio data \cite{2014A&A...571A..61G} at 5--8 GHz and claimed a possible presence of DM contribution in the central emission, with the former being well-fitted by the thermal WIMP with $m_x \approx 30$ GeV. However, the least understandable fact in the story is that the authors seem to ignore, that they robustly ruled out such WIMP in their work \cite{2019ApJ...872..177C} couple of years ago! Thus, their two papers seem to require different WIMPs for the inner and outer parts of M31.. Besides this, their model of observational uncertainties is not transparent, e.g. a map noise and systematics are not mentioned. Also, the authors assumed the stellar density profile to be the tracer of non-DM radio emission. However, CR synchrotron must not necessarily follow the stellar density, since the former is defined by the distribution of CR accelerators and $e^\pm$ diffusion. Also, such frequency range is not optimal for WIMP detection as shown below.

Thus, this section pedagogically demonstrated that the current situation is not really satisfactory -- all the works discussed above have certain inaccuracies. Therefore, M31 still lacks the comprehensive and precise analysis. This created the motivation to build the full 2D model and derive more precise DM constraints. The comparison of newly derived constraints with those from papers above is discussed in secs. \ref{sec:gamma},\ref{sec:summary}.

\section{\label{sec:dm}Modeling the emission due to DM annihilation by GALPROP}

In this section I describe the first key step -- the computation of theoretically expected emission maps and spectra due to WIMPs annihilation. In general, there are two possible ways of this computation: one is to create a new code specifically for this purpose like we did in our previous work \cite{2013PhRvD..88b3504E}; or to adapt widely-used codes, which compute CR propagation and emission. The latter way has a big advantage: we do not need to do an extensive calibration and checking of the whole code, instead we only have to check the part modified for our purpose. Moreover, in the past we had already developed the extension for GALPROP package, which accurately computes DM annihilation. This was done in the frame of our work \cite{2016JCAP...03..060E} on DM searches in MW for version 54 of GALPROP. Later the new version 56 was released, which has many new functionalities including an arbitrary observer placement in the galaxy \cite{GP,2019ICRC...36..111M}. The latter opportunity allows in principle to apply GALPROP to \textit{any} galaxy -- not just to MW. These circumstances evidently defined my choice to employ GALPROP v56 for the task. Only two essential procedures were needed: adaptation of our DM extension for v56, and then -- v56 for M31. Our DM extension for both versions of GALPROP is made public at \cite{github} for convenient usage by astrophysical community. The detailed explanations are also provided there. The extension for v56 includes some features specifically related to M31. 

The source term in the transport equation incorporated by our extension defines injection of $e^\pm$ produced by WIMP annihilation and has the following form:
\begin{equation}\label{eq:dm-q}
q(R,E) = \frac{1}{2} \langle \sigma v \rangle \left(\frac{\rho(R)}{m_x}\right)^2 \xi(R) \frac{dN_e}{dE}(E),
\end{equation} 
where $\langle \sigma v \rangle$ denotes WIMP annihilation cross section, $\rho(R)$ is DM density distribution, $m_x$ is WIMP mass, $\xi(R)$ is the DM annihilation rate boost factor due to substructures and $\frac{dN_e}{dE}(E)$ represents the energy spectrum of $e^\pm$ per annihilation. For the latter the same spectra were used as in our previous works cited above. These spectra were taken from PPPC 4 DM ID resource \cite{PPPC,2011JCAP...03..051C}.

Modeling the emission due to DM annihilation indeed suffers from various significant uncertainties, which include and are not limited to those in DM density and MF distributions, propagation parameters of galactic environment. And it is very important to include and quantify all significant uncertainties in the procedure of DM constraints derivation. To quantify uncertainties, the popular MIN-MED-MAX paradigm (like e.g. in \cite{2021PhRvD.104h3005G}) was utilized, when MIN and MAX model setups provide respectively the lowest and highest DM emission intensities and, hence, the weakest and strongest constraints. And MED setup represents some middle or "average expectation" scenario. It is not constructed to yield exactly the average intensity between MIN and MAX. In this respect, MED is probably less informative, since MIN and MAX represent rather hard limits, and the real intensity lives somewhere between them. MIN-MED-MAX models were built separately for DM density profile and MF distribution together with the propagation parameters (MF/prop. with "/" meaning "and") -- i.e. three independent density profiles and three independent MF/prop. configurations were tested. MF and propagation parameters were treated jointly in order to reduce the computational heaviness of the task. Also, it is not so necessary to vary propagation parameters independently, since the impact of their uncertainties is much smaller than that of DM and MF distributions. Thus, for each WIMP mass and annihilation channel 3 DM density profiles $\times$ 3 MF/prop. configurations = 9 independent models were computed, providing a good coverage of the parameter space. DM intensities were computed for 13 log-evenly spaced WIMP masses over our range of interest 10--1000 GeV, which provides sufficiently dense coverage for further interpolation. And so far I computed only two representative and popular annihilation channels -- $\chi\chi \rightarrow b\overline{b}$ and $\chi\chi \rightarrow \tau^+\tau^-$. Just with two channels the total number of computed models, i.e. separate GALPROP runs, is thus $13\times 2\times 9 = 234$. Such model grid required a significant amount of computational time. For this reason, other annihilation channels were left for future work. Also, as can be seen e.g. at \cite[fig. 9]{2016JCAP...07..041C}, the exclusion limits for other channels can be expected to lie approximately between those for $b\overline{b}$ and $\tau^+\tau^-$. The next subsections describe in details all the essential ingredients needed for modeling of the emission due to DM annihilation. The dependence of thermal annihilation cross section on WIMP mass was gathered from the updated work on the subject \cite{2020JCAP...08..011S,sv}.

\subsection{\label{sec:dm-gp}Adaptation of GALPROP for M31}
The main difference between MW and M31 cases in the frame of modeling by GALPROP is the location of observer. Fortunately, GALPROP from v56 allows to set arbitrary location. However, if one would set the true distance to M31, then the map pixels would be too large to achieve an acceptable angular resolution. For this reason M31 was placed at the fake distance $d_f = 50$ kpc, and all the angular quantities were rescaled by the following relation from a simple geometry: $d\tan\alpha \approx d_f\tan\alpha_f$. The fake distance value was chosen from two obvious requirements: on one hand, it must be much larger than the relevant ROIs; on the other hand, it must be small enough to provide sufficient angular resolution about the angular size of GALPROP spatial grid step. It was checked empirically that the rescaling by relation above with the chosen value of $d_f$ does not introduce any significant intensity distortions. Thus, the introduced $d_f$ allowed to have a comfortable (not too large) number of pixels on the produced HEALPix maps.

Another aspect is the choice between 2D and 3D solution of the transport equation. Since practically all the model ingredients are 2D, this was an obvious choice to speed up computation. Also, the difference between 2D and 3D solutions was checked to be negligible for the small enough spatial grid step in 2D. The numerical values of general GALPROP parameters, which are the same for all MIN-MED-MAX setups, are written out in the Appendix.  

Then it was very important to establish/check the consistency between our previous setups \cite{2013PhRvD..88b3504E,2016JCAP...03..060E} and the current one. After long and deep investigations it was concluded that all three setups agree at the level not worse than $\approx$ 10\%, which is a very good agreement. \cite{2013PhRvD..88b3504E} used the code, which is completely independent w.r.t. GALPROP. Hence, the obtained agreement provides the mutual check between all three setups and their results.

\subsection{\label{sec:dm-rho}Modeling DM density distribution}
The first important question here one may have is whether traditional spherically-symmetric density profiles would describe M31 dark halo adequately. Besides many spherically-symmetric models, two papers were found, which claimed both significantly oblate and prolate halo shapes -- \cite{2008ApJ...685..254B} and \cite{2014ApJ...789...62H} respectively. Thus, no systematic and robust deviation from the spherical symmetry seems to be detected. Hence the spherically-symmetric DM density distributions were employed.

Many studies concerning DM distribution in M31 were considered, and \cite{2006MNRAS.366..996G,2008MNRAS.389.1911S,2009ApJ...705.1395C,2010A&A...511A..89C,2012A&A...546A...4T,2018MNRAS.481.3210B,2021ApJ...919...86B} were utilized among them. Overall, they have quite a good agreement with each other. As was outlined above, I aimed to choose three representative density profiles among many ones derived in the above papers: one "average expectation" case (MED) and two boundary cases around it, which would majorly enclose the uncertainty range (MIN/MAX). Usually, stellar and gas kinematics do not allow to deduce reliably DM distribution in the inner regions of big galaxies; since baryons dominate the potential there, non-circular motions arise etc. Thus, almost all the above studies derived DM density distribution at distances above 7-8 kpc, leaving it very poorly constrained in the inner kiloparsecs. At the same time the density in this inner region majorly defines DM constraints we are looking for, since the signal from DM is very concentrated towards the center (see e.g. \cite[fig. 5]{2013PhRvD..88b3504E}). Hence, our constraints would be very uncertain, if we would just formally extrapolate the profiles derived for an outer region into the inner region. Thus, the estimates showed that the most cored (Burkert) profile from \cite{2012A&A...546A...4T} allows thermal WIMPs with almost \textit{any} mass in our range of interest (for MED MF/prop.), while the most cuspy/contracted (Moore) profile would exclude WIMP masses up to almost 1 TeV! Fortunately, the new works \cite{2018MNRAS.481.3210B,2021ApJ...919...86B} allow to shrink reasonably such a huge uncertainty range. The team \cite{2018MNRAS.481.3210B} introduced newer data and sophisticated two-bulge 2D-model for M31, which yielded satisfactory estimates of DM density in the inner region -- fig. 12 there. Their estimates reliably exclude both extreme cases -- very shallow profile like Burkert and very contracted one like Moore. The authors of \cite{2021ApJ...919...86B} pointed out that "the giant stellar stream and the shell-like features of M31 are probably results of accretion of the satellites from highly eccentric orbits". They modeled the influence of such accretion of DM-rich satellite on the central DM cusp and found, that even if the initial DM distribution in M31 was very cuspy, the central cusp would be unavoidably washed out after the accretion. Thus, the final profile is expected to be rather cored, although with quite high central densities -- see \cite[fig. 5,8]{2021ApJ...919...86B}. Based on these results, both extremely cuspy and cored profiles were excluded from consideration. Other profiles derived in \cite{2006MNRAS.366..996G,2008MNRAS.389.1911S,2009ApJ...705.1395C,2010A&A...511A..89C,2012A&A...546A...4T,2018MNRAS.481.3210B} (with sane fit qualities) form the corresponding uncertainty range. I ranked all these profiles by central densities and chose MIN-MED-MAX, which have approximately minimal-medium-maximum central densities and, hence, would generate respective intensities of DM emission. 

The minimal profile, which would yield the most conservative DM constraints, appeared to be Einasto from \cite{2018MNRAS.481.3210B}. I took for MIN their lowest allowed density profile rather than the best fit to stay conservative. The chosen profile can be seen at \cite[fig. 12]{2018MNRAS.481.3210B} as the bottom edge of violet-shaded family of acceptable Einasto profiles. The exact definition of Einasto profile is the following:
\begin{equation}\label{eq:dm-ein}
	\rho_{\text{Ein}}(R) = \rho_{-2}\exp\left(-\frac{2}{\alpha}\left(\left(\frac{R}{R_{-2}}\right)^{\alpha}-1\right)\right),
\end{equation}
where $R_{-2}$ and $\rho_{-2}$ are the radius and density, where the profile slope (logarithmic hereafter) goes over --2 (more details can be seen in \cite[appendix B]{2012A&A...546A...4T}). The numerical values of these parameters are provided in table \ref{tab:dm}. I note that in fact the density profiles in \cite{2018MNRAS.481.3210B} are 2D and have a mild flattening q = 0.85 as a fiducial value (the definition of flattening parameter q can be seen in \cite[eqs. (22)-(23)]{2018MNRAS.481.3210B}). Such flattened DM halos are physically motivated by the gravitational pull from baryons towards the galactic plane, since the baryonic matter dominates the potential in the central region and is concentrated around the plane. However, according to estimates q = 0.85 and q = 1 (i.e. spherically-symmetric) profiles yield very close J-factors for the relevant ROIs -- the former differ just by few percents. Such small difference is negligible in our context, hence q = 1 was adopted. J-factor by definition means the following quantity:
\begin{equation}\label{eq:dm-J}
	J = \int\limits_{\text{FoV}}\int\limits_{\text{LoS}} \rho^2(\vec{r'}) d\Omega dl',
\end{equation}
i.e. this is the squared DM density integrated over the solid angle of ROI and the line of sight (primed coordinates originate at the observer's position). J-factor exactly defines the flux of prompt gamma rays due to DM annihilation (see e.g. \cite{2021arXiv210902696S,2020JCAP...02..012H} for details). The radio flux can be assumed to be roughly proportional to J-factor, too.

I would like to note that the profile above does not reproduce correctly the total halo mass, which is profile-independent and well-constrained to lie in the range $(0.7-1.3) \cdot 10^{12} M_{\odot}$ (e.g. \cite[table 5]{2012A&A...546A...4T}). The reason is that the authors \cite{2018MNRAS.481.3210B} aimed to reconstruct the density profile only up to radius of 15 kpc, and their profiles are not precise beyond this distance. However, it does not hamper to use it for our purposes; since, as shown below, DM emissivity is too weak at radial distances above $\sim$ 10 kpc and, hence, does not render much relevance there.

As the opposite MAX case, the Einasto profile derived in \cite{2012A&A...546A...4T} was employed. This profile has approximately the biggest central density allowed by \cite{2018MNRAS.481.3210B,2021ApJ...919...86B}. Also, it is well-known that besides the smooth DM component, numerous subhalos with a very wide mass spectrum may reside in the host halo and boost the signal from annihilating DM significantly. For this reason I included the signal boost due to substructure on top of the regular smooth component according to the boost factor model \cite{2010PhRvD..81d3532K} in order to get a bit optimistic DM constraints in MAX case. However, we do not expect a significant signal increase due to substructure, since the latter do not survive much in the central regions of galaxies due to tidal disruption. And DM signal is mainly localized around the center.

\begin{figure}[h]
\includegraphics[width=1\linewidth]{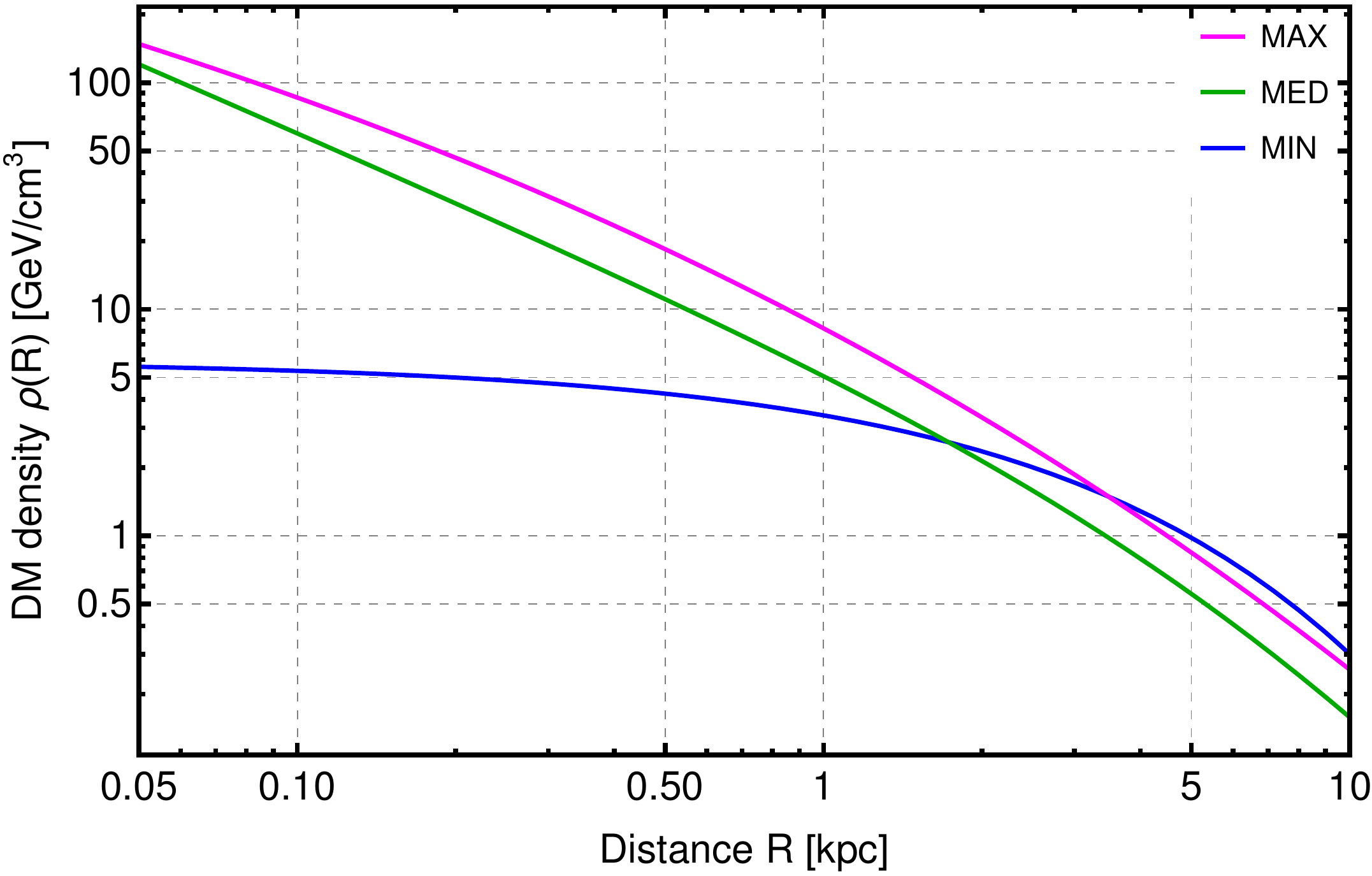}
\caption{\label{fig:dmd}DM density distributions selected as MIN-MED-MAX models. Their parameters are listed in table \ref{tab:dm}.}
\end{figure}

Then finally, we have to choose the MED density profile, which would represent approximately medium constraints between the margins provided by MIN and MAX. Such MED model can be naturally considered as more probable/expected case, or fiducial model. It was found that Navarro-Frenck-White (NFW) best-fitting profile in \cite{2018MNRAS.481.3210B} is the closest to such MED case. The exact functional form for NFW, which is employed in GALPROP, is the following: 
\begin{equation}\label{eq:dm-nfw}
	\rho_{\text{NFW}}(R) = \frac{\rho_s}{\text{Max}[R,R_{tr}]/R_s (1+\text{Max}[R,R_{tr}]/R_s)^2}.
\end{equation}
It is the traditional NFW (firstly introduced by \cite{1997ApJ...490..493N}) truncated at $R=R_{tr}$, below which the density stays constant. $R_{tr} = 0.01$ kpc was chosen, since no N-body simulations are able to resolve anything below such scale (e.g. \cite{2021ApJ...919...86B}). Hence, this truncation radius was introduced in order to stay conservative and avoid unphysical divergence at the central point. However, such small scale does not have any significant influence on the final constraints. The parameter values of all selected profiles are provided in table \ref{tab:dm}, and fig. \ref{fig:dmd} shows the density radial dependencies. As a sanity check, we can see from the figure, that all the accepted profiles reproduce well DM density at the solar radius (one may expect rather similar DM halos in M31 and MW). 

\subsection{\label{sec:dm-mf}Modeling the magnetic field distribution}
As was already mentioned in sec. \ref{sec:cr}, it is very important to model MF properly for the computation of synchrotron emission. Indeed, knowledge about MF distribution in M31 is much more limited than that in MW, which implies corresponding uncertainties for the expected emission. As with modeling of DM density, various available literature concerning MF in M31 was analyzed. Also, a similarity between M31 and MW was naturally utilized: some aspects, which were unclear for M31, were extrapolated from MW.

A basic structure of the utilized 2D MF model comprises piecewise-linear dependence on the radius $r$ multiplied by the exponential dependence along the vertical coordinate $z$. Then MED and MAX models have additional central cusps. Such a choice of the functional form was motivated by the fact that quite all MF measurements in M31 were done inside its disk, hence the typical simple exponential $z$-dependence was assumed for the halo. To model the radial profile of MF in the plane, the results of mainly two papers \cite{1998IAUS..184..351H,2004A&A...414...53F} were utilized. Thus, \cite{1998IAUS..184..351H} reported the following equipartition total field estimates in the central region: $B(0.2<r<0.4\text{ kpc}) = 15 \pm 3~\mu$G, $B(0.8<r<1.0\text{ kpc}) = 19 \pm 3~\mu$G. I took uncertainties of these and other field values in the disk as one of the factors, which defined the MIN-MED-MAX range of models. Thus, for example, for MIN $B(0.2<r<0.4\text{ kpc}) = 15-3=12~\mu$G was used, for MED -- $B(0.2<r<0.4\text{ kpc}) = 15~\mu$G and for MAX -- $B(0.2<r<0.4\text{ kpc}) = 15+3=18~\mu$G. The field value in the very center $0<r<0.2$ kpc was assumed to be equal to that in the range $0.2<r<0.4$ kpc. Then \cite[table 1]{2004A&A...414...53F} provides the field values in the radial range 6--14 kpc (so-called star-forming ring, with 4 sub-ranges) at the level $B \approx 6 \div 7~\mu G \pm 20\%$. And no reliable estimates in the range $1<r<6$ kpc seem to be available. Regarding this gap between the nucleus and ring \cite{2014A&A...571A..61G} states, that MF exists there too, but it is just not illuminated by CR $e^\pm$, which are confined in the nucleus and ring. But it is hard to measure the field value in the gap; since it does not show a radio emission, and really few background sources shine through the gap (for Faraday rotation measurements) \cite{1998A&A...335.1117H}.    

The field does not differ much among the measured ranges and, hence, changes with $r$ relatively slowly. Considering this fact, a simple connection of the measured ranges by linear interpolation was applied in order to build the full radial dependence. Such procedure yielded the continuous radial profile. Another important circumstance, which is usually not taken into account, is that the cited field values must not be considered as values at a point or plane. Instead, these are always values effectively averaged over some volume, which is inevitably non-zero due to, e.g., a finite size of telescope beam, which is used for MF measurement (through the non-thermal emission intensity or rotation measure). The size of this volume along $r$ coordinate is explicitly provided by the results cited above. Considering that the cited radial ranges are quite narrow, and the field changes slowly with $r$, this averaging effect along $r$ was neglected. However, the field changes much faster along $z$ in any spiral galaxy, i.e. the field falls down exponentially (in first approximation) with the scale $\sim$ 1 kpc. Hence, it is important to take into account this averaging effect along $z$ in order to model properly the values in the galactic plane. This requires to choose the effective averaging volume size $\pm z_{av}$ along $z$. A natural choice, also suggested by \cite{2004A&A...414...53F}, would be the characteristic thickness of the non-thermal emission layer, which can be considered as the vertical beam path length in the layer. \cite[table 1]{2004A&A...414...53F} provides the value of this thickness $z_{av} \approx 0.3$ kpc. Then by the definition of volumic average, which in our case comes down to the average over $z$ due to $B(r) \approx $ const locally, we have 
\begin{multline}\label{eq:dm-mf-av}
\langle B \rangle = \int\limits_{-z_{av}}^{z_{av}} B_{pl}\exp\left(-\frac{|z|}{z_B}\right)dz \Bigg/ \int\limits_{-z_{av}}^{z_{av}}dz = \\ = B_{pl}\frac{z_B}{z_{av}}\left(1-\exp\left(-\frac{z_{av}}{z_B}\right)\right).
\end{multline}            
Here $z_B$ means the vertical scale height, $\langle B \rangle$ -- the effectively measured field in a certain radial range, and $B_{pl}$ -- the physical value in galactic plane, which we look for to construct MF distribution. Such averaging correction in practice is relatively small. Thus, for example, the measured MED MF value cited above for the central region $\langle B \rangle = 15~\mu$G yields $B_{pl} = 17~\mu$G with the corresponding $z_B$ value. Having conducted the whole procedure described above, we can write out our basic MIN MF model, which is described by eq. (\ref{eq:dm-mf-min}) below.

Our MIN model is based on the assumption of equipartition between the field energy density and that of CR $e^\pm$. Such a model represents the minimal anticipated MF. However, according to \cite{2012SSRv..166..215B} the equipartition assumption may significantly underestimate the real field in irregular dense environments like galactic nuclei with fast $e^\pm$ cooling. This was evidently demonstrated in the work \cite{2010Natur.463...65C} for the nucleus of our Galaxy. These authors considered an approximately ellipsoidal region around the GC with semi- major and minor axes $(r_1,z_1)$ = (0.42, 0.14) kpc. The simple equipartition assumption gives the field strength of just $\sim 10~\mu$G inside for fiducial model parameter values. However, the independent comprehensive multi-wavelength analysis revealed the necessary minimal field strength of 50 $\mu$G (effective volumic average), with a possibility for the real field to be even at the level $\approx 100~\mu$G (see also \cite{2011MNRAS.411L..11C})! In principle, as \cite{2010Natur.463...65C} says, equipartition may also yield 100 $\mu$G, but with extremal parameter values. Such situation may naturally exist in M31 nucleus too. For this reason it was decided to introduce the same MF "cusps" as in MW into MED and MAX distributions. Such MF profiles are somewhat analogous to DM density profiles in the aspect of cusp or core presence in the central region. MF cusps were embedded into MED and MAX profiles through the following way. MF was assumed to be constant with the strength 50 and 100 $\mu$G for MED and MAX profiles correspondingly inside the ellipsoidal region with semiaxes $(r_1,z_1)$ = (0.42, 0.14) kpc similar to MW. Indeed, in reality, MF likely has some internal non-uniform structure inside this region. However, it would be too speculative to assume any specific non-uniformity. For this reason, the effective volumic-average values according to the results of \cite{2010Natur.463...65C} were set. This cusp is embedded into MF outside the central region, where the environment is less dense, and the equipartition field is expected to be a better estimate. Hence, this MF outside can be described by the same piecewise-linear on $r$ and exponential on $z$ model discussed above for MIN profile. However, it is necessary to connect smoothly the central cusp with its surroundings avoiding unphysical field discontinuity. To do this the transition zone was introduced, where the field is naturally assumed to transform gradually from the central non-equipartition regime down to the equipartition "state" outside. This transition zone was assumed to be restricted from outside by the ellipsoidal surface too with semiaxes $(r_2,z_2) = 2 \times (r_1,z_1)$ = (0.84, 0.28) kpc. From inside it is indeed restricted by the central cusp of the constant field. Thus, the transition zone can be imagined as the ellipsoidal layer around the concentrical ellipsoid with constant field. As for the field distribution in the transition zone, it was defined through the linear interpolation between field values on the boundary surfaces of the zone. Such interpolation reflects the smooth transition mentioned above. The internal surface possesses indeed the central field value 50 or 100 $\mu$G. And the external surface in turn -- the corresponding piecewise-linear on $r$ and exponential on $z$ values. Eqs. (\ref{eq:dm-mf-med})-(\ref{eq:dm-mf-max}) below summarize the structure of MED and MAX models with the formal mathematical description of the ellipsoidal surfaces. 
\begin{widetext}
\begin{equation}\label{eq:dm-mf-min}
	B_{\mu \text{G}}^{\text{MIN}}(r_\text{kpc},z) = \exp \left(-\frac{|z|}{z_B}\right) \times \begin{cases}
	14, \text{ if } r < 0.4; \\
	12r+9, \text{ if } 0.4 \leqslant r < 0.8; \\
	19, \text{ if } 0.8 \leqslant r < 1; \\
	-2.1r+21, \text{ if } 1 \leqslant r < 6; \\
	-0.36r+11,	\text{ if } r \geqslant 6.
	\end{cases}
\end{equation} 
\begin{equation}\label{eq:dm-mf-med}
B_{\mu \text{G}}^{\text{MED}}(r_\text{kpc},z) = \begin{cases}
50, \text{ if } r < r_1,~|z| < z_1\sqrt{1-(r/r_1)^2}; \\
\text{transition zone}~\downarrow, \text{ if } r < r_2,~\text{Max}[0,z_1^2(1-(r/r_1)^2)] \leqslant z^2 < z_2^2(1-(r/r_2)^2); \\
~\exp \left(-\frac{|z|}{z_B}\right) \times \begin{cases}
17, \text{ if } r < r_1, |z| \geqslant z_2\sqrt{1-(r/r_2)^2}; \\
9.1r+13, \text{ if } r_1 \leqslant r < r_2, |z| \geqslant z_2\sqrt{1-(r/r_2)^2}; \\
21,	\text{ if } r_2 \leqslant r < 1; \\
-2.4r+23, \text{ if } 1 \leqslant r < 6; \\
-0.27r+11, \text{ if } r \geqslant 6. 
\end{cases}
\end{cases}
\end{equation}
\begin{equation}\label{eq:dm-mf-max}
B_{\mu \text{G}}^{\text{MAX}}(r_\text{kpc},z) = \begin{cases}
100, \text{ if } r < r_1,~|z| < z_1\sqrt{1-(r/r_1)^2}; \\
\text{transition zone}~\downarrow, \text{ if } r < r_2,~\text{Max}[0,z_1^2(1-(r/r_1)^2)] \leqslant z^2 < z_2^2(1-(r/r_2)^2); \\
~\exp \left(-\frac{|z|}{z_B}\right) \times \begin{cases}
20, \text{ if } r < r_1, |z| \geqslant z_2\sqrt{1-(r/r_2)^2}; \\
9.5r+16, \text{ if } r_1 \leqslant r < r_2, |z| \geqslant z_2\sqrt{1-(r/r_2)^2}; \\
24,	\text{ if } r_2 \leqslant r < 1; \\
-2.8r+27, \text{ if } 1 \leqslant r < 6; \\
-0.35r+12, \text{ if } r \geqslant 6. 
\end{cases}
\end{cases}
\end{equation}  
\end{widetext}
"$\downarrow$" sign above means the transition between the zones described by the rows above and below the sign.

Now let us discuss the choice of vertical scale height $z_B$. As a typical rule  \cite{2008A&A...480...45S,2012SSRv..166..133H} suggest that $z_B \approx 4h_s$ with $h_s$ being the synchrotron emission scale height. The latter is provided by \cite[table 1]{2004A&A...414...53F}. Then naturally I set the following $z_B$ values for MIN-MED-MAX MF models: $z_B = 4 \times$(Min[$h_s$],$\langle h_s \rangle$,Max[$h_s$]) = (0.88, 1.2, 1.6) kpc, where Min/Max attributes to the radial and frequency variations of $h_s$. Thus, M31 has the relatively thin synchrotron and magnetic disks in comparison with other galaxies, as it is also stated in e.g. \cite{2019Galax...8....4B} as a consequence of the peculiar absence of so-called thick synchrotron disk.

GALPROP internally averages $e^\pm$ synchrotron emission intensity over directions regardless of the field direction. This makes the synchrotron emission isotropic from any volume of interest. Thus, such model effectively assumes MF to be purely random, i.e. the total field has the strength (\ref{eq:dm-mf-min})--(\ref{eq:dm-mf-max}) and a random direction at every point. Indeed, the real field has both regular and random components. However, this is a very challenging task to model both these components realistically, and the outcome will be likely model-dependent. A direct application of the comprehensive MW MF models to M31 like Jansson and Farrar one \cite{2012ApJ...757...14J}, which include both regular and random components, is not particularly helpful here, since such models reconstruct the field on large spatial scales and have a relatively poor information about the central kiloparsec, which plays a key role for our purposes.

At this point we have defined the MF distribution in M31. Indeed, the real distribution is likely more sophisticated. Thus, an obvious sophistication could be, for example, the dependence of scale height $z_B$ on $r$. However, the chosen model should at least cover reasonably the possible uncertainty range for the finally expected DM constraints. The key MF model parameters are written out in table \ref{tab:dm}.

\subsection{\label{sec:dm-prop}Propagation parameters}
Another important set of parameters attributes to the diffusion of DM $e^\pm$ in both coordinate and energy space. Let us begin the modeling of this set from defining the size of diffusion cylinder, inside which we solve the transport equation by GALPROP. The cylinder radius is usually assumed to be about the radius of a stellar disk. However, in our case the $e^\pm$ source is very concentrated towards the center, and $r_{\text{max}}$ = 15 kpc was checked and set as absolutely sufficient radius. As for the vertical size $z_{\text{max}}$ -- i.e. the half-height of the diffusion cylinder, it is naturally linked to the vertical extent of MF, since the latter holds $e^\pm$ inside the diffusion region. For this reason usually a linear proportionality is assumed between $z_{\text{max}}$ and $z_B$, and I set $z_B = \delta \times z_{\text{max}}$ according to e.g. \cite[eq. (4.5)]{2012JCAP...01..005F}. $\delta$ here denotes the energy-dependence slope of the diffusion coefficient defined as follows:
\begin{equation}\label{eq:dm-D}
D(E) = D_0 \left(\frac{E}{\text{GeV}}\right)^\delta. 	
\end{equation}
Thus, now we have to determine the diffusion coefficient parameters $D_0$ and $\delta$. Here the results of detailed study \cite{2013MNRAS.435.1598B} were utilized. This study examined the CR $e^\pm$ propagation in M31 and M33. These authors estimated the diffusion coefficient at several GeV and reported the results in their table 5. They found a slight anisotropy of the diffusion coefficient: it is larger by $\approx$ 2 times along the galactic plane than in the orthogonal direction. However, GALPROP v56 allows only isotropic diffusion, hence the coefficients were averaged over directions. The estimates in \cite{2013MNRAS.435.1598B} are too approximate to recover $\delta$ reliably. For this reason, the data on diffusion parameters in MW were also employed -- particularly from \cite{2021PhRvD.104h3005G}. And $\delta$ values were extrapolated from there. This work derived the updated diffusion parameters for DM searches in MW in the frame of MIN-MED-MAX paradigm. These authors outlined the parameter values for various propagation setups. These setups differ by the functional form of the diffusion coefficient, presence of reacceleration and convection. The closest setup to our GALPROP is called "QUAINT" in their notation. The respective parameter values are provided in \cite[table VII]{2021PhRvD.104h3005G}, which yields $\delta$ = 0.4--0.5. These values reasonably agree with the results of \cite{2013MNRAS.435.1598B} and, therefore, were adopted for M31. Having set $\delta$, $D_0$ values were derived by eq. (\ref{eq:dm-D}) using $D(E)$ values from \cite[table 5]{2013MNRAS.435.1598B}. MIN-MED-MAX $D_0$ values were calculated respectively as (Min[$D_0$],$\langle D_0 \rangle$,Max[$D_0$]), i.e. the range is formed by all possible values obtained in \cite{2013MNRAS.435.1598B} by various methods and observational frequencies. All the accepted parameter values for M31 are written out in table \ref{tab:dm}. The accepted $D_0$ values are rather close to those for MW cited in \cite[table VII]{2021PhRvD.104h3005G} -- the values for M31 are lower by 2--3 times.

Another important aspect of the propagation model is a possible reacceleration of $e^\pm$ by magnetohydrodynamic (MHD) waves. The reacceleration together with energy losses defines the diffusion in energy space. GALPROP defines the reacceleration coefficient through essentially the single parameter $V_A/\sqrt{w}$ with $V_A$ being Alfven speed and $w$ -- the ratio of energy densities of the MHD waves and MF. However, it is very difficult to model such fine effects as reacceleration in M31 -- no papers were found on this topic. Another difficulty is a lack of knowledge about $w$ and the thickness of layer around the galactic plane, where reacceleration works. Also, there are some doubts whether the reacceleration is modeled fully correctly by GALPROP, since substitution of the typically recommended value $V_A$ = 30 km/s (e.g. \cite{2017A&A...597A.117D}) yields a very abnormal increase of our synchrotron emission intensity. Taking all these into account, it would be too rough to assume any specific reacceleration in our model. At the same time "QUAINT" setup for MW mentioned above requires significant $V_A$ for all MIN-MED-MAX cases. In such uncertain situation the following conservative choice was made: the reacceleration was switched off in our MIN and MED models, and was included only into MAX model with the minimal possible $V_A$ = 5 km/s according to the recipe in \cite[appendix B]{2021PhRvD.104h3005G} (assuming $w$ = 1). Such a conservative choice ensures that our emission intensities of interest would not be overestimated. And the convection of $e^\pm$ was not included too, as uncertain and insignificant effect: "QUAINT" setup suggests tiny convective speeds $V_c \sim 0.1$ km/s. $\eta$ parameter from \cite[eq. (B1)]{2021PhRvD.104h3005G} was fixed to $\eta = -2$ and checked to be irrelevant at the energies of interest. 

Finally, we have to discuss probably the most important propagation aspect -- energy losses of $e^\pm$ in the galactic environment. These losses are composed by synchrotron, ICS, bremsstrahlung, Coulomb and ionization losses. As was discussed above in sec. \ref{sec:cr}, many studies used incorrect expressions for these losses. Therefore, below I provide the verified expressions for the key loss components $b \equiv -dE/dt$ in \textbf{GeV/s} as they are implemented in GALPROP based mainly on \cite{Ginzburg}:
\begin{align}\label{eq:dm-loss-s}
	b_{\text{synch}} &= 6.60\cdot 10^{-25}B_{\mu\text{G}}^2(\gamma^2-1), \\ \label{eq:dm-loss-i}
	b_{\text{ICS}} &= 2.66\cdot 10^{-23}U_{\text{eV/cm}^3}(\gamma^2-1), \\ \label{eq:dm-loss-b}
	b_{\text{brems}} &\simeq 7.1\cdot 10^{-20}\gamma n_e(\ln\gamma+0.36), \\ \label{eq:dm-loss-c}
	b_{\text{Coul}} &\simeq 7.6\cdot 10^{-18}n_e(\ln(\gamma/n_e)+73);	
\end{align}	
where $\gamma$ is $e^\pm$ Lorentz factor, $U$ is the energy density of ambient radiation field and $n_e$ is the concentration of ambient (fully-) ionized gas in $\mathbf{\textbf{cm}^{-3}}$. MF strength for the synchrotron losses is defined by eqs. (\ref{eq:dm-mf-min})--(\ref{eq:dm-mf-max}). As for the ISRF energy density $U$ in (\ref{eq:dm-loss-i}), its determination is highly non-trivial. In general, ISRF comprises three major components: CMB emission, which is the same everywhere with $U_{\text{CMB}}$ = 0.26 eV/cm$^3$; the infrared emission from a galactic dust and the starlight. As for the spatial profiles of energy densities of the last two components, they were adopted from MW (taken from GALPROP). Then ISRF densities were globally rescaled (i.e., the profiles were renormalized) according to the individual properties of M31. GALPROP easily allows to do such rescaling of each component by an arbitrary factor w.r.t. the typical density in MW. Let us begin this discussion from modeling the infrared rescaling factor. Generally, as e.g. \cite{2012IAUS..284..112G} says, M31 is a relatively dust-poor galaxy with low infrared luminosity. Hence, we expect a lower infrared radiation density in comparison with that in MW. In order to estimate the rescaling factor quantitatively, the infrared intensities of the central regions ($(r,|z|)\leqslant 0.5$ kpc) of MW and M31 were obtained. The central region was chosen as the most important again, since DM is highly confined there. The infrared intensity of the GC region was extracted from DIRBE-COBE and IRIS missions data in the frequency range 1--5 THz. The intensity of M31 nucleus was taken from the dedicated paper \cite{2015A&A...582A..28P} by Planck collaboration. Then the ratio of intensities was obtained -- M31 appeared to be fainter by 10--30 times depending on frequency. Assuming approximate proportionality between the emission intensity and energy density, I set the infrared ISRF energy density factor for M31 to be 0.1 for all MIN-MED-MAX cases. Overall, this factor has a relatively little influence (as well as CMB) on the total ICS losses. But this is opposite for the optical ISRF density, which dominates in ICS losses around the galactic center -- i.e. in the bulge region. To compare the optical bulge intensities of MW and M31, the information from \cite{1989AJ.....97.1614K,1999A&ARv...9..273V,2009A&A...505..497Y,2015ApJ...809...96L} on the bulge luminosities were utilized. The ratio of luminosities would be equal approximately to the ratio of intensities, since according to \cite[table 5]{1999A&ARv...9..273V} and \cite[table 1]{2009A&A...505..497Y} the bulges of MW and M31 have very similar size. These tables clearly demonstrate that M31 bulge is more luminous than MW bulge. Combining the values of M31 and MW bulge masses, their mass-to-light ratios, optical luminosities of both galaxies and their bulge light fractions collected from \cite{1989AJ.....97.1614K,1999A&ARv...9..273V,2009A&A...505..497Y,2015ApJ...809...96L}; the ratio of M31 to MW bulge luminosity (and hence intensity) was obtained to be $\approx 2.0 \pm 0.5$ depending on a filter band. In the absolute units it corresponds to $U_{\text{opt}} \approx (12-25)$ eV/cm$^3$ in M31 center. Thus, we see that the optical radiation energy density in M31 central region is significantly higher than that in MW, which would imply in turn faster cooling of DM $e^\pm$, therefore lower synchrotron intensity and, unfortunately, weaker DM constraints. The optical ISRF density factor was set to 2.5 for MIN scenario, 2.0 for MED and 1.6 for MAX to cover the potential uncertainty range. This choice provides the same increment 1.25 between MIN and MED, MED and MAX: $2.5=2.0\times 1.25=1.6\times 1.25^2$.

At this point we have defined the synchrotron and ICS energy losses, which play the main role in the overall loss budget. Then in order to determine bremsstrahlung, Coulomb and ionization losses (which are subdominant) I had to model and substitute into GALPROP the spatial distributions of the ionized, atomic and molecular gas in M31. As for the first one, combining the relevant information from \cite{2004A&A...414...53F,2008A&A...480...45S}, the following approximate profile was set:
\begin{equation}\label{eq:dm-ne}
	n_e(r_{\text{kpc}},z_{\text{kpc}}) = 0.03 \exp\left(-\left(\frac{r}{10}\right)^2-\frac{|z|}{1}\right) \text{cm}^{-3}.
\end{equation}
The description of neutral gas distributions and other secondary parameters are placed into Appendix in order to not overload the main body of paper.
\begin{table}[h]
	\caption{\label{tab:dm}Parameter values of the employed models of DM density and MF distributions, and propagation.}
	\begin{ruledtabular}
		\centering
		\begin{tabular}{cccc}
			Model & MIN & MED & MAX \\
			\hline
			\multicolumn{4}{c}{DM density profile} \\
			\hline
			Type & Einasto & NFW & Einasto \\
			Scale radius $R_{-2(s)}$, kpc & 8.9 & 10.4 & 17.4 \\                     
			Scale density $\rho_{-2(s)}$, $\frac{\text{GeV}}{\text{cm}^3}$ & 0.383 & 0.585 & 0.0892 \\
			$\alpha$-slope & 0.73 & -- & 0.17 \\
			Central cusp & -- & + & + \\
			Substructure boost & -- & -- & + \\
			\hline
			\multicolumn{4}{c}{MF distribution} \\
			\hline
			Central cusp & -- & + & + \\
			Central field value, $\mu$G & 14 & 50 & 100 \\
			Vertical scale $z_B$, kpc & 0.88 & 1.2 & 1.6 \\ 
			\hline
			\multicolumn{4}{c}{Propagation parameters} \\
			\hline
			Half-height of diff. & & & \\
			cylinder $z_{\text{max}}$, kpc & 2.2 & 2.7 & 3.2 \\
			Diffusion coefficient & & & \\
			at 1 GeV $D_0$, cm$^2$/s & $10 \cdot 10^{27}$ & $6 \cdot 10^{27}$ & $2 \cdot 10^{27}$ \\
			Diff. coeff. slope $\delta$ & 0.4 & 0.45 & 0.5 \\
			Alfven speed $V_A$ for & & & \\			
			reacc. coeff., km/s & 0 & 0 & 5 \\			
			Optical ISRF density & & & \\
			factor w.r.t. MW & 2.5 & 2.0 & 1.6 \\
		\end{tabular}
	\end{ruledtabular}
\end{table}

\subsection{\label{sec:dm-a}Emission absorption}
It is important to take into account possible emission absorption along the line of sight. We estimated in \cite[appendix A]{2013PhRvD..88b3504E} that the synchrotron self-absorption is not relevant in the circumnuclear environment of M31 at the frequencies of interest. Then we have to estimate the thermal free-free absorption both in M31 and MW. For both GALPROP was used -- it has the functionality to model both free-free emission and absorption by an ionized gas. In order to estimate the absorption in MW by GALPROP the observer was placed at the Sun's position, and the test source was generated outside MW in the direction of M31. Apriori, we do not expect a significant free-free absorption in this direction, since M31 is located quite far from the GC in the sky -- M31 Galactic coordinates can be seen in table \ref{tab:i-par}.  The level of absorption was checked with a couple of representative ionized gas distributions for MW contained in GALPROP. The emission intensity decreases by no more than $\approx 1.5\%$ at our lowest frequency of interest 74 MHz. The free-free absorption coefficient indeed decreases with frequency increase. Therefore, the absorption inside MW was justifiably neglected.

The free-free absorption inside M31 was basically modeled by three entities: the thermal electrons' spatial distribution, their temperature and clumping factor. The first entity is our ionized gas distribution (\ref{eq:dm-ne}) -- this is fixed. The electrons' temperature was fixed to the value 7000 K according to \cite[fig. 1]{2013A&A...557A.129T}. Then the electrons' clumping factor was adjusted as a free parameter to reproduce the global M31 free-free emission flux density, which is indeed produced by the same ionized gas. Thus, the thermal emission measurements help to model the thermal absorption. The former was taken from \cite[fig. 15]{2015A&A...582A..28P} and \cite[fig. 3]{2019ApJ...877L..31B} at the frequency about 0.5 GHz. These independent studies found approximately the following flux densities of free-free emission from the whole galaxy at the mentioned frequency: 0.9 Jy and 0.5 Jy respectively. Then the electrons' clumping factor in GALPROP was fitted to the value 21, which provides approximately the average free-free flux between the cited values. This value of clumping factor lies inside the expected range 10--100 suggested by GALPROP manual \cite{GPs}. The corresponding free-free emission flux density generated by GALPROP can be seen in fig. \ref{fig:sed}. The described procedure yielded the synchrotron emission absorption at the level 0.5--1.0\% for the relevant ROIs with $R$ = 1--2 kpc at 74 MHz. Thus, the free-free absorption inside M31 can be neglected too, though it was still included in calculations at no cost. However, in general, it is important to model it carefully at $\nu \lesssim 50$ MHz. Another very important aspect here is that a significant absorption may break the linear proportionality between DM $e^\pm$ synchrotron intensity and annihilation cross section. This linearity in turn is fundamental for the whole procedure of DM constraints derivation. Here the absorption was included into the calculations by the simple single model, since the absorption is tiny and, hence, does not require much attention in the frequency range of interest. Our range lies essentially in the domain, where the optically transparent medium approximation works. However, one should be very careful in the case of going outside the optical transparency regime.

At this point we have discussed all essential galactic model parameters and their uncertainties. The constructed MIN-MED-MAX configurations are expected to form the comprehensive uncertainty range for the final DM constraints. Uncertainties of the model parameters, which are not varied by MIN-MED-MAX, have a negligible contribution into the final uncertainties.

\section{\label{sec:maps}Obtained emission maps and spectra}

\begin{figure*}[t]
	\includegraphics[width=1\linewidth]{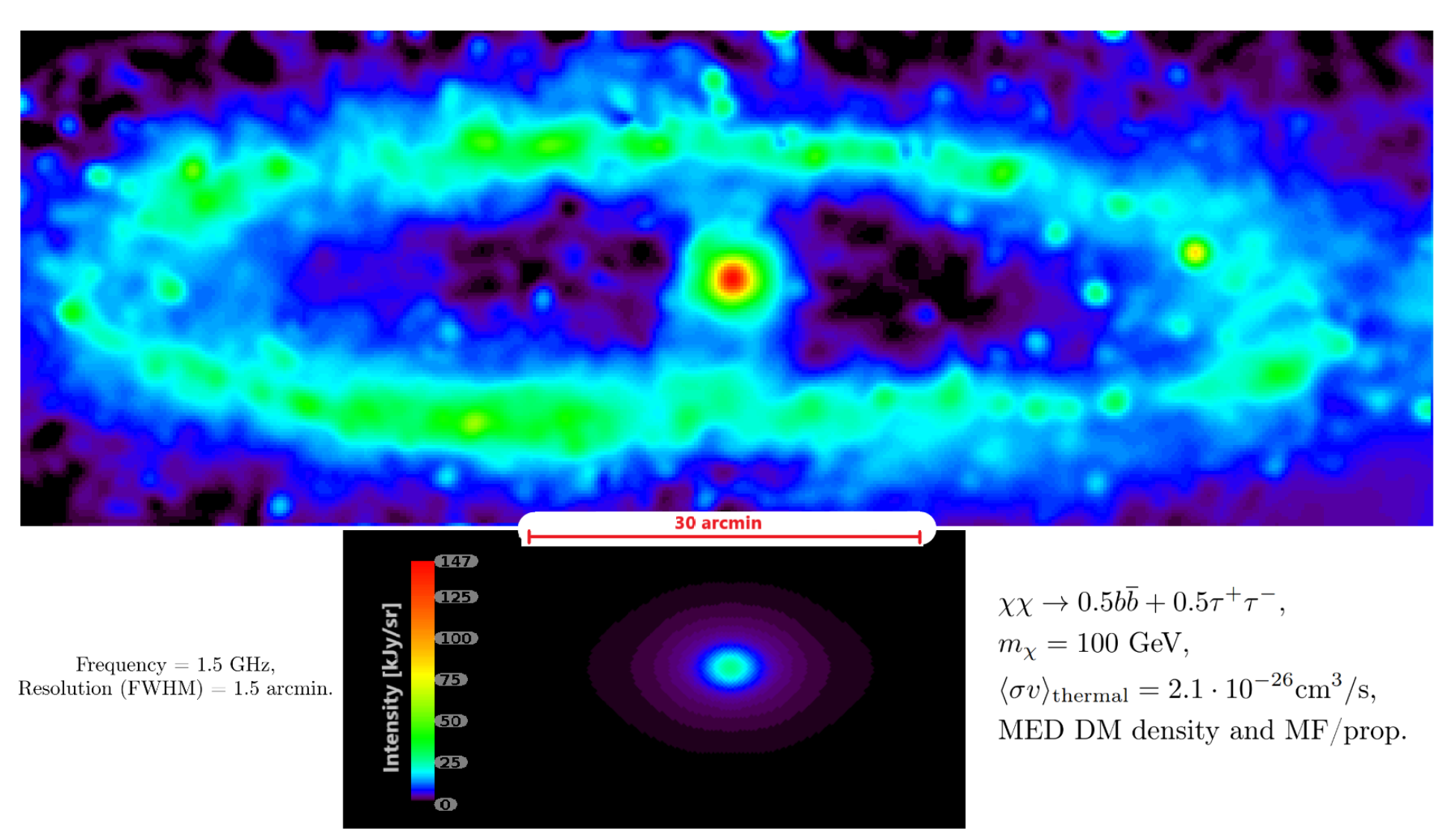}
	\caption{\label{fig:map}\textit{Top:} the observed non-thermal emission (source-subtracted) map of M31 at 1.5 GHz from \cite[+ priv. comm.]{1998A&AS..129..329B}. \textit{Bottom:} the intensity map due to DM annihilation computed by GALPROP for the specified configuration of parameters. 1 arcmin corresponds to $R \approx 0.22$ kpc.}
\end{figure*}

Combining together all the ingredients described above GALPROP solved the transport equation for each of 234 DM models and generated the respective synchrotron intensity maps in HEALPix format at 8 frequencies of interest listed in table \ref{tab:radio}. Fig. \ref{fig:map} depicts the example of representative DM intensity map in comparison with the observed non-thermal diffuse emission map of M31 at 1.5 GHz. A benchmark thermal WIMP with $m_x = 100$ GeV annihilating evenly to $b\overline{b}$ and $\tau^+\tau^-$ is assumed there (MED configuration). At the non-thermal (top) image we see very bright central region and the well-known star-forming ring around $r \approx 10$ kpc. The central emission has a remarkably circular shape. This can not be said about DM emission morphology: it clearly shows a mild ellipticity, which reflects primarily the asymmetry of MF distribution in $r$ and $z$ directions. The observed intensity in the center is larger by $\approx$ 5 times than that on DM map, which preliminary indicates that it is rather tricky to constrain even 100 GeV WIMP mass scale. DM emission intensity decreases fast with the angular distance from center waning down completely beyond $\approx 15'$. This fast decrease reflects the steepness of annihilation rate profile, which is proportional to $\rho^2$.

\begin{figure}[H]
	\includegraphics[width=1\linewidth]{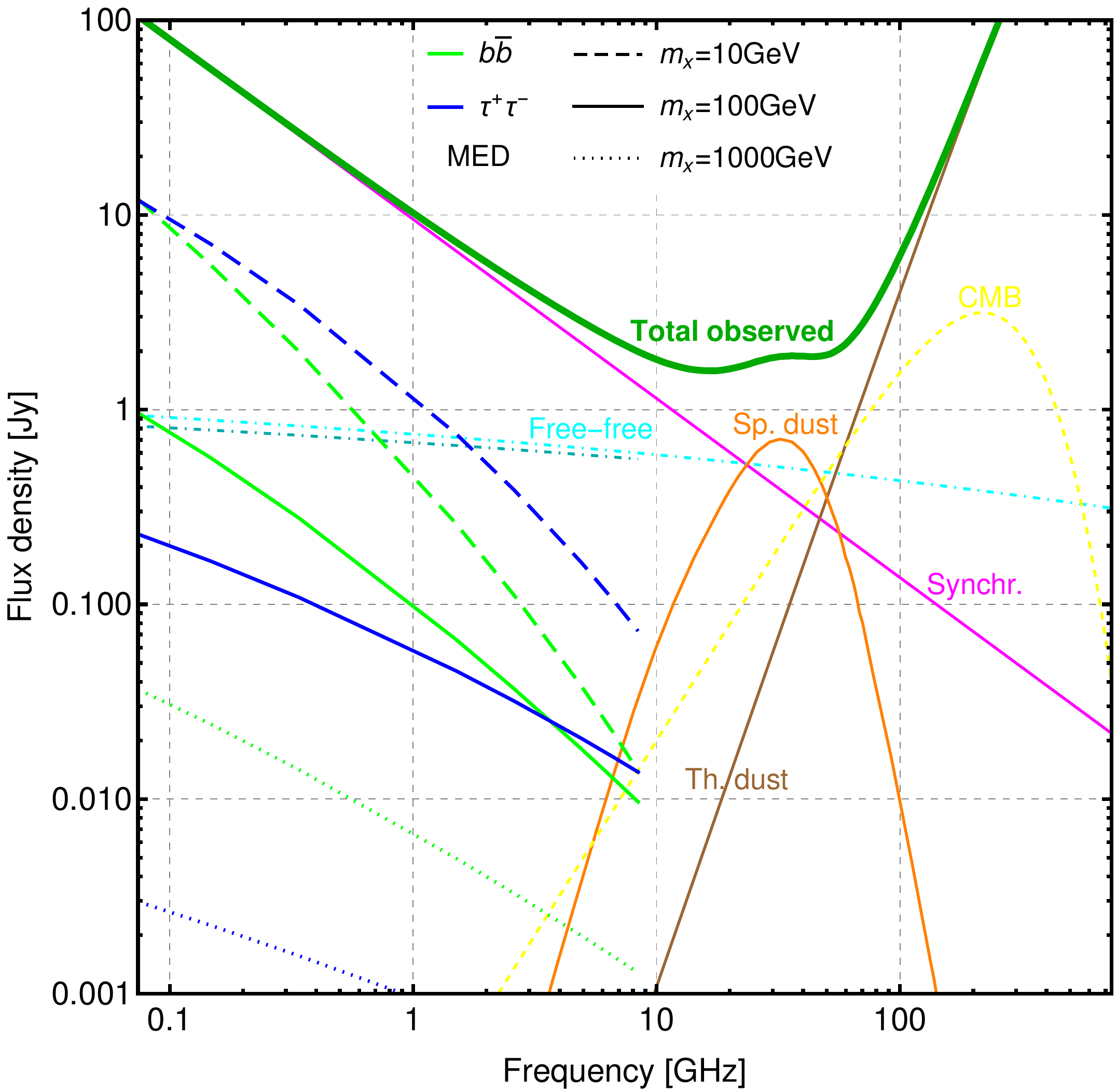}
	\caption{\label{fig:sed}The flux density spectrum of the whole M31 galaxy. The total observed emission spectrum and all its specified components (except DM) are taken from \cite[fig. 15]{2015A&A...582A..28P}. The green and blue lines show our representative DM models commented on the plot. DM annihilation cross section is thermal, the density profile and MF/prop. model are MED. The truncated dot-dashed line shows the free-free spectrum obtained by GALPROP in our calculations.}
\end{figure}

Fig. \ref{fig:sed} displays the emission spectra integrated over the whole galaxy (ROI radius is $\approx 2\degree$). The separation of all regular emission components was done in the dedicated study of M31 by Planck collaboration \cite{2015A&A...582A..28P}. Their results generally agree with the similar independent study \cite{2019ApJ...877L..31B}. Then the spectra of specified benchmark DM models were superimposed on the plot to compare them with various emission components. First of all, we can see that DM component expectedly decreases fast with the mass increase. The ratio of fluxes between $b\overline{b}$ and $\tau^+\tau^-$ channels depends both on the mass and frequency. Also, we see that the slope of DM spectra is rather similar to that of the total synchrotron. DM component (if it exists) would contribute to the synchrotron spectrum drawn by the magenta line. Hence, it is optimal to search DM at frequencies, where the synchrotron dominates the total emission, in order to avoid a contamination by other components. The synchrotron absolutely dominates up to $\approx$ 1 GHz, then free-free adds slightly between 1 and 10 GHz. Thus, we can infer approximately the optimal frequency range for DM detection/constraining: below $\approx$ 1 GHz is the best situation, where we can work with the total emission; then between 1 and 10 GHz it is preferential to remove the free-free component; and it does not make sense to go beyond 10 GHz, since many other components arise there, but DM rapidly faints. This justifies my choice of the frequency range $\approx$(0.1--10) GHz for derivation of constraints: below this range the free-free absorption hinders, above -- DM signal drowns among other emissions. Also, we see that working with the emission integrated over the whole galaxy would not be able to yield any useful constraints: even the lightest WIMP with $m_x = 10$ GeV produces the flux by an order of magnitude lower than the observed synchrotron. Hence, the ROI optimization is necessary.

\begin{figure}[H]
	\includegraphics[width=1\linewidth]{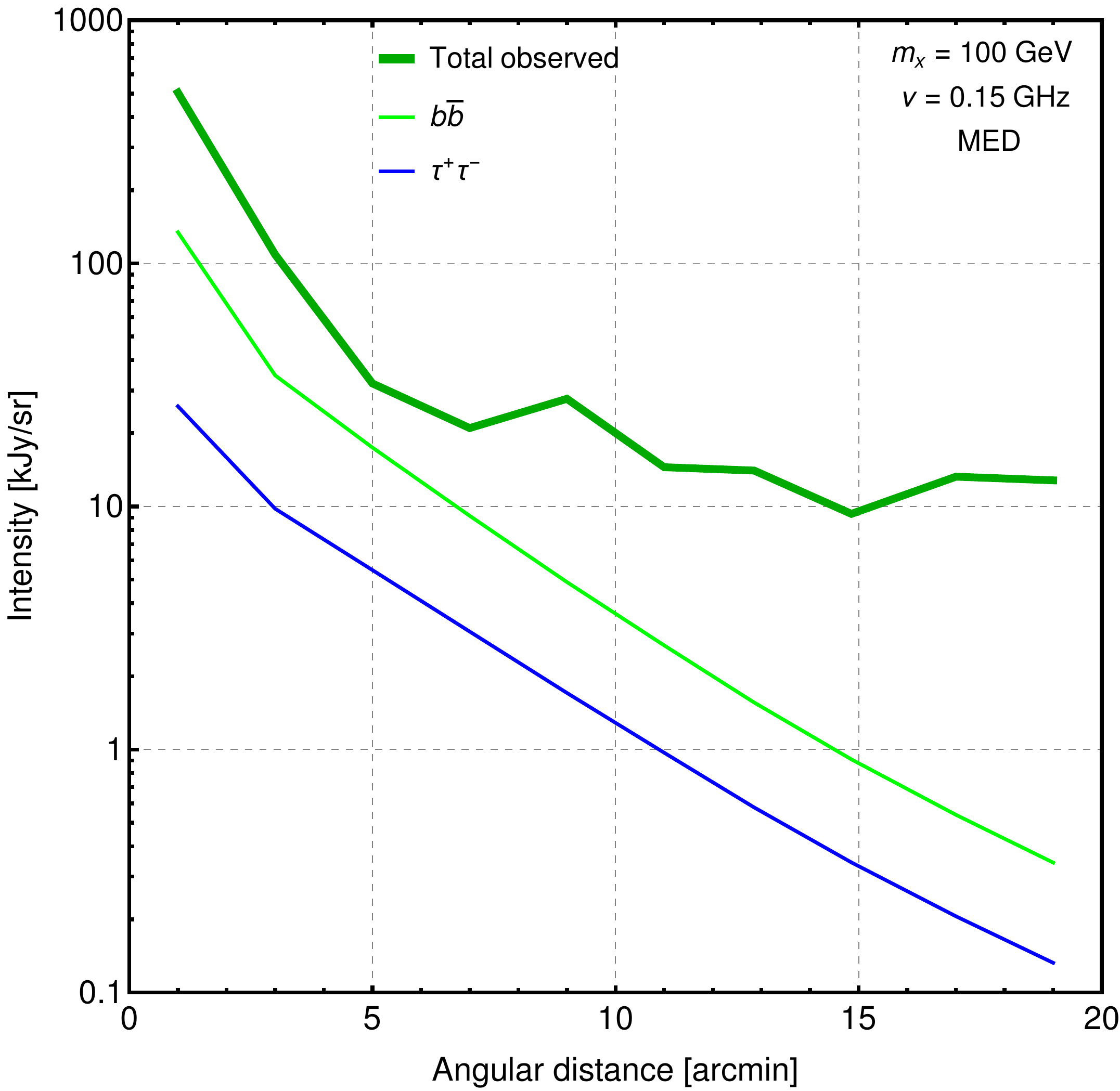}
	\caption{\label{fig:I(a)} The radial dependence of the emission intensity at the frequency 0.15 GHz. The intensity is averaged over concentric annuli with 2$'$ width. An estimate of the total observed emission was extracted from LoTSS maps \cite{2022A&A...659A...1S}. DM annihilation cross section is thermal, the density profile and MF/prop. model are MED. Beyond 5$'$--10$'$ the estimated observed emission intensity is very approximate due to the limitations of LOFAR ability to see large scales. }
\end{figure}

Fig. \ref{fig:I(a)} shows the azimuthally-averaged radial emission intensity profiles at the frequency of LOFAR Two-meter Sky Survey (LoTSS). As pointed out above, only the synchrotron exists at such low frequencies. At angular radii above 5$'$--10$'$ the observed emission intensity is an approximate estimate, since LOFAR does not see completely diffuse structures with such big sizes according to \cite{2022A&A...659A...1S}. We see that DM emission is expectedly highly-concentrated around the center; and its potential contribution into the total emission decreases strongly beyond 10$'$--15$'$, especially counting that LOFAR does not see all the intensity there. Then at large radii $\gtrsim 20'$, the bright star-forming ring enters the ROI, and the DM signal -- to -- non-thermal noise ratio drops even further. Thus, the plot suggests that the optimal ROI radii for DM searches are likely not larger than 10$'$--15$'$. Also, we may notice that the radial profiles of different channels are very parallel.

\begin{figure}[h]
	\includegraphics[width=1\linewidth]{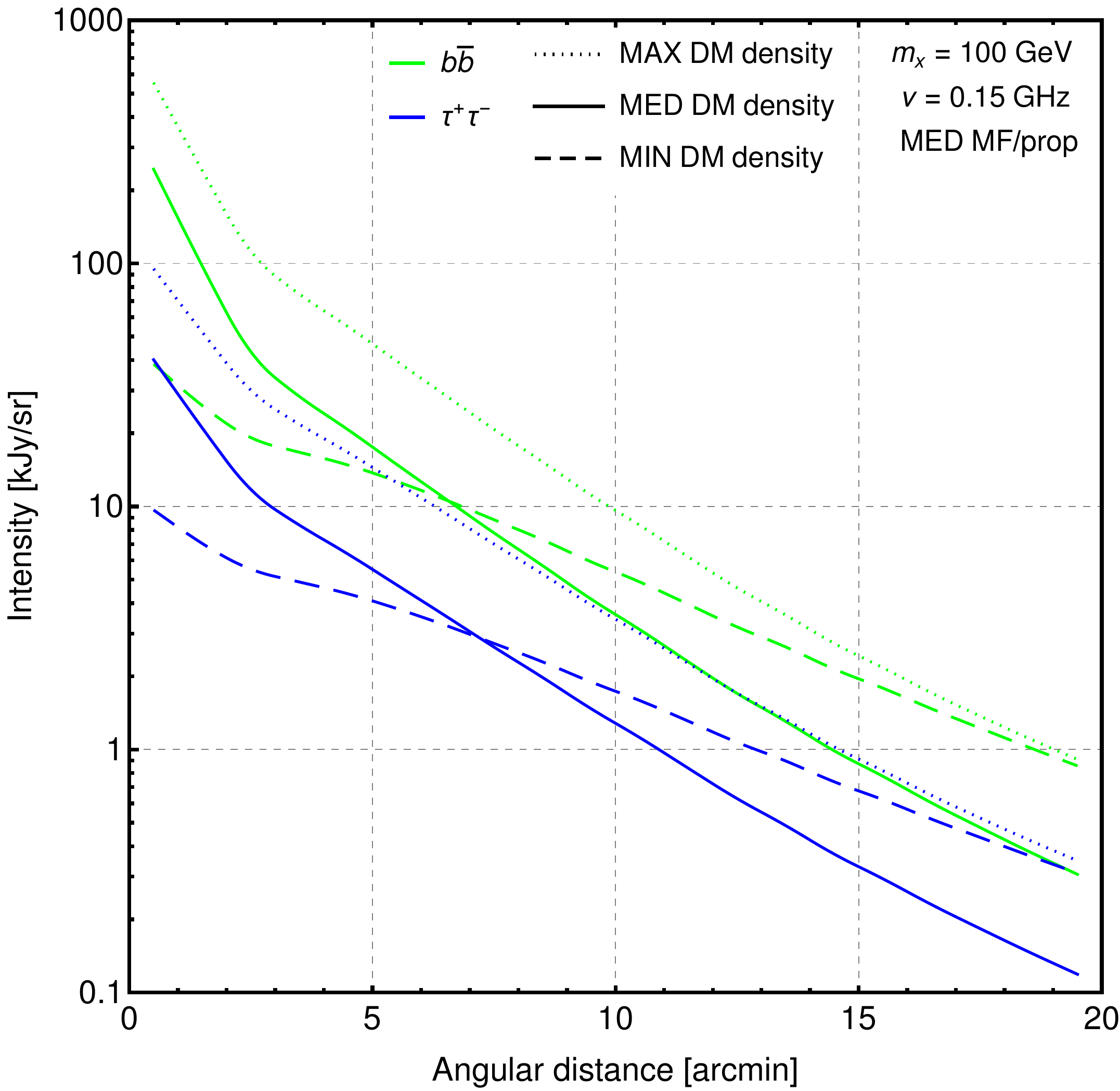}
	\caption{\label{fig:I(a)dmd}The radial DM intensity distribution dependence on DM density profile for the specified benchmark configurations.}
\end{figure}

\begin{figure}[t]
	\includegraphics[width=1\linewidth]{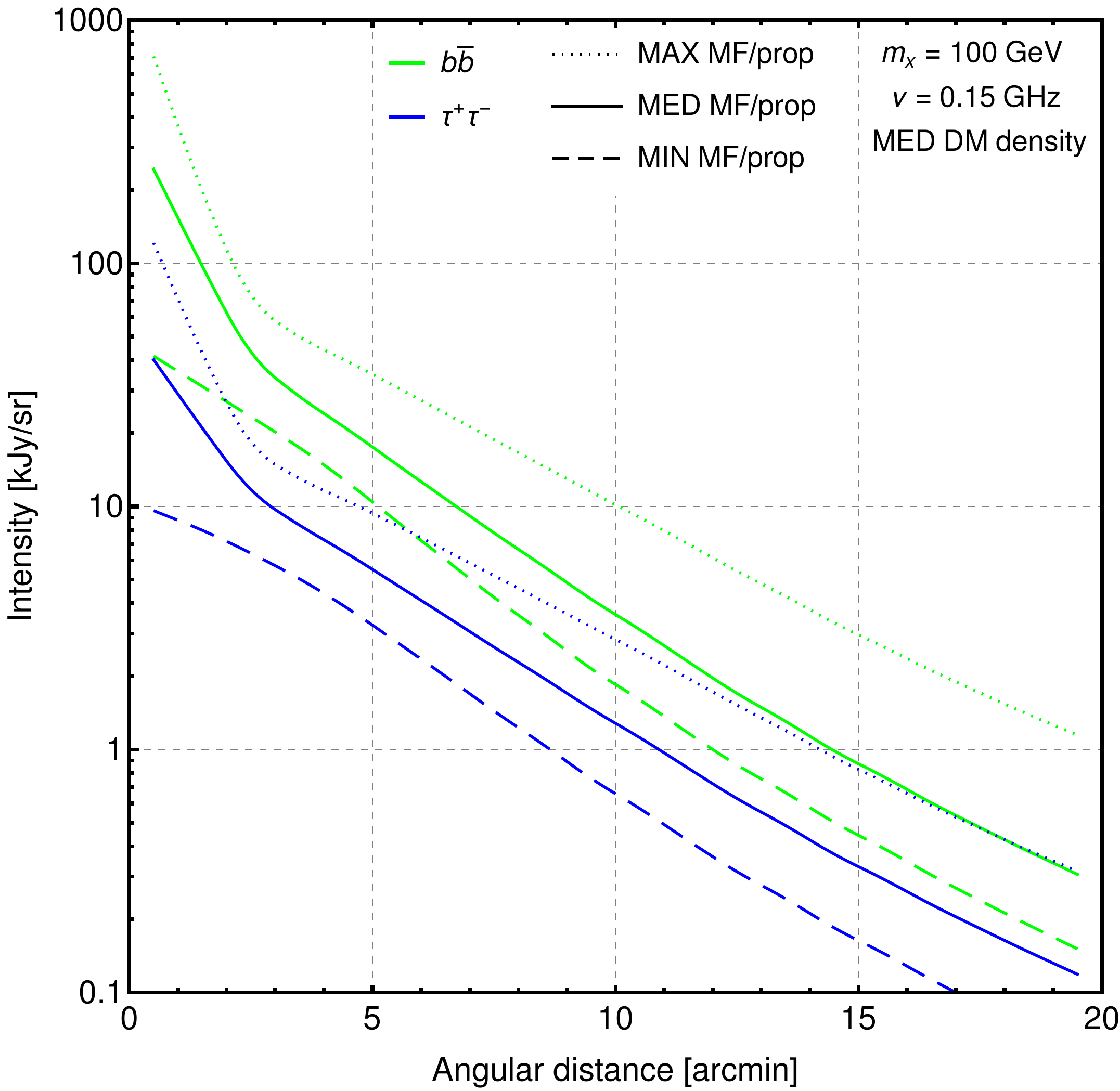}
	\caption{\label{fig:I(a)mfp}The radial DM intensity distribution dependence on MF/prop. model for the specified benchmark configurations.}
\end{figure}

Then figs. \ref{fig:I(a)dmd} and \ref{fig:I(a)mfp} illustrate the dependence of radial profiles on DM density distribution and MF/prop. setup respectively for both channels. 
\noindent Regarding the dependence on density profile, MED and MAX curves are quite parallel and differ mainly by their amplitudes. But MIN intensity decreases slower: in the center it is the lowest, and at 20$'$ it is almost the highest among three. This likely reflects the fact that MIN density overtakes both MED and MAX at $R \gtrsim 3$ kpc, which is visible at fig. \ref{fig:dmd}. Regarding the dependence on MF/prop. model, all three curves here are rather parallel. The main difference is due to the presence or absence of the central MF cusp: MED and MAX cases elevate up the intensity steeply in the cusp zone. And MIN evolves with the distance more smoothly. The ratio of MAX to MED intensity is comparable to the ratio of MED to MIN.

The main inference from this section is the optimal frequency and spatial domains for the derivation of DM constraints, which is described in the next section.

\section{\label{sec:constr}Derivation of DM constraints from various radio data}

For comparison of the theoretically calculated DM maps with the observed maps, all the relevant radio data on M31 was thoroughly collected in the frequency range of interest $\approx$(0.1--10) GHz. The selected radio surveys and targeted observations are listed in table \ref{tab:radio} with their characteristics. We can see that a large variety of high-quality data is currently available, which evenly covers our frequency range. Ideally, one would prefer to have an image, which is cleared off both projected discrete sources and thermal diffuse emission (especially above 1 GHz) by the image authors. Availability of such images among the selected sample is marked by "+" in the last column of the table. Just for 3 (out of 8) frequency channels such images were found. However, in principle the presence of nuisance emissions would spoil the produced constraints towards a weaker side (i.e., more conservative), hence it is not dangerous. Also, as was explained above, the free-free component is negligible below $\approx$ 1 GHz. Let us now briefly discuss the data at each frequency. 0.074 GHz survey (VLSSr) represents the redux of VLSS (VLA Low-frequency Sky Survey). Being interferometric sky survey, it inevitably has a limitation of the largest visible angular scale, which is 36$'$ according to the survey description \cite{2014MNRAS.440..327L}. Such a scale is sufficient for our purposes. Then 0.15 GHz is very new and high-quality LOFAR Two-meter Sky Survey (LoTSS). Its largest scale is estimated to be $\approx 10'$ from the description \cite{2022A&A...659A...1S}. Another peculiarity of this survey is a very wide band. In order to count the finite bandwidth effect, DM maps around this frequency were averaged over the band assuming the power-law spectra. The rest of data is targeted imaging by (mainly) single-dish antennas. These maps have a typical systematic intensity uncertainty around 5\%. When the thermal emission subtraction is performed, it inevitably introduces additional systematics. Thus, at 1.5 and 4.9 GHz, the systematic uncertainty is assumed to be 10\% according to \cite[table 3]{2020A&A...633A...5B} (for small ROIs). And at 6.6 GHz I utilized only the synchrotron flux from the central disk with 6$'$ radius provided in \cite[table 7]{2021A&A...651A..98F}, and the large systematics of 28\% here reflects the total uncertainty of their component separation procedure. At 4.9 and 8.4 GHz there are two independent data sets -- with low and high resolution. The latter is produced by combining the observations made on Effelsberg telescope and VLA.
\begin{table*}[t]
	\caption{\label{tab:radio}Characteristics of the radio images of M31 used for derivation of constraints. The last column indicates the availability of emission intensity cleaned of the thermal emission.}
	\begin{ruledtabular}
		\centering
		\begin{tabular}{cccccccc}
	Frequency, & Ref. & Bandwidth, & Targeted obs. or & Beam FWHM, & RMS noise, & Syst. uncert., & Non-thermal  \\
	    GHz    &      & GHz        & sky survey name  & arcmin     &  mJy/beam  & \%             & int. avail.  \\
	\hline
	0.074 & \cite{2007AJ....134.1245C,2014MNRAS.440..327L} & 0.0016 & VLSSr & 1.25 & 100 & 10 & -- \\
	0.15 & {\cite{2022A&A...659A...1S}} & 0.048 & LoTSS & 0.33 & $\approx$ 0.25 & 10 & -- \\
    0.34 & {\cite{2013A&A...559A..27G}} & $\approx$ 0.07 & Targeted & 5 & 3 & 8 & -- \\
    1.5 & {\cite{1998A&AS..129..329B}} & $\approx$ 0.02 & Targeted & 1.5 & 0.3 & $\approx$ 10 & + \\
    2.6 & {\cite{2020A&A...633A...5B}} & 0.07 & Targeted & 4.4 & 1.2 & $\approx$ 5 & -- \\
    4.9 & {\cite{2020A&A...633A...5B,2014A&A...571A..61G}} & 0.3 & Targeted & 0.25, 3 & 0.0075, 0.35 & $\approx$   5,10 & --,+ \\        
    6.6 & {\cite{2021A&A...651A..98F}} & 1.2 & Targeted & 2.9 & 0.43 & 28 & + \\ 
    8.4 & {\cite{2020A&A...633A...5B,2014A&A...571A..61G}} & 1.1 & Targeted & 0.25, 1.4 & 0.0055, 0.25 & $\approx$ 5 & -- \\    
		\end{tabular}
	\end{ruledtabular}
\end{table*}

Then let us discuss the construction of ROIs for the derivation of constraints. ROIs must satisfy several basic requirements: 1) they must not overlap in order to enter the likelihood function independently; 2) their size along any direction must not be smaller than the image resolution (beam FWHM); 3) and they must not be larger than the largest survey visible scale discussed above. According to these requirements, the circular disk around the center and several concentric annuli around it were utilized at each frequency. Such structure looks like $\circledcirc$ as an example of disk and one annulus. The circular ROIs are very convenient in practice, since they are invariant on the position (azimuthal) angle in the sky plane. The angular radius of disk and the width of annuli were set to be about the respective image resolution at each frequency in order to cover the radial emission profile densely, since apriori it is not known which radial range yields the strongest constraints. The radius of the biggest (most outer) annulus is defined by either the survey scale limitation or the distance, where the ratio of DM intensity to the observed intensity becomes too low. Thus, for example, 0.15 GHz (LOFAR) channel has the following ROIs: the central disk with 1$'$ radius and 4 concentric annuli with 1$'$ widths covering up to 5$'$. In total 33 ROIs were utilized (from 1 to 5 at each frequency) covering the distances up to 15$'$ from the center. Beyond 15$'$ the bright star-forming ring begins to enter, worsening significantly the signal-to-noise ratio. Meanwhile, the inner 6$'$ is almost free from the projected nuisance sources. One may also use in principle the off-center ROIs. The dark gaps between the nucleus and ring well visible at fig. \ref{fig:map} (top) are particularly attractive in this respect. It was calculated that these gaps (especially the right one) actually provide very high signal-to-noise ratio; i.e. DM constraints, which are even stronger than those derived from the nuclear region. However, as was explained in details in sec. \ref{sec:dm-mf}, these regions lack any direct MF measurements. MF there was defined in our model just by linear interpolation between the nuclear and ring regions. Hence, it would be too speculative and unreliable to claim strong constraints based on the gap region only. And I made the decision to utilize only the central ROIs, where MF was more or less certainly measured. 

Meanwhile, the following alternative selection of the central ROIs was also tested. This selection uses the disks only. Firstly, at each frequency the optimal disk is being searched, i.e. the disk with the radius, which provides the best sensitivity. Then one such optimal ROI for each frequency is being substituted in the joint likelihood. In this case the latter would have just 8 ROIs in total instead of 33 as in the method described above. However, this strategy with disks only demonstrated a worse sensitivity for all DM models. Hence, the strategy with 33 ROIs was employed.

Another important step in the preparation of ROI intensities for substitution into the final likelihood was the map smoothing in order to take into account the angular resolution at each frequency. The effective pixel size of DM maps generated by GALPROP before smoothing is $\approx 0.3'$. Then smoothing by Gaussian beam was applied to each map (except very high-resolution maps) with the corresponding beam FWHM (listed in the 5th column of table \ref{tab:radio}). 

Now we can move to the construction of the likelihood function $L(\vec{c}|\vec{\Theta})$, which represents the probability density for the parameter vector $\vec{\Theta}$ in case of the observational data $\vec{c}$. $\vec{c}$ in our case is the observed mean intensities inside ROIs, and $\vec{\Theta}$ contains the annihilation cross section and various nuisance parameters. The likelihood (assumed to be Gaussian) takes into account both the statistical and systematical uncertainties of the observed intensities -- $L = L_{\text{stat}}\times L_{\text{syst}}$:
\begin{equation}\label{eq:L}
L \propto \prod_{i=1}^{8\text{freq.}} \prod_{j=1}^{\#\text{ROI}_i} \exp\left(-\frac{n_{ij}^2}{2\sigma_{ij}^2}\right)\times \exp\left(-\frac{(s_{ij}-c_{ij})^2}{2\sigma_{s,ij}^2}\right),
\end{equation}
where $n_{ij}$ is the map noise inside $j$-th ROI at $i$-th frequency with RMS $\sigma_{ij}$, $s_{ij}-c_{ij}$ is the systematical offset between the measured mean intensity $c_{ij}$ and the true one $s_{ij}$. The systematic map uncertainties listed in the 7th column of table \ref{tab:radio} (let us denote them by $u_i$) represent essentially the sky-averaged values. However, the systematic offset in any specific ROI is not known apriori. Hence, it is essentially the random variable, which is being modeled by the Gaussian distribution with $\sigma_{s,ij} = u_i c_{ij}$. Then the total offset-corrected intensity is composed by the following terms: $s_{ij} = n_{ij}+a_{ij}+w_{ij}$, where $a$ is all the standard astrophysical emissions (mainly CR synchrotron and free-free) and $w$ is the potential DM contribution. The noise RMS for an arbitrary ROI is calculated as $\sigma_{ij} = \sigma_{m,i}/\sqrt{N_{ij}}$ with $\sigma_{m,i}$ being the map noise in Jy/beam listed in the 6th column of table \ref{tab:radio} and $N_{ij}$ -- the number of beams inside $j$-th ROI. A reminder here just in case: all the calculations are done in terms of the ROI-averaged intensity (i.e. in Jy/sr), not the flux density [Jy]. Then in the absolute units $\sigma_{ij} = \sigma_{m,i}/\sqrt{\Omega_{b,i}\Omega_{ij}}$, where $\Omega$ is the ROI size in sr. Thus, in order to convert all the map intensities and noises into telescope-independent units, one has to calculate the effective beam sizes $\Omega_{b,i}$. For this purpose the Gaussian telescope beam was assumed: 
\begin{multline}\label{eq:omega}
\Omega_{b,i} = \int\limits_{4\pi}\exp\left(-\frac{\theta^2}{2\sigma_{b,i}^2}\right)d\Omega = \\ = 2\pi\int\limits_0^\pi \exp\left(-\frac{4\ln2~\theta^2}{\text{ FWHM}_i^2}\right)\sin\theta d\theta 
\end{multline}
with the beam FWHM being listed in the 5th column of table \ref{tab:radio}. 

After likelihood construction, it was an important step to decide which statistical framework to apply for the derivation of constraints -- bayesian or frequentist. The latter employs Wilks' theorem, and the constraints are derived from the test statistic TS = $2\ln(L/L_0)$. However, as was meaningfully pointed out in \cite{2020JCAP...02..012H}, Wilks' theorem strictly applies only if the hypothesis being tested does not lie at the boundary of parameter space, meaning that the null likelihood $L_0$ with $\langle \sigma v \rangle = 0$ is not suitable. A violation of this applicability condition causes a deviation of TS from $\chi^2$ distribution and finally may spoil the constraints by several times towards weaker side, as was shown in \cite{2020JCAP...02..012H}. Hence, I employed the bayesian inference, when the likelihood function is being marginalized over nuisance parameters inferring the probability density for the variables of interest. Similar methodology was utilized in our previous work \cite[sec. III]{2013PhRvD..88b3504E}. The nuisance parameters in our case are the intensities of usual astrophysical emission $a_{ij}$ and the systematic-free observed intensities $s_{ij}$. An important decision is what to assume for $a_{ij}$. As was intended initially, the constraints being derived here must be conservative and minimally model-dependent. Hence, the flat prior uncorrelated between frequencies was assumed for $a_{ij}$, meaning that \textit{any} astrophysical intensities are equally allowed. The prior for $s_{ij}$ is flat too. With flat priors, we can straightly marginalize the likelihood (\ref{eq:L}) to get the probability density for the annihilation cross section (other DM parameters are fixed):
\begin{multline}\label{eq:Lm}
L(\langle \sigma v \rangle | \vec{c}) \propto \int\limits_{-\infty}^\infty \int\limits_{0}^\infty \prod_{i,j} \exp\left(-\frac{(s_{ij}-a_{ij}-w_{ij}(\langle \sigma v \rangle))^2}{2\sigma_{ij}^2}\right) \\ \times \exp\left(-\frac{(s_{ij}-c_{ij})^2}{2\sigma_{s,ij}^2}\right)ds_{ij}da_{ij}.	
\end{multline}
The integral above was calculated analytically yielding:
\begin{equation}\label{eq:Lmi}
L(\langle \sigma v \rangle | \vec{c}) \propto \prod_{i=1}^{8\text{freq.}} \prod_{j=1}^{\#\text{ROI}_i} \left(  1+\text{erf}\left(\frac{c_{ij}-w_{ij}(\langle \sigma v \rangle)}{\sqrt{2(\sigma_{ij}^2+\sigma_{s,ij}^2)}}\right)\right),  	
\end{equation}
where erf$(x)=2/\sqrt{\pi}\int_0^x e^{-t^2}dt$. The double product above has 33 multipliers in total. Then the following equation derives the constraint on $\langle \sigma v \rangle$ at the typical 95\% confidence level:
\begin{equation}\label{eq:95}
\int\limits_0^{\langle \sigma v \rangle _{\text{lim}}} L(\langle \sigma v \rangle | \vec{c}) d\langle \sigma v \rangle \Bigg/ \int\limits_0^\infty L(\langle \sigma v \rangle | \vec{c}) d\langle \sigma v \rangle = 0.95.
\end{equation}
The limiting cross section $\langle \sigma v \rangle _{\text{lim}}$ was found numerically from the eq. above individually for each of 234 DM models computed by GALPROP. These models enter (\ref{eq:Lmi}) through the mean intensities $w_{ij}$, which are linearly proportional to $\langle \sigma v \rangle$.
\begin{figure*}[t]
	\includegraphics[width=0.497\linewidth]{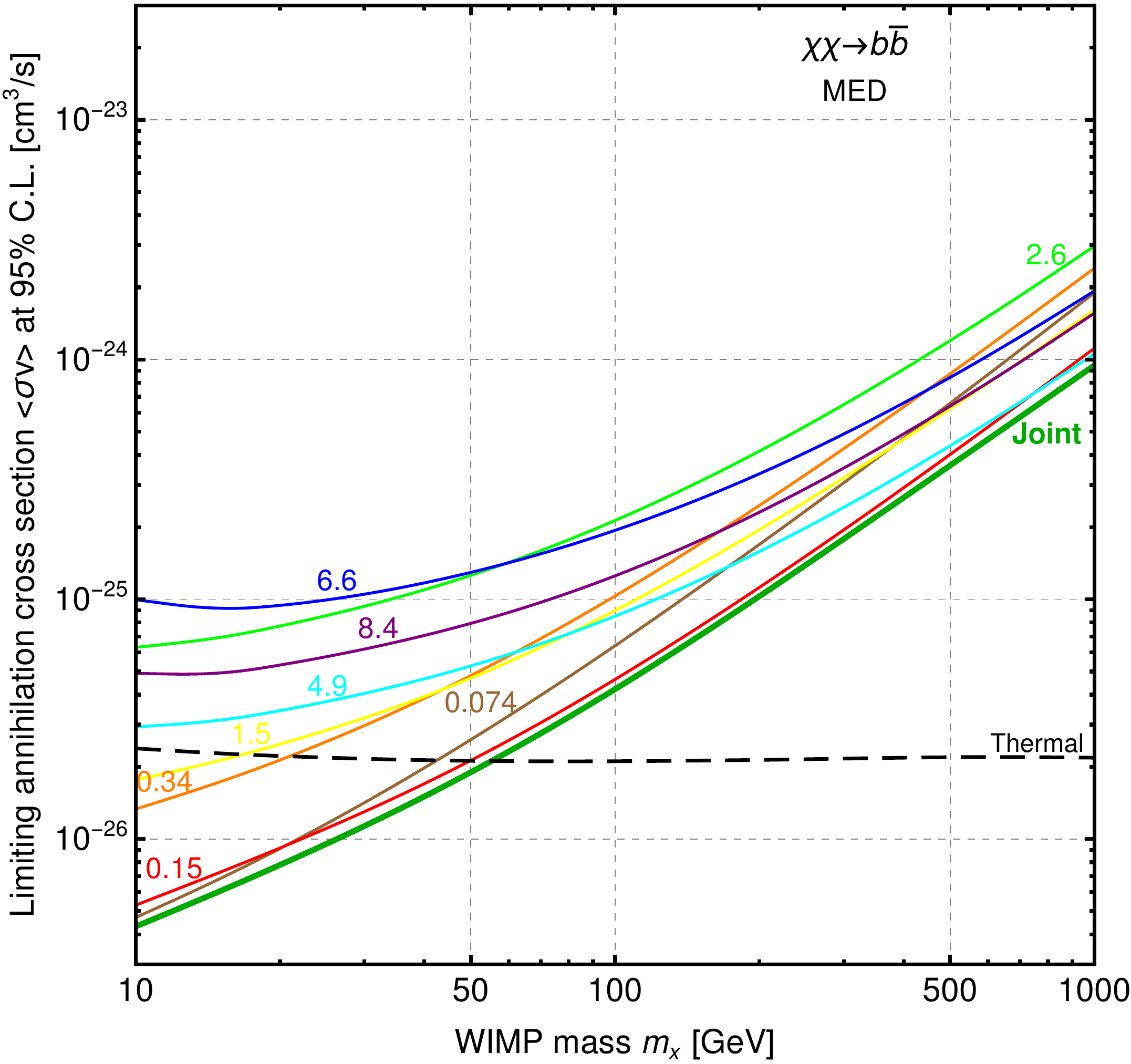}
	\includegraphics[width=0.497\linewidth]{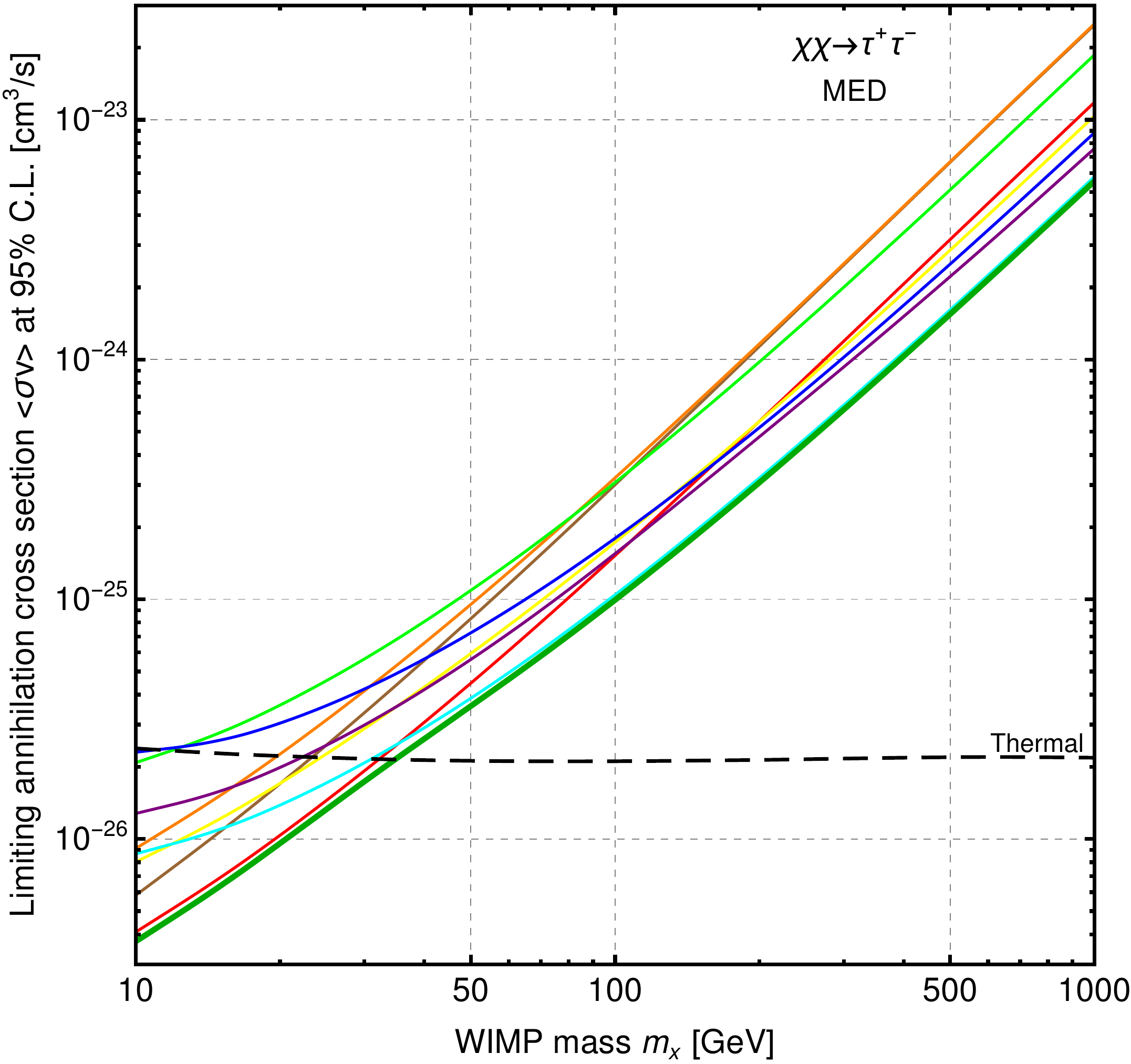}
	\caption{\label{fig:svf}The limits on annihilation cross section vs. WIMP mass for both annihilation channels (MED DM density and MF/prop.) split by frequency, which is marked in GHz near each curve. The thick green line represents the joint exclusion from all frequencies together. The excluded cross section values are above the lines. The (almost) horizontal dashed line shows the thermal relic cross section.}
\end{figure*}
\begin{figure*}[t]
	\includegraphics[width=0.497\linewidth]{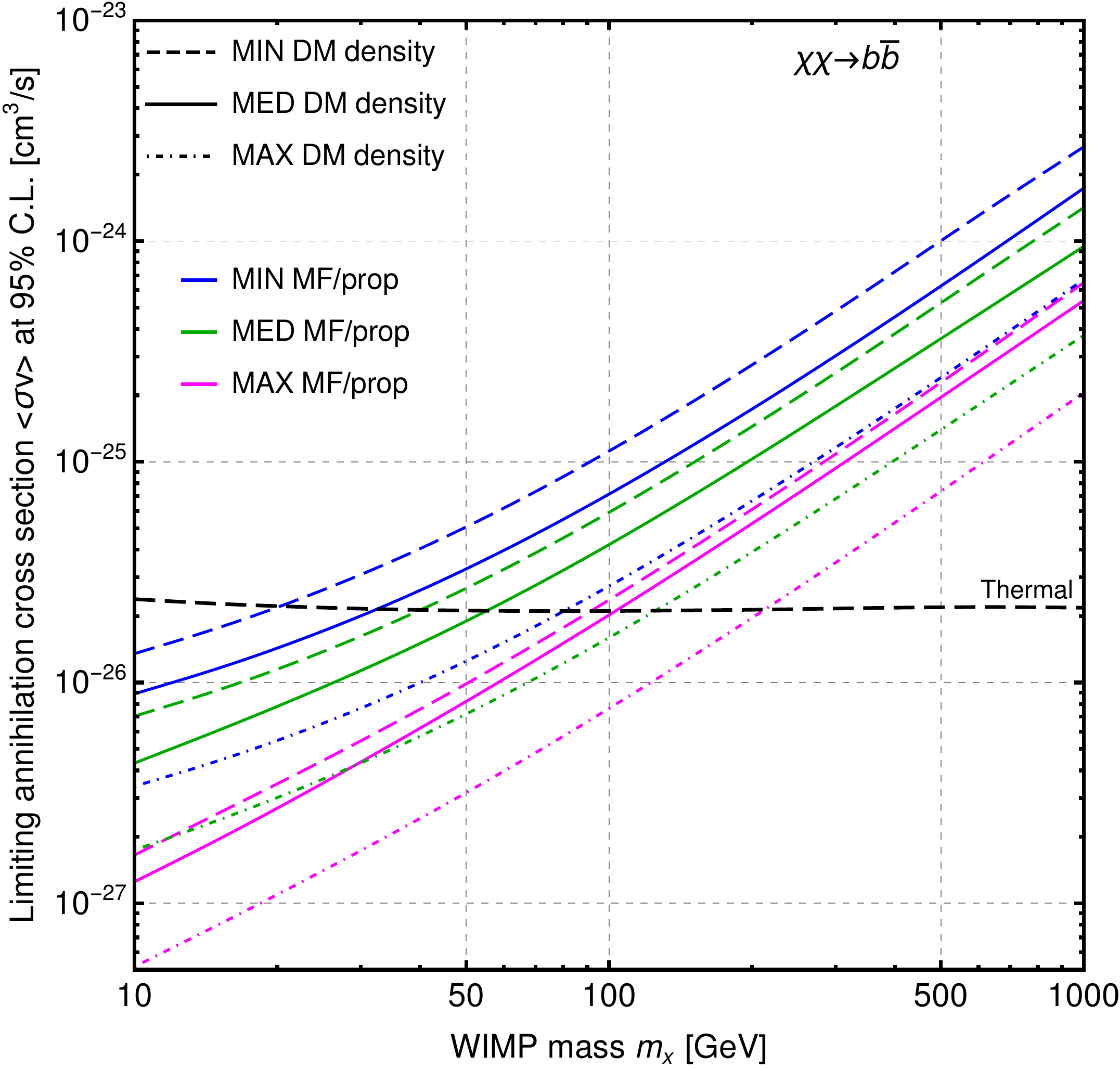}
	\includegraphics[width=0.497\linewidth]{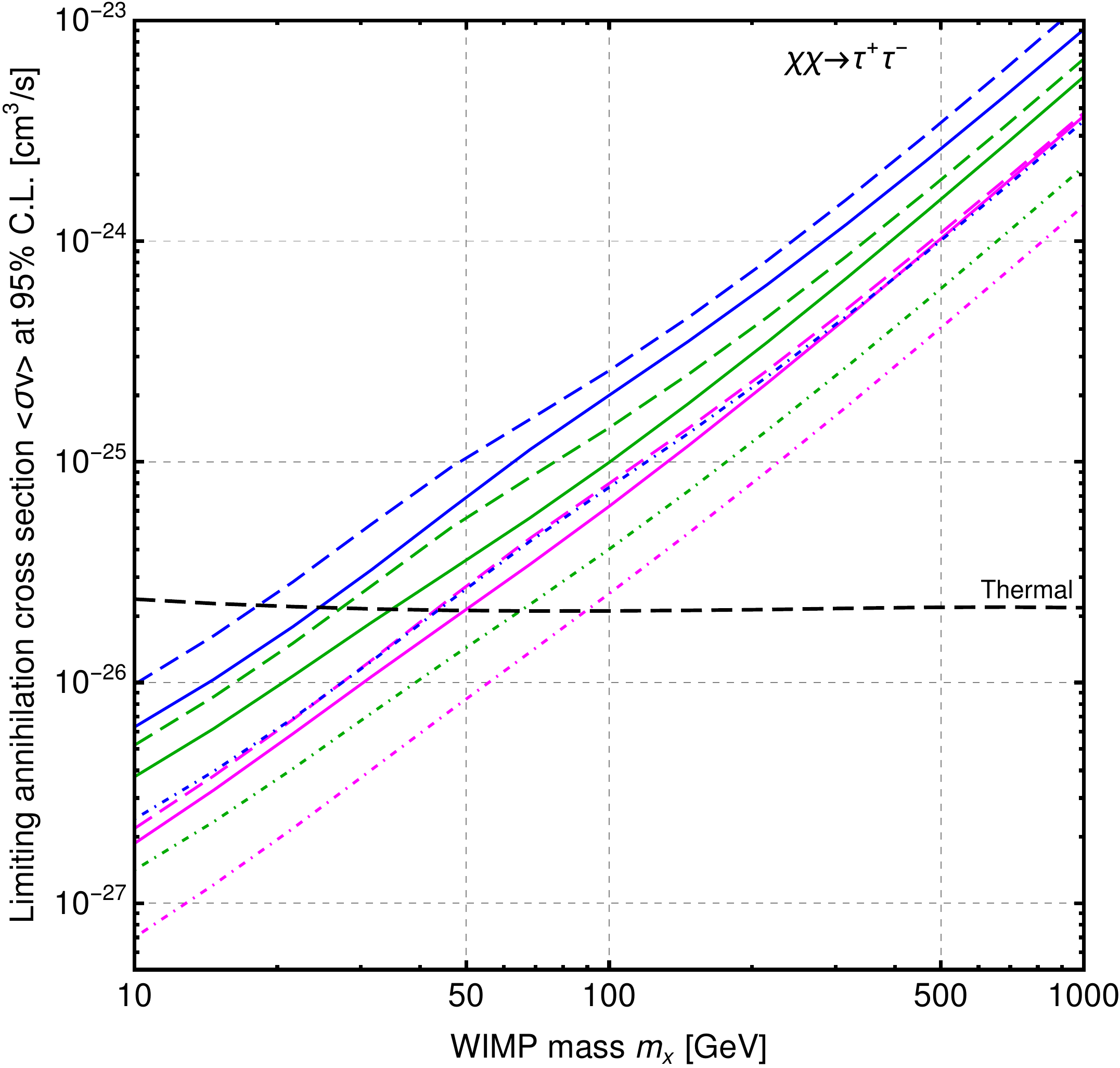}
	\caption{\label{fig:sv}The joint constraints for all 9 DM density and MF/prop. models for both annihilation channels.}
\end{figure*}
\begin{figure*}[t]
	\includegraphics[width=0.497\linewidth]{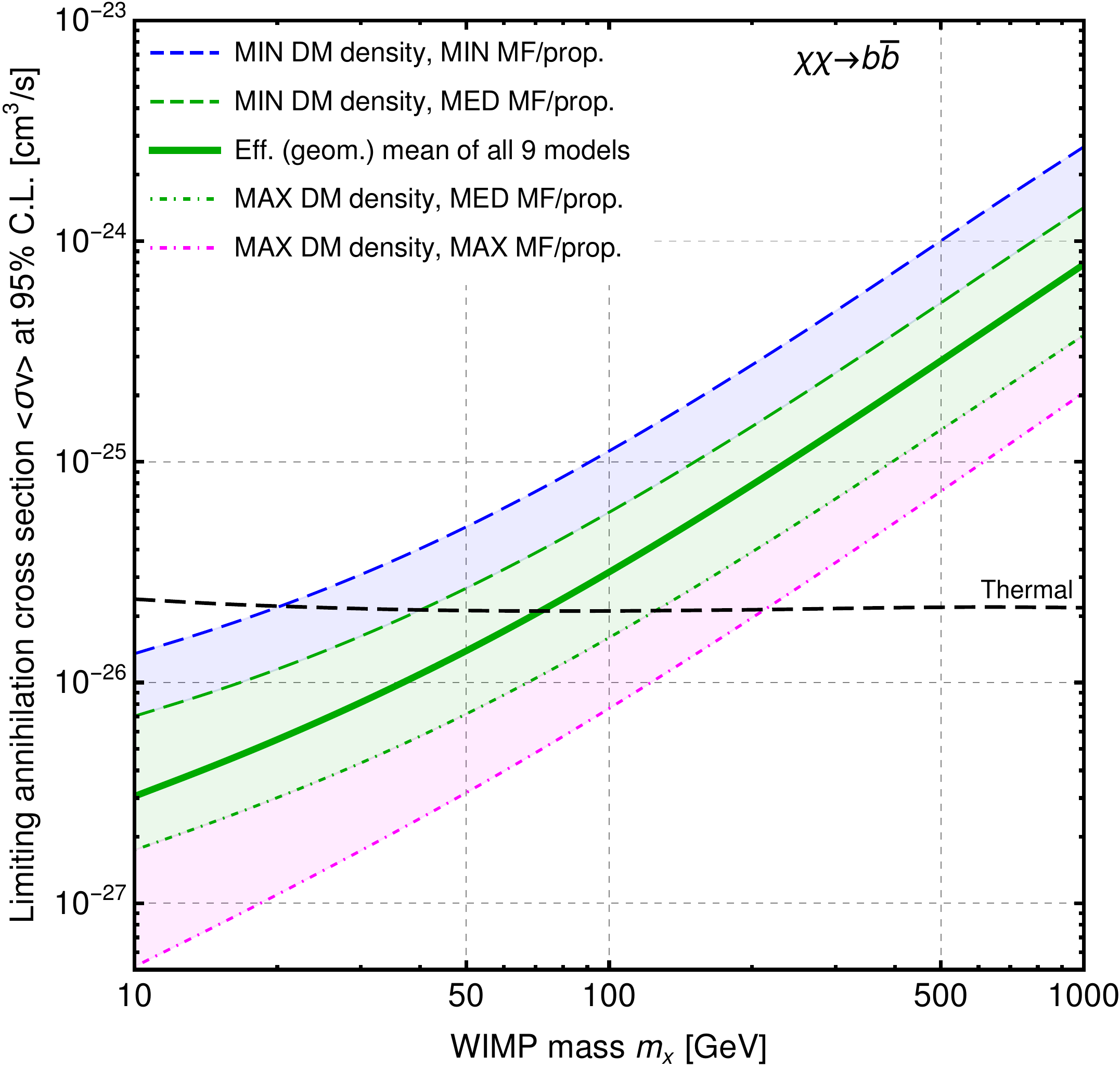}
	\includegraphics[width=0.497\linewidth]{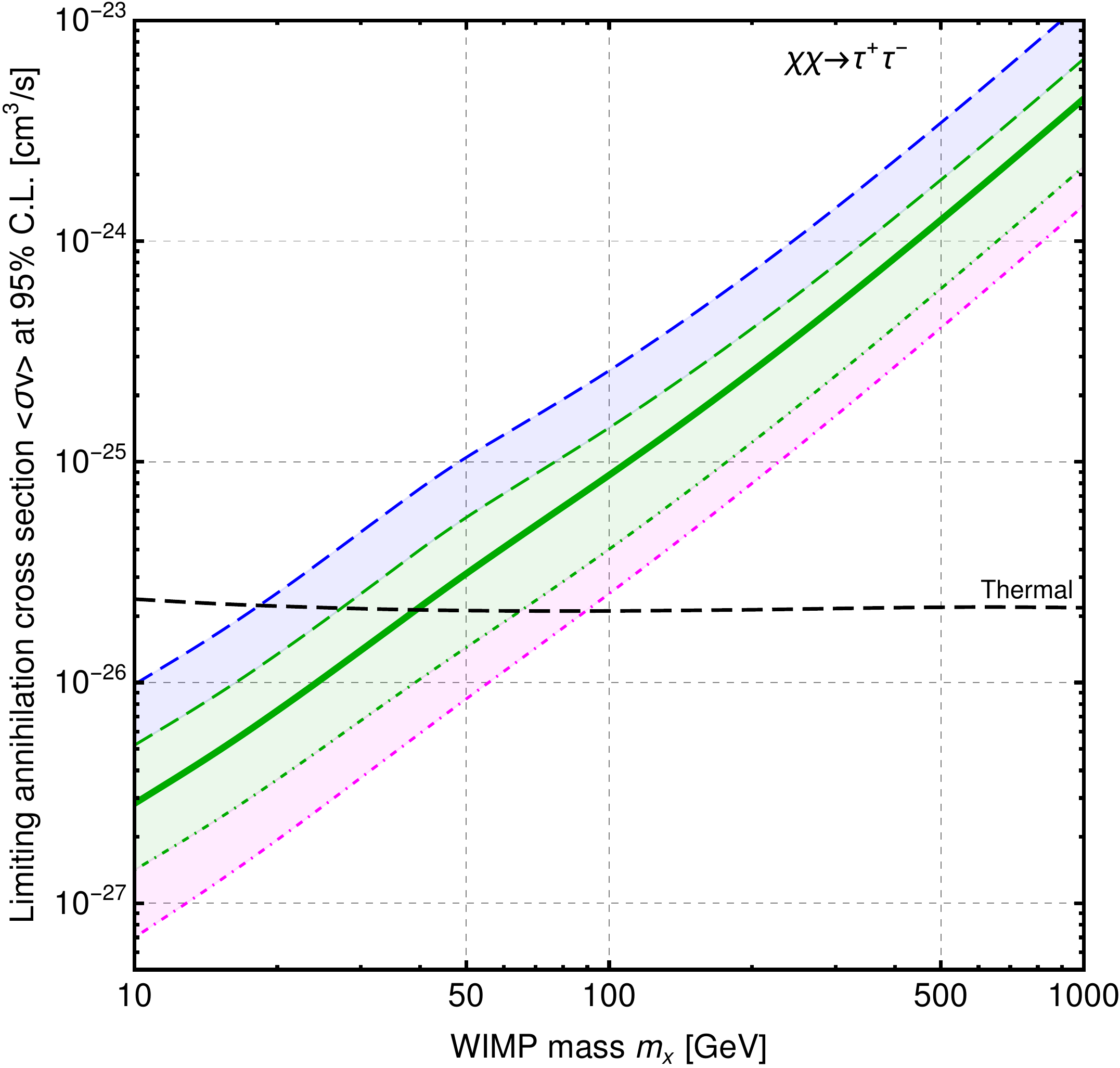}
	\caption{\label{fig:sva}The constraints' uncertainty ranges formed by the uncertainties of DM density profile only (the green-shaded zone), and both DM density and MF/prop. (the blue- and magenta-shaded zones). The green thick line represents the geometric average of all 9 density/MF/prop. models and can be considered as the effective average limit.}
\end{figure*}

Before drawing the final constraints, which join all the frequencies together, the frequency-individual constraints were derived for MED parameter configuration in order to see the influence of each frequency. In this case the product over $i$ in (\ref{eq:Lmi}) is absent. The result is shown in fig. \ref{fig:svf} for both annihilation channels. We see that in $\chi\chi \rightarrow b\overline{b}$ case the LoTSS data dominates the constraints for almost all WIMP masses considered. At low masses, VLSSr also helps. LoTSS domination was naturally anticipated: this is the highest sensitivity data at low frequencies, where DM synchrotron is expected to be bright. Other frequencies influence the likelihood weakly, since for a strong influence the exclusion lines should be packed densely to each other, creating the multiplication effect, as the plot demonstrates. For $\chi\chi \rightarrow \tau^+\tau^-$ channel, the "cloud" of lines is narrower. Here LoTSS defines the exclusion until $m_x \approx 30$ GeV, then the non-thermal map at 4.9 GHz dominates absolutely. However, the picture could be different for MIN and MAX setups.

Then fig. \ref{fig:sv} demonstrates the exclusions for each of our 9 DM density and MF/prop. setups. We can see that they generate quite wide uncertainty range -- its vertical width slightly exceeds an order of magnitude, especially for $b\overline{b}$. In general, the obtained constrains are meaningful in a sense that some thermal WIMP masses inside our range are excluded, even in the most conservative MIN case for both channels. We can also note that the separation between the MIN and MED curves is smaller than that between MED and MAX with our log axes. This can be explained mainly by the presence of substructure boost and reacceleration in MAX models, which cause a significant pull towards larger DM signal intensities and, hence, stronger constraints.

More meaningful exclusions are probably those which enclose the whole uncertainty band, since they represent quite hard limits for possible reality. These hard limits are set by MIN and MAX density/MF/prop. setups. However, one would prefer also to have some representative (or effective) average exclusion. Selection of MED scenario for the role of such exclusion would not be correct, since MED reflects just a particular parameter configuration. Instead, all 9 possible parameter configurations must be averaged. This was done by geometric averaging, since MIN-MED-MAX tend to form the geometric progression rather than the arithmetic one. The resulting exclusion is shown by the green thick line at fig. \ref{fig:sva} and can be considered as the effective average exclusion. The green-shaded area around it reflects the uncertainties of DM density distribution (with MED MF/prop.). The blue- and magenta-shaded areas add MF/prop. uncertainties. We can notice (also from fig. \ref{fig:sv}) that the width of uncertainty band due to uncertainties in DM density is comparable to that due to uncertainties in MF/prop. in many cases. Thus, both factors have important contributions into the total uncertainty budget. Then WIMP mass limits for the thermal cross section were numerically calculated for the cases displayed in fig. \ref{fig:sva}. The results are written out in table \ref {tab:fc}. The mass for the case of $\tau^+\tau^-$ channel is confined significantly better than for $b\overline{b}$. In the latter case, the total uncertainty reaches an order of magnitude. Assuming that the exclusions for other annihilation channels lie somewhere between the tested cases, it can be claimed that the thermal WIMP with $m_x \lesssim 20$ GeV is robustly excluded under \textit{any} reasonable choice of DM density and MF/prop. models. In the opposite case of the optimistic models, the mass threshold is in the range $\approx$(90--200) GeV. And finally, the \textit{average} expectation for the \textit{thermal WIMP is $m_x \lesssim$ (40--70) GeV}, which reflects essentially the moderately conservative models. Indeed, the conducted procedure does not yield any probabilistic inference for the thermal WIMP mass thresholds inside the obtained uncertainty ranges quoted in table \ref {tab:fc}. The threshold value probability density can be assumed quasi-uniform over these ranges with a slight preference toward the effective averages just from a general intuition.
\begin{table}[h]
	\caption{\label{tab:fc}The obtained lower limits on the thermal WIMP mass and their uncertainty ranges. See the details in sec. \ref{sec:constr}.}
	\begin{ruledtabular}
		\centering
		\begin{tabular}{ccc}
			& $\chi\chi \rightarrow b\overline{b}$ & $\chi\chi \rightarrow \tau^+\tau^-$ \\
			\hline 
			Effective (geometric) average & \textbf{72} & \textbf{39} \\
			Range due to DM density uncert. & 40--120 & 27--65 \\
			Range due to DM density and & & \\
			MF/prop. uncert. & 20--210 & 18--89 \\
		\end{tabular}
	\end{ruledtabular}
\end{table}

\section{\label{sec:gamma}Relation with the gamma-ray band and antiprotons}

\begin{figure*}[t]
	\includegraphics[width=0.8\linewidth]{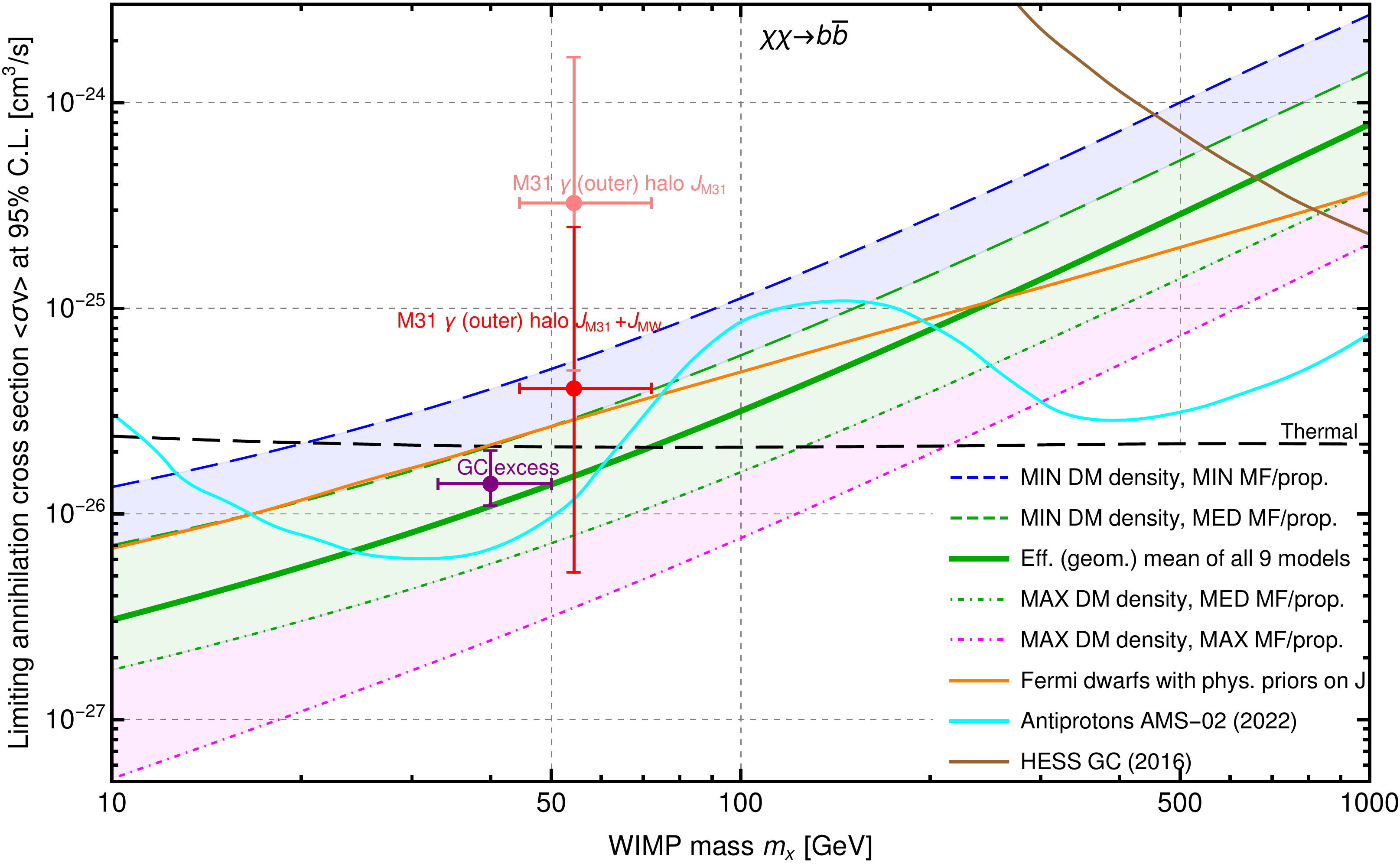}
	\caption{\label{fig:gamma}Our constraints for $b\overline{b}$ annihilation channel in comparison with the WIMP parameter ranges required to fit the gamma-ray emission from the outer halo of M31 (red crosses, taken from \cite[fig. 7]{2021PhRvD.103b3027K}) and GC excess (purple cross, taken from \cite{2021arXiv211209706C}). Also, the constraints from Fermi-LAT observations of dwarfs \cite[fig. 4]{2020PhRvD.102f1302A}, AMS-02 data on antiprotons \cite[fig. 6]{2022arXiv220203076C} and HESS observations of GC \cite[fig. 4S]{2016PhRvL.117k1301A} are shown. }
\end{figure*}

Bedsides $e^\pm$ WIMP annihilation produces other particles including antiprotons, neutrons, antideutrons and photons (more details are in e.g. \cite{2011JCAP...03..051C}). The latter is often called the final state or prompt radiation and falls to the gamma-ray band. In this context \cite{2018PhRvD..97j3021M} analyzed whether M31 radio brightness allows to fit the central gamma-ray emission (seen by Fermi-LAT \cite{2017ApJ...836..208A}) by annihilating DM only. As was mentioned in sec. \ref{sec:cr}, \cite{2018PhRvD..97j3021M} concluded that such fit is not allowed, but their model contains some imprecise assumptions. Having more precise model developed here, we can check whether the conclusion of \cite{2018PhRvD..97j3021M} still holds. According to their fig. 3 WIMP must have the following parameters to fit the gamma emission: $m_x \approx 10$ GeV, $\langle \sigma v \rangle \approx 10^{-25}$ cm$^3$/s. This is for NFW density profile with the inner slope $\gamma_{\text{NFW}} = 1$, which is quite similar to our MED profile. According to our fig. \ref{fig:sva} such WIMP is excluded absolutely robustly. Then for larger values of $\gamma_{\text{NFW}}$, they indeed obtained smaller required cross sections. Our constraints would strengthen for steeper profiles too, down to at least $\langle \sigma v \rangle \approx 10^{-27}$ cm$^3$/s. Therefore, we conclude that the main claim of \cite{2018PhRvD..97j3021M} is verified to be valid in spite of their model approximations. This is generally anticipated, since it is unlikely that the gamma-ray emission around M31 center is generated by DM only -- some contributions from usual astrophysical sources like CR interactions, millisecond pulsars, etc. are expected too.

More interesting situation was revealed for the outer halo of M31, which was claimed to manifest the gamma-ray emission too in \cite{2019ApJ...880...95K}. And DM annihilation was found to be the plausible mechanism for this emission in \cite{2021PhRvD.103b3027K}. A domination of DM source is more probable in the outer halo, since CR interactions with the environment generate gammas mainly inside the disk. Fig. \ref{fig:gamma} shows by the red crosses WIMP parameter ranges required to fit the outer halo emission according to \cite{2021PhRvD.103b3027K}. The top cross reflects the model, in which only M31 dark halo contributes into the signal. The bottom cross takes into account both MW and M31 halo contributions along the line of sight. We see that in both cases the vertical uncertainty range is quite large. However, our constraints robustly exclude the first case, when only M31 J-factor (\ref{eq:dm-J}) is included. This is generally anticipated, since DM must inevitably annihilate in both halos. The second case is allowed very partially; i.e. non-conservative assumptions are needed for the fit, particularly regarding the J-factor values.

Then the purple cross in fig. \ref{fig:gamma} shows the parameter values required to fit the well-known gamma-ray excess around the GC according to the newest study \cite{2021arXiv211209706C}. Our derived constraints in principle allow these values, although for slightly "pessimistic" models of DM density and MF/prop. in M31. Thus, in summary, both the gamma-ray phenomena -- M31 outer halo and GC excess -- can be explained by DM annihilation without an apparent contradiction with the radio constraints, although some restrictions and cautions exist. Hence, these phenomena persist to be promising directions for DM indirect searches. The community needs new higher-quality gamma-ray data on both GC and M31 in order to progress understanding of the nature of both phenomena. Such new data in future may come from (besides other projects) our space-based gamma-ray telescope GAMMA-400 being developed currently. Its anticipated sensitivity to various types of DM signals are reviewed in details in \cite{2020JCAP...11..049E}.

Fig. \ref{fig:gamma} also demonstrates Fermi-LAT constraints from the dwarf MW satellites. These constraints are generally comparable to our effective average exclusion. Fermi-LAT constraints significantly depend on the assumed J-factor values for dwarfs, which lack the direct measurements of J-factors. As pointed out in \cite{2020PhRvD.102f1302A}, the optimistic J-factor values yield the constraints slightly stronger than those derived here and exclude thermal WIMP up to $\approx$ 100 GeV. More realistic scenario shown by the orange line in fig. \ref{fig:gamma} yields the exclusion, which is weaker than our one until $\approx$ 200 GeV.

The constraints for heavy WIMPs come from ground-based gamma-ray telescopes. Thus, the brown line in fig. \ref{fig:gamma} shows the limit from HESS observations of the GC region \cite{2016PhRvL.117k1301A}. This limit was obtained from 254 hours of pure observational time gathered over 10 years of telescope operations. As can be seen from e.g. \cite[fig. 2]{2016PhRvL.117k1301A}-\cite{2022PDU....3500912A}, Imaging Atmospheric Cherenkov Telescopes (IACTs) develop much better sensitivity, when they target GC rather than dwarf satellites. However, the constraints from observations of GC are more model-dependent due to large uncertainties of DM density in the central kiloparsec. IACTs are generally unable to mitigate this uncertainty by observing a bigger part of the Galactic halo due to a relatively small field of view (see e.g. \cite[sec. 5.3]{2021JCAP...01..057A}). Overall, we may conclude that currently IACTs overtake the limits derived here at WIMP masses above $\approx$ 500 GeV. 

And finally fig. \ref{fig:gamma} also contains the limit from AMS-02 antiproton data derived in the newest study \cite{2022arXiv220203076C}, which meanwhile disproved a significance of the antiproton excess attributed to DM. We see that the antiproton exclusion line jumps below and above our exclusion until $\approx$ 200 GeV, where overtakes finally. Thus, overall, our approach is very competitive to both Fermi-LAT observations of dwarfs and CR antiproton measurements.

\section{\label{sec:signal}Possible DM contribution into the non-thermal emission in the central region}

\begin{figure}[t]
	\includegraphics[width=1\linewidth]{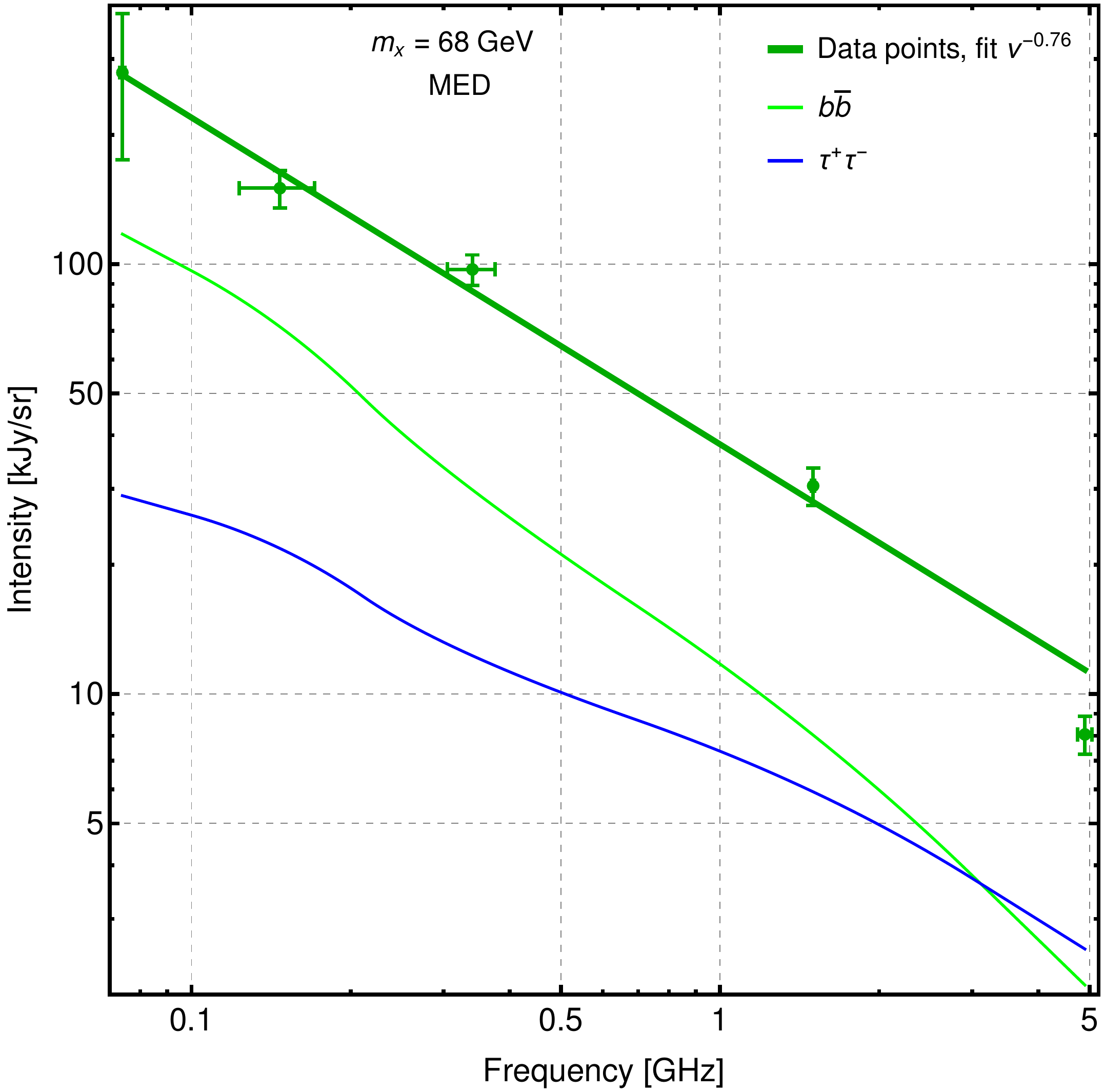}
	\caption{\label{fig:sigf}The observed non-thermal spectrum in the disk with 5$'$ radius around the center and the spectra of annihilating thermal WIMP with $m_x = 68$ GeV. The case of $b\overline{b}$ channel fits the gamma-ray emission from the outer halo of M31.}
\end{figure}

\begin{figure}[t]
	\includegraphics[width=1\linewidth]{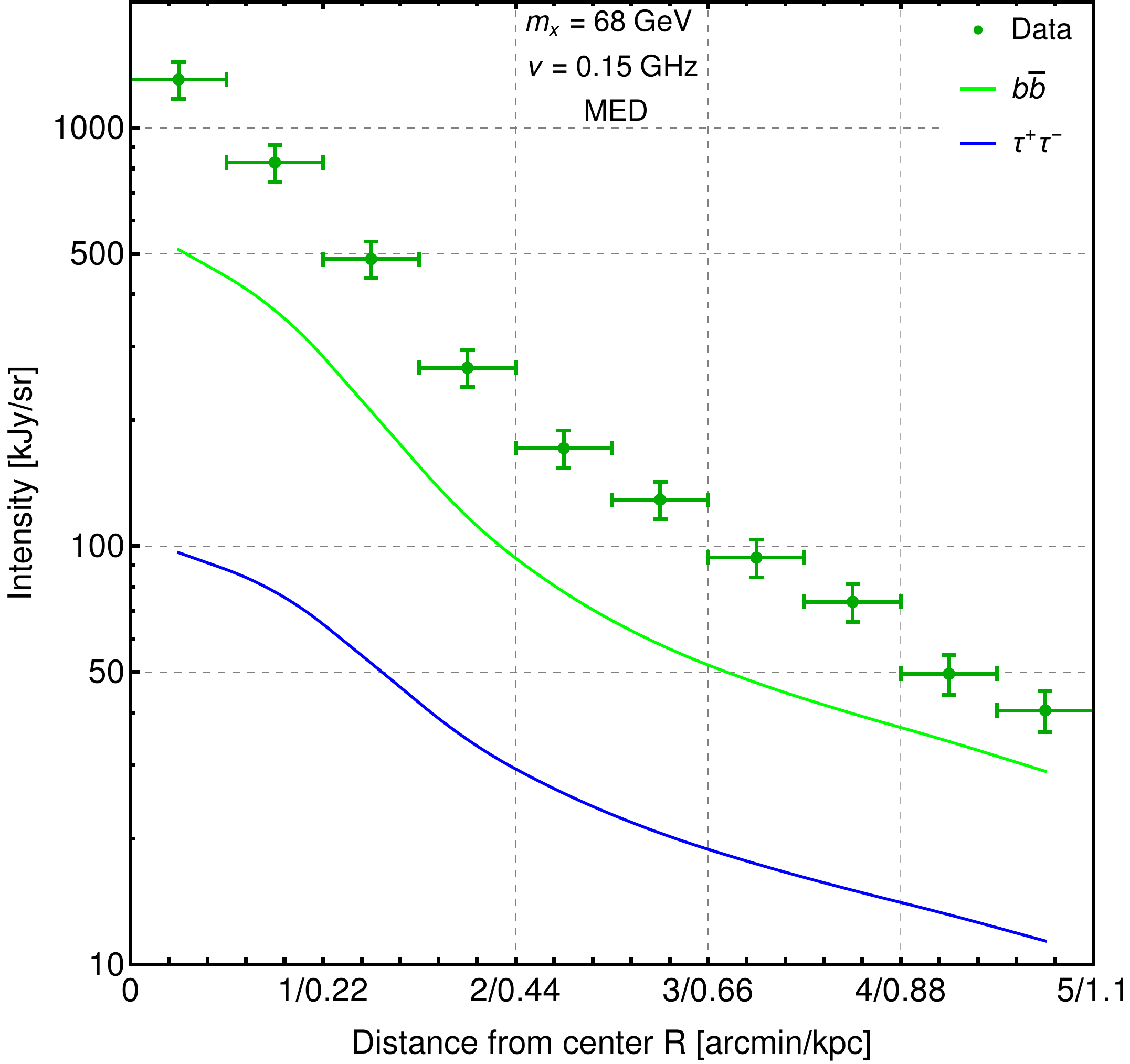}
	\caption{\label{fig:sigr}The observed radial intensity profile from LoTSS data and that from annihilating thermal WIMP with $m_x = 68$ GeV. The case of $b\overline{b}$ channel fits the gamma-ray emission from the outer halo of M31.}
\end{figure}

This section is dedicated to the semi-qualitative discussion on the following interesting question: what kind of a radio signal in M31 nucleus would be produced by WIMP, which fits the gamma-ray outer halo? For this purpose I decided to test WIMP with the thermal annihilation cross section (as the most motivated value) and $m_x = (60-70)$ GeV. This WIMP lies inside the range bounded by the red bottom cross in fig. \ref{fig:gamma}, and it is "partially" (i.e. for certain reasonable parameter configurations) allowed by all the main constraints: our radio, Fermi-LAT gamma by dwarfs and antiprotons. The lighter thermal WIMP is severely constrained, and the heavier one would not fit the outer halo. $m_x = 68$ GeV was set due to practical convenience. Then the spectrum of radio emission due to such WIMP and the radial intensity profile were generated (for MED configuration) for the disk with 5$'$ radius around the center. The results are displayed by the light-green lines at figs. \ref{fig:sigf}-\ref{fig:sigr}. The spectrum was compared with the observed non-thermal spectrum, which was obtained from the data at 0.074, 0.15, 0.34, 1.5 and 4.9 GHz. The observed spectrum was fitted by a simple power law yielding $I_\nu \propto \nu^{-0.76}$ (machine fit) -- it is shown by the thick green line. We see that the slopes of the observed and DM spectra are quite similar. The blue line shows the case of $\tau^+\tau^-$ channel just for comparison ($\tau^+\tau^-$ was not used to fit the outer halo). The radial intensity profile from DM also follows the observed one very well, just from a visual inspection. Thus, such WIMP can easily mimic both the spectrum and morphology of the central emission contributing up to about half of the total emission intensity. Another contribution can naturally come from CR $e^\pm$. And a good consistency of the spectrum and morphology was seen with just first near-at hand parameter configuration without any tuning, i.e. with MED DM density profile and MF/prop. configuration. An intentional tuning would further improve consistency. DM model variation causes a significant variation of the spectral index. As an illustration of this fact, all our computed 234 DM models show the spectral indices in the wide range 0.39--1.77, if to approximate the spectra by the power law inside inner 5$'$. The branching ratio of the annihilation to different channels is another free parameter. Thus, overall there are wide opportunities to achieve the necessary fit for the spectrum, morphology and amplitude.

The obtained good consistency between the considered DM emission spectrum/morphology and the observed ones may in principle indicate indirectly in favor of the presence of DM contribution in the central emission. Meanwhile \cite{2014A&A...571A..61G} noticed, that the source of CRs, which produce the radio emission in the nucleus, is unclear; since the star formation and, hence, CR acceleration sites are located in the outer zone $r \gtrsim$ 3 kpc. The possibility of diffusion of CRs from there to the central region is also doubted. Hence, it is not easy to find a conventional explanation for the relatively bright central radio emission. 

\section{\label{sec:summary}Conclusions and discussion}

The present work aimed to obtain the robust and model-independent WIMP annihilation constraints from radio observations of M31. Various new observational data on M31 were implemented, including very recent high-quality images obtained in LOFAR Two-meter Sky Survey. For the first time, the transport equation for DM $e^\pm$ was properly solved in 2D taking into account the spatial diffusion. Solution of the transport equation and generation of DM synchrotron emission maps were performed by the specific adaptation of GALPROP code (v56). This is the first (known to the author) application of GALPROP to another galaxy. A wide variety of radio data in the frequency range $\approx$(0.1--10) GHz were utilized to derive the constraints. The systematic uncertainties of the derived constraints were quantified by three benchmark MIN-MED-MAX scenarios (individual) for both DM density distribution and MF/prop. parameters. The constraints were obtained for two representative annihilation channels -- into $b\overline{b}$ and $\tau^+\tau^-$. The final results can be viewed in figs. \ref{fig:sv}-\ref{fig:sva} and table \ref{tab:fc}. Thus, WIMP mass lower threshold is constrained to lie in the range 20--210 GeV for $b\overline{b}$ channel and 18--89 GeV for $\tau^+\tau^-$ channel for the case of thermal annihilation cross section. If one would like to have some representative effective average limits, then the exclusion drawn by the thick green line at fig. \ref{fig:sva} is suggested to be used for that. This exclusion presents the geometric mean of all 9 computed parameter configurations and yields $m_x \gtrsim 70$ GeV for $b\overline{b}$, $m_x \gtrsim 40$ GeV for $\tau^+\tau^-$. However, one has to keep in mind that indeed these average limits are somewhat fiducial rather than statistically-hard. Our limits are rather model-independent in the sense that no specific assumptions were made regarding the intensity of usual CR synchrotron emission, i.e. the flat priors were assumed for it at all frequencies.

Let us compare the derived constraints with the results of previous studies on the subject. First of all, it is very interesting to make the comparison with our previous work \cite{2013PhRvD..88b3504E}. There we claimed the (average effective) exclusion of thermal WIMP lighter than 55--100 GeV (the same annihilation channels). The little difference w.r.t. the new results appeared to be fully caused by the change of the thermal cross section value: the older value used 10 years ago is larger by $\approx$ 1.5 times than the modern refined value utilized here. Rescaling the thermal cross section would bring the old and new results into nearly ideal agreement! The supposed reason of such a remarkable reconciliation is the following. On one hand, the new study introduced various refining aspects to the model, which lowered DM intensities (hence weakened the constraints). These aspects include the spatial diffusion, more realistic MF distribution and energy losses, etc. On the other hand, newer and more sensitive radio images were utilized. Some of these images are cleaned of nuisance emissions. A better sensitivity allowed to constrain the observed non-thermal emission intensity better, which strengthened DM constraints. And it looks like these two opposite effects exactly canceled each other! The uncertainty ranges of the final constraints were slightly shrunk by the new model.

Regarding the studies made by other authors, we basically disprove the strong constraints claimed in \cite{2019ApJ...872..177C,2019JCAP...08..019B}, which excluded heavy thermal WIMPs with $m_x \gg 100$ GeV. These studies obtained the constraints utilizing very large ROIs comparable by size to the whole galaxy. Fig. \ref{fig:sed} above indicates semi-intuitively impossibility of such strong constraints, even if MED configuration there would be replaced by MAX, which produces the largest possible DM flux (from the whole galaxy). The latter reaches the level of observed flux just for the lightest WIMP with $m_x \approx 10$ GeV, meaning that only such low masses can be constrained. The natural question is what does cause such a big discrepancy between the results here and in \cite{2019ApJ...872..177C,2019JCAP...08..019B}? Although the precise answer would require to do specific calculations; on a qualitative level, the main supposed reason is the deficient 1D MF model in \cite{2019ApJ...872..177C,2019JCAP...08..019B} together with abandoning the spatial diffusion, as was discussed in sec. \ref{sec:cr}. Their MF models extrapolate spherically-symmetrically significant field values of several $\mu$G up to huge distances of decades kpc. However, no direct field measurements are available at such distances, especially outside the disk. From a general reasoning, one would not expect for MF to extend significantly beyond the disk, since the typical field generation by a dynamo would not work there. For this reason, the exponential field decline is usually assumed in the halo. Thus, \cite{2019ApJ...872..177C,2019JCAP...08..019B} presumably boosted DM flux from the halo abnormally due to filling an unrealistically large volume by the significant MF. More realistic 2D MF models like those employed here (eqs. (\ref{eq:dm-mf-min})--(\ref{eq:dm-mf-max})) are needed instead, especially taking into account the mentioned thinness of M31 magnetic disk.

The obtained exclusions are competitive and comparable to those from Fermi-LAT observations of MW satellites and AMS-02 measurements of CR antiprotons. Rather conservative scenarios of our constraints allow the explanations of the gamma-ray M31 outer halo and GC excess by annihilating DM.

It was found that the thermal WIMP with $m_x \approx 70$ GeV, which fits the gamma-ray outer halo and is not strongly excluded by all the main constraints, would make a significant contribution into the non-thermal emission in the central region of M31 fitting well both the emission spectrum and radial intensity profile. This is demonstrated at figs. \ref{fig:sigf}-\ref{fig:sigr} particularly for MED DM density and MF/prop. configuration. In that case, DM emission intensity reaches up to about half of the total observed intensity.

As a further development of this work it is planned to extend this methodology to all other possible annihilation channels in order to study them in details and obtain the channel-independent exclusions. Then the future data from LOFAR LBA Sky Survey (LoLSS) \cite{2021A&A...648A.104D} is anticipated for further sensitivity improvement. The study similar to the present work can be conducted for MW too, since the most recent such study \cite{2016JCAP...07..041C} utilized rather approximate solution of the transport equation. And the information on the emission polarization can be employed for an ultimate improvement of the sensitivity (for both galaxies). This was already probed in \cite{2022arXiv220404232M}.

\appendix*
\section{\label{sec:a}Additional aspects of modeling of DM $\mathbf{e^\pm}$ propagation and emission from them in GALPROP.}

Here I describe the aspects, which were not covered in the main text. Table \ref{tab:a} below lists the values of various relevant parameters, which were used in GALPROP. These parameters are the same for all MIN-MED-MAX configurations. The spatial and energy grid step sizes, as well as the propagation energy range cited in the table, were empirically adjusted to achieve a good precision of the intensity calculation at the level of few percents. In fact, the step sizes are very small. Such small steps were needed in the numerical procedure, because both DM density and $e^\pm$ injection spectra changes very steeply.
\begin{table}[h]
	\caption{\label{tab:a}Values of various parameters used in GALPROP.}
	\begin{ruledtabular}
		\centering
		\begin{tabular}{cc}
			Parameter & Value \\
			\hline
			Radius of the diffusion cylinder $r_{\text{max}}$ & 15 kpc \\
			Spatial grid step size $\Delta r = \Delta z$ & 0.05 kpc \\
			Propagation energy range & 10MeV--1.2$m_x c^2$ \\
			Energy grid step increment & 1.04 \\
			Infrared ISRF density factor w.r.t. MW & 0.1 \\
			HEALPix resolution of maps $N_{\text{side}}$ & $2^{10}=1024$ \\
			Temperature of free $e^-$ in galactic gas $T_e$ & 7000 K \\
			Free electrons' clumping factor & 21 \\	
			He/H ratio in galactic gas & 0.1 \\		
		\end{tabular}
	\end{ruledtabular}
\end{table}

Then let us discuss the neutral gas concentration profiles, which were needed to calculate $e^\pm$ bremsstrahlung and ionization energy losses mentioned in sec. \ref{sec:dm-prop}. The measured surface density profiles of the atomic and molecular gas were taken from \cite[fig. 1]{2009A&A...505..497Y}. Then I assumed the vertical profiles of gases to be similar to those in MW. Usually these vertical profiles are assumed to be Gaussian -- e.g. \cite{2007A&A...467..611F}. The vertical scale heights (i.e. Gaussian RMS) were adopted from \cite{2007A&A...467..611F,1990ARA&A..28..215D}: $\sigma_{\text{HI}}$ = 0.06 kpc and $\sigma_{\text{H}_2}$ = 0.03 kpc, which are attributed to the bulge region. Then the normalization coefficients for the concentration profiles were calculated, requiring that each profile must reasonably reproduce the total gas mass in M31: $\int n_{\text{gas}}(r,z)m_pdV \approx M_{\text{gas}}$. The total gas masses in M31 were taken from \cite[table 1]{2009A&A...505..497Y} and \cite{2006A&A...453..459N}. This procedure produced the following fitting functions for the gas distributions, which were substituted into GALPROP:
\begin{multline}\label{eq:a-nI}
n_{\text{HI}}(r_{\text{kpc}},z) = 0.269(10\string^(\sin((r/8.23)^{1.35})-0.377) + \\ + (r/23)^2 +0.23/(r+1.8)^{0.7})\exp(-0.5z^2/\sigma_{\text{HI}}^2)~\text{cm}^{-3},
\end{multline}
\begin{multline}\label{eq:a-n2}
n_{\text{H}_2}(r_{\text{kpc}},z) = 0.538(0.073+0.42\exp(-(r-4.35)^2) + \\ +\exp(-(r-10.8)^2/3.5)) \exp(-0.5z^2/\sigma_{\text{H}_2}^2)~\text{cm}^{-3},
\end{multline}
Indeed, these are quite simple distributions, which track only the large-scale structure of the gas over the whole disk and do not contain small-scale irregularities like dense clouds. As an overall sanity check of the gas profiles constructed above (including (\ref{eq:dm-ne})), I compared the synchrotron flux from the relevant ROI ($R \leqslant$ 1 kpc) produced with the profiles above and that with the standard MW profiles. The replacement of $n_e(r,z)$ and $n_{\text{HI}}(r,z)$ has a minor effect on the flux -- less than 10\%. This agrees with the data presented in \cite[table 1, fig. 1]{2009A&A...505..497Y} for HI -- its disk surface density and total mass are very similar in MW and M31. However, the replacement of $n_{\text{H}_2}(r,z)$ profile lowers the flux more -- by 10--30\% depending on frequency. This also agrees with the mentioned source of data -- H$_2$ surface density and mass in M31 are lower than those in MW by several times! Thus, the constructed profiles look realistic in the first approximation. Also, the fact that the emission fluxes depend mildly on the gas concentrations releases a necessity to vary the latter with the purpose of studying the corresponding uncertainty range. I.e., the potential uncertainties in the synchrotron intensity due to the uncertainties in gas distributions are negligible in comparison with those driven by, e.g., the uncertainties in DM density distribution. Thus, I set the same gas distributions for all MIN-MED-MAX scenarios.

Full galdef files for GALPROP runs can be requested from the author by e-mail.

\begin{acknowledgments}
I greatly appreciate the following software, which is very useful in my work: Wolfram Mathematica; "Aladin sky atlas" developed at CDS, Strasbourg Observatory, France \cite{2000A&AS..143...33B}; NumPy \cite{numpy}; Healpy \cite{healpy}; WebPlotDigitizer \cite{WPD}.
 
I am very grateful to Mikhail Zavertyaev for providing the server for long computations, Rainer Beck for providing the radio maps of M31, also to George Heald, Nikolay Topchiev and Mikhail Razumeiko.
\end{acknowledgments}

\bibliography{../../../universal}

\begin{thebibliography}{92}%
\makeatletter
\providecommand \@ifxundefined [1]{%
 \@ifx{#1\undefined}
}%
\providecommand \@ifnum [1]{%
 \ifnum #1\expandafter \@firstoftwo
 \else \expandafter \@secondoftwo
 \fi
}%
\providecommand \@ifx [1]{%
 \ifx #1\expandafter \@firstoftwo
 \else \expandafter \@secondoftwo
 \fi
}%
\providecommand \natexlab [1]{#1}%
\providecommand \enquote  [1]{``#1''}%
\providecommand \bibnamefont  [1]{#1}%
\providecommand \bibfnamefont [1]{#1}%
\providecommand \citenamefont [1]{#1}%
\providecommand \href@noop [0]{\@secondoftwo}%
\providecommand \href [0]{\begingroup \@sanitize@url \@href}%
\providecommand \@href[1]{\@@startlink{#1}\@@href}%
\providecommand \@@href[1]{\endgroup#1\@@endlink}%
\providecommand \@sanitize@url [0]{\catcode `\\12\catcode `\$12\catcode
  `\&12\catcode `\#12\catcode `\^12\catcode `\_12\catcode `\%12\relax}%
\providecommand \@@startlink[1]{}%
\providecommand \@@endlink[0]{}%
\providecommand \url  [0]{\begingroup\@sanitize@url \@url }%
\providecommand \@url [1]{\endgroup\@href {#1}{\urlprefix }}%
\providecommand \urlprefix  [0]{URL }%
\providecommand \Eprint [0]{\href }%
\providecommand \doibase [0]{https://doi.org/}%
\providecommand \selectlanguage [0]{\@gobble}%
\providecommand \bibinfo  [0]{\@secondoftwo}%
\providecommand \bibfield  [0]{\@secondoftwo}%
\providecommand \translation [1]{[#1]}%
\providecommand \BibitemOpen [0]{}%
\providecommand \bibitemStop [0]{}%
\providecommand \bibitemNoStop [0]{.\EOS\space}%
\providecommand \EOS [0]{\spacefactor3000\relax}%
\providecommand \BibitemShut  [1]{\csname bibitem#1\endcsname}%
\let\auto@bib@innerbib\@empty
\bibitem [{\citenamefont {{Rubakov}}(2019)}]{2019arXiv191204727R}%
  \BibitemOpen
  \bibfield  {author} {\bibinfo {author} {\bibfnamefont {V.~A.}\ \bibnamefont
  {{Rubakov}}},\ }\bibfield  {title} {\bibinfo {title} {{Cosmology and Dark
  Matter}},\ }\href@noop {} {\bibfield  {journal} {\bibinfo  {journal} {arXiv}\
  ,\ \bibinfo {eid} {arXiv:1912.04727}} (\bibinfo {year} {2019})},\ \Eprint
  {https://arxiv.org/abs/1912.04727} {arXiv:1912.04727 [hep-ph]} \BibitemShut
  {NoStop}%
\bibitem [{\citenamefont {{Zeldovich}}\ \emph {et~al.}(1980)\citenamefont
  {{Zeldovich}}, \citenamefont {{Klypin}}, \citenamefont {{Khlopov}},\ and\
  \citenamefont {{Chechetkin}}}]{1980SvJNP..31..664Z}%
  \BibitemOpen
  \bibfield  {author} {\bibinfo {author} {\bibfnamefont {Y.~B.}\ \bibnamefont
  {{Zeldovich}}}, \bibinfo {author} {\bibfnamefont {A.}~\bibnamefont
  {{Klypin}}}, \bibinfo {author} {\bibfnamefont {M.~Y.}\ \bibnamefont
  {{Khlopov}}},\ and\ \bibinfo {author} {\bibfnamefont {V.~M.}\ \bibnamefont
  {{Chechetkin}}},\ }\bibfield  {title} {\bibinfo {title} {{Astrophysical
  bounds on the mass of heavy stable neutral leptons}},\ }\href@noop {}
  {\bibfield  {journal} {\bibinfo  {journal} {SvJNP}\ }\textbf {\bibinfo
  {volume} {31}},\ \bibinfo {pages} {664} (\bibinfo {year} {1980})}\BibitemShut
  {NoStop}%
\bibitem [{\citenamefont {{Slatyer}}(2021)}]{2021arXiv210902696S}%
  \BibitemOpen
  \bibfield  {author} {\bibinfo {author} {\bibfnamefont {T.~R.}\ \bibnamefont
  {{Slatyer}}},\ }\bibfield  {title} {\bibinfo {title} {{Les Houches Lectures
  on Indirect Detection of Dark Matter}},\ }\href@noop {} {\bibfield  {journal}
  {\bibinfo  {journal} {arXiv}\ ,\ \bibinfo {eid} {arXiv:2109.02696}} (\bibinfo
  {year} {2021})},\ \Eprint {https://arxiv.org/abs/2109.02696}
  {arXiv:2109.02696 [hep-ph]} \BibitemShut {NoStop}%
\bibitem [{\citenamefont {{Ando}}\ \emph {et~al.}(2020)\citenamefont {{Ando}},
  \citenamefont {{Geringer-Sameth}}, \citenamefont {{Hiroshima}}, \citenamefont
  {{Hoof}}, \citenamefont {{Trotta}},\ and\ \citenamefont
  {{Walker}}}]{2020PhRvD.102f1302A}%
  \BibitemOpen
  \bibfield  {author} {\bibinfo {author} {\bibfnamefont {S.}~\bibnamefont
  {{Ando}}}, \bibinfo {author} {\bibfnamefont {A.}~\bibnamefont
  {{Geringer-Sameth}}}, \bibinfo {author} {\bibfnamefont {N.}~\bibnamefont
  {{Hiroshima}}}, \bibinfo {author} {\bibfnamefont {S.}~\bibnamefont {{Hoof}}},
  \bibinfo {author} {\bibfnamefont {R.}~\bibnamefont {{Trotta}}},\ and\
  \bibinfo {author} {\bibfnamefont {M.~G.}\ \bibnamefont {{Walker}}},\
  }\bibfield  {title} {\bibinfo {title} {{Structure formation models weaken
  limits on WIMP dark matter from dwarf spheroidal galaxies}},\ }\href
  {https://doi.org/10.1103/PhysRevD.102.061302} {\bibfield  {journal} {\bibinfo
   {journal} {PhRvD}\ }\textbf {\bibinfo {volume} {102}},\ \bibinfo {eid}
  {061302} (\bibinfo {year} {2020})},\ \Eprint
  {https://arxiv.org/abs/2002.11956} {arXiv:2002.11956 [astro-ph.CO]}
  \BibitemShut {NoStop}%
\bibitem [{\citenamefont {{Hoof}}\ \emph {et~al.}(2020)\citenamefont {{Hoof}},
  \citenamefont {{Geringer-Sameth}},\ and\ \citenamefont
  {{Trotta}}}]{2020JCAP...02..012H}%
  \BibitemOpen
  \bibfield  {author} {\bibinfo {author} {\bibfnamefont {S.}~\bibnamefont
  {{Hoof}}}, \bibinfo {author} {\bibfnamefont {A.}~\bibnamefont
  {{Geringer-Sameth}}},\ and\ \bibinfo {author} {\bibfnamefont
  {R.}~\bibnamefont {{Trotta}}},\ }\bibfield  {title} {\bibinfo {title} {{A
  global analysis of dark matter signals from 27 dwarf spheroidal galaxies
  using 11 years of Fermi-LAT observations}},\ }\href
  {https://doi.org/10.1088/1475-7516/2020/02/012} {\bibfield  {journal}
  {\bibinfo  {journal} {J. Cosmology Astropart. Phys.}\ }\textbf {\bibinfo
  {volume} {02}},\ \bibinfo {eid} {012} (\bibinfo {year} {2020})},\ \Eprint
  {https://arxiv.org/abs/1812.06986} {arXiv:1812.06986 [astro-ph.CO]}
  \BibitemShut {NoStop}%
\bibitem [{\citenamefont {{Acharyya}}\ \emph {et~al.}(2021)\citenamefont
  {{Acharyya}}, \citenamefont {{Adam}}, \citenamefont {{Adams}}, \citenamefont
  {{Agudo}}, \citenamefont {{Aguirre-Santaella}}, \citenamefont {{Alfaro}},
  \citenamefont {{Alfaro}}, \citenamefont {{Alispach}}, \citenamefont
  {{Aloisio}}, \citenamefont {{Alves Batista}},\ and\ \citenamefont
  {et~al.}}]{2021JCAP...01..057A}%
  \BibitemOpen
  \bibfield  {author} {\bibinfo {author} {\bibfnamefont {A.}~\bibnamefont
  {{Acharyya}}}, \bibinfo {author} {\bibfnamefont {R.}~\bibnamefont {{Adam}}},
  \bibinfo {author} {\bibfnamefont {C.}~\bibnamefont {{Adams}}}, \bibinfo
  {author} {\bibfnamefont {I.}~\bibnamefont {{Agudo}}}, \bibinfo {author}
  {\bibfnamefont {A.}~\bibnamefont {{Aguirre-Santaella}}}, \bibinfo {author}
  {\bibfnamefont {R.}~\bibnamefont {{Alfaro}}}, \bibinfo {author}
  {\bibfnamefont {J.}~\bibnamefont {{Alfaro}}}, \bibinfo {author}
  {\bibfnamefont {C.}~\bibnamefont {{Alispach}}}, \bibinfo {author}
  {\bibfnamefont {R.}~\bibnamefont {{Aloisio}}}, \bibinfo {author}
  {\bibfnamefont {R.}~\bibnamefont {{Alves Batista}}},\ and\ \bibinfo {author}
  {\bibnamefont {et~al.}},\ }\bibfield  {title} {\bibinfo {title} {{Sensitivity
  of the Cherenkov Telescope Array to a dark matter signal from the Galactic
  centre}},\ }\href {https://doi.org/10.1088/1475-7516/2021/01/057} {\bibfield
  {journal} {\bibinfo  {journal} {JCAP}\ }\textbf {\bibinfo {volume}
  {2021}}\bibfield  {number} {\bibinfo  {number} { (1)},\ \bibinfo {eid}
  {057}},\ }\Eprint {https://arxiv.org/abs/2007.16129} {arXiv:2007.16129
  [astro-ph.HE]} \BibitemShut {NoStop}%
\bibitem [{\citenamefont {{Manconi}}\ \emph {et~al.}(2022)\citenamefont
  {{Manconi}}, \citenamefont {{Cuoco}},\ and\ \citenamefont
  {{Lesgourgues}}}]{2022arXiv220404232M}%
  \BibitemOpen
  \bibfield  {author} {\bibinfo {author} {\bibfnamefont {S.}~\bibnamefont
  {{Manconi}}}, \bibinfo {author} {\bibfnamefont {A.}~\bibnamefont {{Cuoco}}},\
  and\ \bibinfo {author} {\bibfnamefont {J.}~\bibnamefont {{Lesgourgues}}},\
  }\bibfield  {title} {\bibinfo {title} {{Dark Matter constraints from Planck
  observations of the Galactic polarized synchrotron emission}},\ }\href@noop
  {} {\bibfield  {journal} {\bibinfo  {journal} {arXiv}\ ,\ \bibinfo {eid}
  {arXiv:2204.04232}} (\bibinfo {year} {2022})},\ \Eprint
  {https://arxiv.org/abs/2204.04232} {arXiv:2204.04232 [astro-ph.HE]}
  \BibitemShut {NoStop}%
\bibitem [{Has()}]{Haslam}%
  \BibitemOpen
  \href@noop {} {}\bibinfo {howpublished} {Haslam 408 MHz all-sky map from
  \url{https://lambda.gsfc.nasa.gov/product/foreground/fg_2014_haslam_408_info.cfm}}\BibitemShut
  {NoStop}%
\bibitem [{\citenamefont {{Reich}}\ \emph {et~al.}(2001)\citenamefont
  {{Reich}}, \citenamefont {{Testori}},\ and\ \citenamefont
  {{Reich}}}]{2001A&A...376..861R}%
  \BibitemOpen
  \bibfield  {author} {\bibinfo {author} {\bibfnamefont {P.}~\bibnamefont
  {{Reich}}}, \bibinfo {author} {\bibfnamefont {J.~C.}\ \bibnamefont
  {{Testori}}},\ and\ \bibinfo {author} {\bibfnamefont {W.}~\bibnamefont
  {{Reich}}},\ }\bibfield  {title} {\bibinfo {title} {{A radio continuum survey
  of the southern sky at 1420 MHz. The atlas of contour maps}},\ }\href
  {https://doi.org/10.1051/0004-6361:20011000} {\bibfield  {journal} {\bibinfo
  {journal} {Astron. Astrophys.}\ }\textbf {\bibinfo {volume} {376}},\ \bibinfo
  {pages} {861} (\bibinfo {year} {2001})}\BibitemShut {NoStop}%
\bibitem [{\citenamefont {{Walterbos}}\ and\ \citenamefont
  {{Graeve}}(1985)}]{1985A&A...150L...1W}%
  \BibitemOpen
  \bibfield  {author} {\bibinfo {author} {\bibfnamefont {R.~A.~M.}\
  \bibnamefont {{Walterbos}}}\ and\ \bibinfo {author} {\bibfnamefont
  {R.}~\bibnamefont {{Graeve}}},\ }\bibfield  {title} {\bibinfo {title} {{Radio
  continuum emission from the nuclear region of M 31 : evidence for a nuclear
  radio spiral.}},\ }\href@noop {} {\bibfield  {journal} {\bibinfo  {journal}
  {Astron. Astrophys.}\ }\textbf {\bibinfo {volume} {150}},\ \bibinfo {pages}
  {L1} (\bibinfo {year} {1985})}\BibitemShut {NoStop}%
\bibitem [{\citenamefont {{Egorov}}\ and\ \citenamefont
  {{Pierpaoli}}(2013)}]{2013PhRvD..88b3504E}%
  \BibitemOpen
  \bibfield  {author} {\bibinfo {author} {\bibfnamefont {A.~E.}\ \bibnamefont
  {{Egorov}}}\ and\ \bibinfo {author} {\bibfnamefont {E.}~\bibnamefont
  {{Pierpaoli}}},\ }\bibfield  {title} {\bibinfo {title} {{Constraints on dark
  matter annihilation by radio observations of M31}},\ }\href
  {https://doi.org/10.1103/PhysRevD.88.023504} {\bibfield  {journal} {\bibinfo
  {journal} {Phys. Rev. D}\ }\textbf {\bibinfo {volume} {88}},\ \bibinfo {eid}
  {023504} (\bibinfo {year} {2013})},\ \Eprint
  {https://arxiv.org/abs/1304.0517} {arXiv:1304.0517 [astro-ph.CO]}
  \BibitemShut {NoStop}%
\bibitem [{NED()}]{NED}%
  \BibitemOpen
  \href@noop {} {}\bibinfo {howpublished} {NASA/IPAC Extragalactic Database
  \url{https://ned.ipac.caltech.edu/}}\BibitemShut {NoStop}%
\bibitem [{\citenamefont {{Li}}\ \emph {et~al.}(2021)\citenamefont {{Li}},
  \citenamefont {{Riess}}, \citenamefont {{Busch}}, \citenamefont
  {{Casertano}}, \citenamefont {{Macri}},\ and\ \citenamefont
  {{Yuan}}}]{2021ApJ...920...84L}%
  \BibitemOpen
  \bibfield  {author} {\bibinfo {author} {\bibfnamefont {S.}~\bibnamefont
  {{Li}}}, \bibinfo {author} {\bibfnamefont {A.~G.}\ \bibnamefont {{Riess}}},
  \bibinfo {author} {\bibfnamefont {M.~P.}\ \bibnamefont {{Busch}}}, \bibinfo
  {author} {\bibfnamefont {S.}~\bibnamefont {{Casertano}}}, \bibinfo {author}
  {\bibfnamefont {L.~M.}\ \bibnamefont {{Macri}}},\ and\ \bibinfo {author}
  {\bibfnamefont {W.}~\bibnamefont {{Yuan}}},\ }\bibfield  {title} {\bibinfo
  {title} {{A Sub-2\% Distance to M31 from Photometrically Homogeneous
  Near-infrared Cepheid Period-Luminosity Relations Measured with the Hubble
  Space Telescope}},\ }\href {https://doi.org/10.3847/1538-4357/ac1597}
  {\bibfield  {journal} {\bibinfo  {journal} {ApJ}\ }\textbf {\bibinfo {volume}
  {920}},\ \bibinfo {eid} {84} (\bibinfo {year} {2021})},\ \Eprint
  {https://arxiv.org/abs/2107.08029} {arXiv:2107.08029 [astro-ph.CO]}
  \BibitemShut {NoStop}%
\bibitem [{\citenamefont {{McDaniel}}\ \emph {et~al.}(2018)\citenamefont
  {{McDaniel}}, \citenamefont {{Jeltema}},\ and\ \citenamefont
  {{Profumo}}}]{2018PhRvD..97j3021M}%
  \BibitemOpen
  \bibfield  {author} {\bibinfo {author} {\bibfnamefont {A.}~\bibnamefont
  {{McDaniel}}}, \bibinfo {author} {\bibfnamefont {T.}~\bibnamefont
  {{Jeltema}}},\ and\ \bibinfo {author} {\bibfnamefont {S.}~\bibnamefont
  {{Profumo}}},\ }\bibfield  {title} {\bibinfo {title} {{Multiwavelength
  analysis of annihilating dark matter as the origin of the gamma-ray emission
  from M31}},\ }\href {https://doi.org/10.1103/PhysRevD.97.103021} {\bibfield
  {journal} {\bibinfo  {journal} {PhRvD}\ }\textbf {\bibinfo {volume} {97}},\
  \bibinfo {eid} {103021} (\bibinfo {year} {2018})},\ \Eprint
  {https://arxiv.org/abs/1802.05258} {arXiv:1802.05258 [astro-ph.HE]}
  \BibitemShut {NoStop}%
\bibitem [{\citenamefont {{Chan}}\ \emph {et~al.}(2019)\citenamefont {{Chan}},
  \citenamefont {{Cui}}, \citenamefont {{Liu}},\ and\ \citenamefont
  {{Leung}}}]{2019ApJ...872..177C}%
  \BibitemOpen
  \bibfield  {author} {\bibinfo {author} {\bibfnamefont {M.~H.}\ \bibnamefont
  {{Chan}}}, \bibinfo {author} {\bibfnamefont {L.}~\bibnamefont {{Cui}}},
  \bibinfo {author} {\bibfnamefont {J.}~\bibnamefont {{Liu}}},\ and\ \bibinfo
  {author} {\bibfnamefont {C.~S.}\ \bibnamefont {{Leung}}},\ }\bibfield
  {title} {\bibinfo {title} {{Ruling Out {\ensuremath{\sim}}100-300 GeV Thermal
  Relic Annihilating Dark Matter by Radio Observation of the Andromeda
  Galaxy}},\ }\href {https://doi.org/10.3847/1538-4357/aafe0b} {\bibfield
  {journal} {\bibinfo  {journal} {ApJ}\ }\textbf {\bibinfo {volume} {872}},\
  \bibinfo {eid} {177} (\bibinfo {year} {2019})},\ \Eprint
  {https://arxiv.org/abs/1901.04638} {arXiv:1901.04638 [astro-ph.GA]}
  \BibitemShut {NoStop}%
\bibitem [{\citenamefont {{Beck}}(2019)}]{2019JCAP...08..019B}%
  \BibitemOpen
  \bibfield  {author} {\bibinfo {author} {\bibfnamefont {G.}~\bibnamefont
  {{Beck}}},\ }\bibfield  {title} {\bibinfo {title} {{An excess of excesses
  examined via dark matter radio emissions from galaxies}},\ }\href
  {https://doi.org/10.1088/1475-7516/2019/08/019} {\bibfield  {journal}
  {\bibinfo  {journal} {JCAP}\ }\textbf {\bibinfo {volume} {2019}}\bibfield
  {number} {\bibinfo  {number} { (8)},\ \bibinfo {eid} {019}},\ }\Eprint
  {https://arxiv.org/abs/1905.05599} {arXiv:1905.05599 [astro-ph.HE]}
  \BibitemShut {NoStop}%
\bibitem [{\citenamefont {{Chan}}\ \emph {et~al.}(2021)\citenamefont {{Chan}},
  \citenamefont {{Yeung}}, \citenamefont {{Cui}},\ and\ \citenamefont
  {{Leung}}}]{2021MNRAS.501.5692C}%
  \BibitemOpen
  \bibfield  {author} {\bibinfo {author} {\bibfnamefont {M.~H.}\ \bibnamefont
  {{Chan}}}, \bibinfo {author} {\bibfnamefont {C.~F.}\ \bibnamefont {{Yeung}}},
  \bibinfo {author} {\bibfnamefont {L.}~\bibnamefont {{Cui}}},\ and\ \bibinfo
  {author} {\bibfnamefont {C.~S.}\ \bibnamefont {{Leung}}},\ }\bibfield
  {title} {\bibinfo {title} {{Analysing the radio flux density profile of the
  M31 galaxy: a possible dark matter interpretation}},\ }\href
  {https://doi.org/10.1093/mnras/staa4004} {\bibfield  {journal} {\bibinfo
  {journal} {MNRAS}\ }\textbf {\bibinfo {volume} {501}},\ \bibinfo {pages}
  {5692} (\bibinfo {year} {2021})},\ \Eprint {https://arxiv.org/abs/2101.00372}
  {arXiv:2101.00372 [astro-ph.HE]} \BibitemShut {NoStop}%
\bibitem [{\citenamefont {{Colafrancesco}}\ \emph {et~al.}(2006)\citenamefont
  {{Colafrancesco}}, \citenamefont {{Profumo}},\ and\ \citenamefont
  {{Ullio}}}]{2006A&A...455...21C}%
  \BibitemOpen
  \bibfield  {author} {\bibinfo {author} {\bibfnamefont {S.}~\bibnamefont
  {{Colafrancesco}}}, \bibinfo {author} {\bibfnamefont {S.}~\bibnamefont
  {{Profumo}}},\ and\ \bibinfo {author} {\bibfnamefont {P.}~\bibnamefont
  {{Ullio}}},\ }\bibfield  {title} {\bibinfo {title} {{Multi-frequency analysis
  of neutralino dark matter annihilations in the Coma cluster}},\ }\href
  {https://doi.org/10.1051/0004-6361:20053887} {\bibfield  {journal} {\bibinfo
  {journal} {A\&A}\ }\textbf {\bibinfo {volume} {455}},\ \bibinfo {pages} {21}
  (\bibinfo {year} {2006})},\ \Eprint {https://arxiv.org/abs/astro-ph/0507575}
  {arXiv:astro-ph/0507575 [astro-ph]} \BibitemShut {NoStop}%
\bibitem [{\citenamefont {{Chan}}\ \emph {et~al.}(2020)\citenamefont {{Chan}},
  \citenamefont {{Lee}}, \citenamefont {{Ng}},\ and\ \citenamefont
  {{Leung}}}]{2020ApJ...900..126C}%
  \BibitemOpen
  \bibfield  {author} {\bibinfo {author} {\bibfnamefont {M.~H.}\ \bibnamefont
  {{Chan}}}, \bibinfo {author} {\bibfnamefont {C.~M.}\ \bibnamefont {{Lee}}},
  \bibinfo {author} {\bibfnamefont {C.~Y.}\ \bibnamefont {{Ng}}},\ and\
  \bibinfo {author} {\bibfnamefont {C.~S.}\ \bibnamefont {{Leung}}},\
  }\bibfield  {title} {\bibinfo {title} {{Constraining Annihilating Dark Matter
  Mass by the Radio Continuum Spectral Data of a High-redshift Galaxy
  Cluster}},\ }\href {https://doi.org/10.3847/1538-4357/aba74b} {\bibfield
  {journal} {\bibinfo  {journal} {ApJ}\ }\textbf {\bibinfo {volume} {900}},\
  \bibinfo {eid} {126} (\bibinfo {year} {2020})},\ \Eprint
  {https://arxiv.org/abs/2007.06547} {arXiv:2007.06547 [astro-ph.HE]}
  \BibitemShut {NoStop}%
\bibitem [{\citenamefont {{Chan}}\ and\ \citenamefont
  {{Lee}}(2019)}]{2019PDU....2600355C}%
  \BibitemOpen
  \bibfield  {author} {\bibinfo {author} {\bibfnamefont {M.~H.}\ \bibnamefont
  {{Chan}}}\ and\ \bibinfo {author} {\bibfnamefont {C.~M.}\ \bibnamefont
  {{Lee}}},\ }\bibfield  {title} {\bibinfo {title} {{Fitting dark matter mass
  with the radio continuum spectral data of the Ophiuchus cluster}},\ }\href
  {https://doi.org/10.1016/j.dark.2019.100355} {\bibfield  {journal} {\bibinfo
  {journal} {PDU}\ }\textbf {\bibinfo {volume} {26}},\ \bibinfo {eid} {100355}
  (\bibinfo {year} {2019})},\ \Eprint {https://arxiv.org/abs/1908.03712}
  {arXiv:1908.03712 [astro-ph.HE]} \BibitemShut {NoStop}%
\bibitem [{\citenamefont {{Storm}}\ \emph {et~al.}(2013)\citenamefont
  {{Storm}}, \citenamefont {{Jeltema}}, \citenamefont {{Profumo}},\ and\
  \citenamefont {{Rudnick}}}]{2013ApJ...768..106S}%
  \BibitemOpen
  \bibfield  {author} {\bibinfo {author} {\bibfnamefont {E.}~\bibnamefont
  {{Storm}}}, \bibinfo {author} {\bibfnamefont {T.~E.}\ \bibnamefont
  {{Jeltema}}}, \bibinfo {author} {\bibfnamefont {S.}~\bibnamefont
  {{Profumo}}},\ and\ \bibinfo {author} {\bibfnamefont {L.}~\bibnamefont
  {{Rudnick}}},\ }\bibfield  {title} {\bibinfo {title} {{Constraints on Dark
  Matter Annihilation in Clusters of Galaxies from Diffuse Radio Emission}},\
  }\href {https://doi.org/10.1088/0004-637X/768/2/106} {\bibfield  {journal}
  {\bibinfo  {journal} {ApJ}\ }\textbf {\bibinfo {volume} {768}},\ \bibinfo
  {eid} {106} (\bibinfo {year} {2013})},\ \Eprint
  {https://arxiv.org/abs/1210.0872} {arXiv:1210.0872 [astro-ph.CO]}
  \BibitemShut {NoStop}%
\bibitem [{\citenamefont {{Chan}}(2021)}]{2021Galax...9...11C}%
  \BibitemOpen
  \bibfield  {author} {\bibinfo {author} {\bibfnamefont {M.~H.}\ \bibnamefont
  {{Chan}}},\ }\bibfield  {title} {\bibinfo {title} {{Radio Constraints of Dark
  Matter: A Review and Some Future Perspectives}},\ }\href
  {https://doi.org/10.3390/galaxies9010011} {\bibfield  {journal} {\bibinfo
  {journal} {Galax}\ }\textbf {\bibinfo {volume} {9}},\ \bibinfo {pages} {11}
  (\bibinfo {year} {2021})}\BibitemShut {NoStop}%
\bibitem [{\citenamefont {{McDaniel}}\ \emph {et~al.}(2017)\citenamefont
  {{McDaniel}}, \citenamefont {{Jeltema}}, \citenamefont {{Profumo}},\ and\
  \citenamefont {{Storm}}}]{2017JCAP...09..027M}%
  \BibitemOpen
  \bibfield  {author} {\bibinfo {author} {\bibfnamefont {A.}~\bibnamefont
  {{McDaniel}}}, \bibinfo {author} {\bibfnamefont {T.}~\bibnamefont
  {{Jeltema}}}, \bibinfo {author} {\bibfnamefont {S.}~\bibnamefont
  {{Profumo}}},\ and\ \bibinfo {author} {\bibfnamefont {E.}~\bibnamefont
  {{Storm}}},\ }\bibfield  {title} {\bibinfo {title} {{Multiwavelength analysis
  of dark matter annihilation and RX-DMFIT}},\ }\href
  {https://doi.org/10.1088/1475-7516/2017/09/027} {\bibfield  {journal}
  {\bibinfo  {journal} {JCAP}\ }\textbf {\bibinfo {volume} {2017}}\bibfield
  {number} {\bibinfo  {number} { (9)},\ \bibinfo {eid} {027}},\ }\Eprint
  {https://arxiv.org/abs/1705.09384} {arXiv:1705.09384 [astro-ph.HE]}
  \BibitemShut {NoStop}%
\bibitem [{\citenamefont {{Ackermann}}\ \emph {et~al.}(2017)\citenamefont
  {{Ackermann}}, \citenamefont {{Ajello}}, \citenamefont {{Albert}},
  \citenamefont {{Baldini}}, \citenamefont {{Ballet}}, \citenamefont
  {{Barbiellini}}, \citenamefont {{Bastieri}}, \citenamefont {{Bellazzini}},
  \citenamefont {{Bissaldi}}, \citenamefont {{Bloom}}, \citenamefont
  {{Bonino}}, \citenamefont {{Bottacini}}, \citenamefont {{Brandt}},
  \citenamefont {{Bregeon}}, \citenamefont {{Bruel}}, \citenamefont
  {{Buehler}}, \citenamefont {{Cameron}}, \citenamefont {{Caputo}},
  \citenamefont {{Caragiulo}}, \citenamefont {{Caraveo}}, \citenamefont
  {{Cavazzuti}}, \citenamefont {{Cecchi}}, \citenamefont {{Charles}},
  \citenamefont {{Chekhtman}}, \citenamefont {{Chiaro}}, \citenamefont
  {{Ciprini}}, \citenamefont {{Costanza}}, \citenamefont {{Cutini}},
  \citenamefont {{D'Ammando}}, \citenamefont {{de Palma}}, \citenamefont
  {{Desiante}}, \citenamefont {{Digel}}, \citenamefont {{Di Lalla}},
  \citenamefont {{Di Mauro}}, \citenamefont {{Di Venere}}, \citenamefont
  {{Favuzzi}}, \citenamefont {{Funk}}, \citenamefont {{Fusco}}, \citenamefont
  {{Gargano}}, \citenamefont {{Giglietto}}, \citenamefont {{Giordano}},
  \citenamefont {{Giroletti}}, \citenamefont {{Glanzman}}, \citenamefont
  {{Green}}, \citenamefont {{Grenier}}, \citenamefont {{Guillemot}},
  \citenamefont {{Guiriec}}, \citenamefont {{Hayashi}}, \citenamefont {{Hou}},
  \citenamefont {{J{\'o}hannesson}}, \citenamefont {{Kamae}}, \citenamefont
  {{Kn{\"o}dlseder}}, \citenamefont {{Kong}}, \citenamefont {{Kuss}},
  \citenamefont {{La Mura}}, \citenamefont {{Larsson}}, \citenamefont
  {{Latronico}}, \citenamefont {{Li}}, \citenamefont {{Longo}}, \citenamefont
  {{Loparco}}, \citenamefont {{Lubrano}}, \citenamefont {{Maldera}},
  \citenamefont {{Malyshev}}, \citenamefont {{Manfreda}}, \citenamefont
  {{Martin}}, \citenamefont {{Mazziotta}}, \citenamefont {{Michelson}},
  \citenamefont {{Mirabal}}, \citenamefont {{Mitthumsiri}}, \citenamefont
  {{Mizuno}}, \citenamefont {{Monzani}}, \citenamefont {{Morselli}},
  \citenamefont {{Moskalenko}}, \citenamefont {{Negro}}, \citenamefont
  {{Nuss}}, \citenamefont {{Ohsugi}}, \citenamefont {{Omodei}}, \citenamefont
  {{Orlando}}, \citenamefont {{Ormes}}, \citenamefont {{Paneque}},
  \citenamefont {{Persic}}, \citenamefont {{Pesce-Rollins}}, \citenamefont
  {{Piron}}, \citenamefont {{Porter}}, \citenamefont {{Principe}},
  \citenamefont {{Rain{\`o}}}, \citenamefont {{Rando}}, \citenamefont
  {{Razzano}}, \citenamefont {{Reimer}}, \citenamefont {{S{\'a}nchez-Conde}},
  \citenamefont {{Sgr{\`o}}}, \citenamefont {{Simone}}, \citenamefont
  {{Siskind}}, \citenamefont {{Spada}}, \citenamefont {{Spandre}},
  \citenamefont {{Spinelli}}, \citenamefont {{Tanaka}}, \citenamefont
  {{Tibaldo}}, \citenamefont {{Torres}}, \citenamefont {{Troja}}, \citenamefont
  {{Uchiyama}}, \citenamefont {{Wang}}, \citenamefont {{Wood}}, \citenamefont
  {{Wood}}, \citenamefont {{Zaharijas}},\ and\ \citenamefont
  {{Zhou}}}]{2017ApJ...836..208A}%
  \BibitemOpen
  \bibfield  {author} {\bibinfo {author} {\bibfnamefont {M.}~\bibnamefont
  {{Ackermann}}}, \bibinfo {author} {\bibfnamefont {M.}~\bibnamefont
  {{Ajello}}}, \bibinfo {author} {\bibfnamefont {A.}~\bibnamefont {{Albert}}},
  \bibinfo {author} {\bibfnamefont {L.}~\bibnamefont {{Baldini}}}, \bibinfo
  {author} {\bibfnamefont {J.}~\bibnamefont {{Ballet}}}, \bibinfo {author}
  {\bibfnamefont {G.}~\bibnamefont {{Barbiellini}}}, \bibinfo {author}
  {\bibfnamefont {D.}~\bibnamefont {{Bastieri}}}, \bibinfo {author}
  {\bibfnamefont {R.}~\bibnamefont {{Bellazzini}}}, \bibinfo {author}
  {\bibfnamefont {E.}~\bibnamefont {{Bissaldi}}}, \bibinfo {author}
  {\bibfnamefont {E.~D.}\ \bibnamefont {{Bloom}}}, \bibinfo {author}
  {\bibfnamefont {R.}~\bibnamefont {{Bonino}}}, \bibinfo {author}
  {\bibfnamefont {E.}~\bibnamefont {{Bottacini}}}, \bibinfo {author}
  {\bibfnamefont {T.~J.}\ \bibnamefont {{Brandt}}}, \bibinfo {author}
  {\bibfnamefont {J.}~\bibnamefont {{Bregeon}}}, \bibinfo {author}
  {\bibfnamefont {P.}~\bibnamefont {{Bruel}}}, \bibinfo {author} {\bibfnamefont
  {R.}~\bibnamefont {{Buehler}}}, \bibinfo {author} {\bibfnamefont {R.~A.}\
  \bibnamefont {{Cameron}}}, \bibinfo {author} {\bibfnamefont {R.}~\bibnamefont
  {{Caputo}}}, \bibinfo {author} {\bibfnamefont {M.}~\bibnamefont
  {{Caragiulo}}}, \bibinfo {author} {\bibfnamefont {P.~A.}\ \bibnamefont
  {{Caraveo}}}, \bibinfo {author} {\bibfnamefont {E.}~\bibnamefont
  {{Cavazzuti}}}, \bibinfo {author} {\bibfnamefont {C.}~\bibnamefont
  {{Cecchi}}}, \bibinfo {author} {\bibfnamefont {E.}~\bibnamefont {{Charles}}},
  \bibinfo {author} {\bibfnamefont {A.}~\bibnamefont {{Chekhtman}}}, \bibinfo
  {author} {\bibfnamefont {G.}~\bibnamefont {{Chiaro}}}, \bibinfo {author}
  {\bibfnamefont {S.}~\bibnamefont {{Ciprini}}}, \bibinfo {author}
  {\bibfnamefont {F.}~\bibnamefont {{Costanza}}}, \bibinfo {author}
  {\bibfnamefont {S.}~\bibnamefont {{Cutini}}}, \bibinfo {author}
  {\bibfnamefont {F.}~\bibnamefont {{D'Ammando}}}, \bibinfo {author}
  {\bibfnamefont {F.}~\bibnamefont {{de Palma}}}, \bibinfo {author}
  {\bibfnamefont {R.}~\bibnamefont {{Desiante}}}, \bibinfo {author}
  {\bibfnamefont {S.~W.}\ \bibnamefont {{Digel}}}, \bibinfo {author}
  {\bibfnamefont {N.}~\bibnamefont {{Di Lalla}}}, \bibinfo {author}
  {\bibfnamefont {M.}~\bibnamefont {{Di Mauro}}}, \bibinfo {author}
  {\bibfnamefont {L.}~\bibnamefont {{Di Venere}}}, \bibinfo {author}
  {\bibfnamefont {C.}~\bibnamefont {{Favuzzi}}}, \bibinfo {author}
  {\bibfnamefont {S.}~\bibnamefont {{Funk}}}, \bibinfo {author} {\bibfnamefont
  {P.}~\bibnamefont {{Fusco}}}, \bibinfo {author} {\bibfnamefont
  {F.}~\bibnamefont {{Gargano}}}, \bibinfo {author} {\bibfnamefont
  {N.}~\bibnamefont {{Giglietto}}}, \bibinfo {author} {\bibfnamefont
  {F.}~\bibnamefont {{Giordano}}}, \bibinfo {author} {\bibfnamefont
  {M.}~\bibnamefont {{Giroletti}}}, \bibinfo {author} {\bibfnamefont
  {T.}~\bibnamefont {{Glanzman}}}, \bibinfo {author} {\bibfnamefont
  {D.}~\bibnamefont {{Green}}}, \bibinfo {author} {\bibfnamefont {I.~A.}\
  \bibnamefont {{Grenier}}}, \bibinfo {author} {\bibfnamefont {L.}~\bibnamefont
  {{Guillemot}}}, \bibinfo {author} {\bibfnamefont {S.}~\bibnamefont
  {{Guiriec}}}, \bibinfo {author} {\bibfnamefont {K.}~\bibnamefont
  {{Hayashi}}}, \bibinfo {author} {\bibfnamefont {X.}~\bibnamefont {{Hou}}},
  \bibinfo {author} {\bibfnamefont {G.}~\bibnamefont {{J{\'o}hannesson}}},
  \bibinfo {author} {\bibfnamefont {T.}~\bibnamefont {{Kamae}}}, \bibinfo
  {author} {\bibfnamefont {J.}~\bibnamefont {{Kn{\"o}dlseder}}}, \bibinfo
  {author} {\bibfnamefont {A.~K.~H.}\ \bibnamefont {{Kong}}}, \bibinfo {author}
  {\bibfnamefont {M.}~\bibnamefont {{Kuss}}}, \bibinfo {author} {\bibfnamefont
  {G.}~\bibnamefont {{La Mura}}}, \bibinfo {author} {\bibfnamefont
  {S.}~\bibnamefont {{Larsson}}}, \bibinfo {author} {\bibfnamefont
  {L.}~\bibnamefont {{Latronico}}}, \bibinfo {author} {\bibfnamefont
  {J.}~\bibnamefont {{Li}}}, \bibinfo {author} {\bibfnamefont {F.}~\bibnamefont
  {{Longo}}}, \bibinfo {author} {\bibfnamefont {F.}~\bibnamefont {{Loparco}}},
  \bibinfo {author} {\bibfnamefont {P.}~\bibnamefont {{Lubrano}}}, \bibinfo
  {author} {\bibfnamefont {S.}~\bibnamefont {{Maldera}}}, \bibinfo {author}
  {\bibfnamefont {D.}~\bibnamefont {{Malyshev}}}, \bibinfo {author}
  {\bibfnamefont {A.}~\bibnamefont {{Manfreda}}}, \bibinfo {author}
  {\bibfnamefont {P.}~\bibnamefont {{Martin}}}, \bibinfo {author}
  {\bibfnamefont {M.~N.}\ \bibnamefont {{Mazziotta}}}, \bibinfo {author}
  {\bibfnamefont {P.~F.}\ \bibnamefont {{Michelson}}}, \bibinfo {author}
  {\bibfnamefont {N.}~\bibnamefont {{Mirabal}}}, \bibinfo {author}
  {\bibfnamefont {W.}~\bibnamefont {{Mitthumsiri}}}, \bibinfo {author}
  {\bibfnamefont {T.}~\bibnamefont {{Mizuno}}}, \bibinfo {author}
  {\bibfnamefont {M.~E.}\ \bibnamefont {{Monzani}}}, \bibinfo {author}
  {\bibfnamefont {A.}~\bibnamefont {{Morselli}}}, \bibinfo {author}
  {\bibfnamefont {I.~V.}\ \bibnamefont {{Moskalenko}}}, \bibinfo {author}
  {\bibfnamefont {M.}~\bibnamefont {{Negro}}}, \bibinfo {author} {\bibfnamefont
  {E.}~\bibnamefont {{Nuss}}}, \bibinfo {author} {\bibfnamefont
  {T.}~\bibnamefont {{Ohsugi}}}, \bibinfo {author} {\bibfnamefont
  {N.}~\bibnamefont {{Omodei}}}, \bibinfo {author} {\bibfnamefont
  {E.}~\bibnamefont {{Orlando}}}, \bibinfo {author} {\bibfnamefont {J.~F.}\
  \bibnamefont {{Ormes}}}, \bibinfo {author} {\bibfnamefont {D.}~\bibnamefont
  {{Paneque}}}, \bibinfo {author} {\bibfnamefont {M.}~\bibnamefont {{Persic}}},
  \bibinfo {author} {\bibfnamefont {M.}~\bibnamefont {{Pesce-Rollins}}},
  \bibinfo {author} {\bibfnamefont {F.}~\bibnamefont {{Piron}}}, \bibinfo
  {author} {\bibfnamefont {T.~A.}\ \bibnamefont {{Porter}}}, \bibinfo {author}
  {\bibfnamefont {G.}~\bibnamefont {{Principe}}}, \bibinfo {author}
  {\bibfnamefont {S.}~\bibnamefont {{Rain{\`o}}}}, \bibinfo {author}
  {\bibfnamefont {R.}~\bibnamefont {{Rando}}}, \bibinfo {author} {\bibfnamefont
  {M.}~\bibnamefont {{Razzano}}}, \bibinfo {author} {\bibfnamefont
  {O.}~\bibnamefont {{Reimer}}}, \bibinfo {author} {\bibfnamefont
  {M.}~\bibnamefont {{S{\'a}nchez-Conde}}}, \bibinfo {author} {\bibfnamefont
  {C.}~\bibnamefont {{Sgr{\`o}}}}, \bibinfo {author} {\bibfnamefont
  {D.}~\bibnamefont {{Simone}}}, \bibinfo {author} {\bibfnamefont {E.~J.}\
  \bibnamefont {{Siskind}}}, \bibinfo {author} {\bibfnamefont {F.}~\bibnamefont
  {{Spada}}}, \bibinfo {author} {\bibfnamefont {G.}~\bibnamefont {{Spandre}}},
  \bibinfo {author} {\bibfnamefont {P.}~\bibnamefont {{Spinelli}}}, \bibinfo
  {author} {\bibfnamefont {K.}~\bibnamefont {{Tanaka}}}, \bibinfo {author}
  {\bibfnamefont {L.}~\bibnamefont {{Tibaldo}}}, \bibinfo {author}
  {\bibfnamefont {D.~F.}\ \bibnamefont {{Torres}}}, \bibinfo {author}
  {\bibfnamefont {E.}~\bibnamefont {{Troja}}}, \bibinfo {author} {\bibfnamefont
  {Y.}~\bibnamefont {{Uchiyama}}}, \bibinfo {author} {\bibfnamefont {J.~C.}\
  \bibnamefont {{Wang}}}, \bibinfo {author} {\bibfnamefont {K.~S.}\
  \bibnamefont {{Wood}}}, \bibinfo {author} {\bibfnamefont {M.}~\bibnamefont
  {{Wood}}}, \bibinfo {author} {\bibfnamefont {G.}~\bibnamefont
  {{Zaharijas}}},\ and\ \bibinfo {author} {\bibfnamefont {M.}~\bibnamefont
  {{Zhou}}},\ }\bibfield  {title} {\bibinfo {title} {{Observations of M31 and
  M33 with the Fermi Large Area Telescope: A Galactic Center Excess in
  Andromeda?}},\ }\href {https://doi.org/10.3847/1538-4357/aa5c3d} {\bibfield
  {journal} {\bibinfo  {journal} {ApJ}\ }\textbf {\bibinfo {volume} {836}},\
  \bibinfo {eid} {208} (\bibinfo {year} {2017})},\ \Eprint
  {https://arxiv.org/abs/1702.08602} {arXiv:1702.08602 [astro-ph.HE]}
  \BibitemShut {NoStop}%
\bibitem [{\citenamefont {{Ruiz-Granados}}\ \emph {et~al.}(2010)\citenamefont
  {{Ruiz-Granados}}, \citenamefont {{Rubi{\~n}o-Mart{\'\i}n}}, \citenamefont
  {{Florido}},\ and\ \citenamefont {{Battaner}}}]{2010ApJ...723L..44R}%
  \BibitemOpen
  \bibfield  {author} {\bibinfo {author} {\bibfnamefont {B.}~\bibnamefont
  {{Ruiz-Granados}}}, \bibinfo {author} {\bibfnamefont {J.~A.}\ \bibnamefont
  {{Rubi{\~n}o-Mart{\'\i}n}}}, \bibinfo {author} {\bibfnamefont
  {E.}~\bibnamefont {{Florido}}},\ and\ \bibinfo {author} {\bibfnamefont
  {E.}~\bibnamefont {{Battaner}}},\ }\bibfield  {title} {\bibinfo {title}
  {{Magnetic Fields and the Outer Rotation Curve of M31}},\ }\href
  {https://doi.org/10.1088/2041-8205/723/1/L44} {\bibfield  {journal} {\bibinfo
   {journal} {ApJL}\ }\textbf {\bibinfo {volume} {723}},\ \bibinfo {pages}
  {L44} (\bibinfo {year} {2010})},\ \Eprint {https://arxiv.org/abs/1010.0270}
  {arXiv:1010.0270 [astro-ph.CO]} \BibitemShut {NoStop}%
\bibitem [{\citenamefont {{Gie{\ss}{\"u}bel}}\ and\ \citenamefont
  {{Beck}}(2014)}]{2014A&A...571A..61G}%
  \BibitemOpen
  \bibfield  {author} {\bibinfo {author} {\bibfnamefont {R.}~\bibnamefont
  {{Gie{\ss}{\"u}bel}}}\ and\ \bibinfo {author} {\bibfnamefont
  {R.}~\bibnamefont {{Beck}}},\ }\bibfield  {title} {\bibinfo {title} {{The
  magnetic field structure of the central region in M 31}},\ }\href
  {https://doi.org/10.1051/0004-6361/201323211} {\bibfield  {journal} {\bibinfo
   {journal} {A\&A}\ }\textbf {\bibinfo {volume} {571}},\ \bibinfo {eid} {A61}
  (\bibinfo {year} {2014})},\ \Eprint {https://arxiv.org/abs/1408.4582}
  {arXiv:1408.4582 [astro-ph.GA]} \BibitemShut {NoStop}%
\bibitem [{\citenamefont {{Egorov}}\ \emph {et~al.}(2016)\citenamefont
  {{Egorov}}, \citenamefont {{Gaskins}}, \citenamefont {{Pierpaoli}},\ and\
  \citenamefont {{Pietrobon}}}]{2016JCAP...03..060E}%
  \BibitemOpen
  \bibfield  {author} {\bibinfo {author} {\bibfnamefont {A.~E.}\ \bibnamefont
  {{Egorov}}}, \bibinfo {author} {\bibfnamefont {J.~M.}\ \bibnamefont
  {{Gaskins}}}, \bibinfo {author} {\bibfnamefont {E.}~\bibnamefont
  {{Pierpaoli}}},\ and\ \bibinfo {author} {\bibfnamefont {D.}~\bibnamefont
  {{Pietrobon}}},\ }\bibfield  {title} {\bibinfo {title} {{Dark matter
  implications of the WMAP-Planck Haze}},\ }\href
  {https://doi.org/10.1088/1475-7516/2016/03/060} {\bibfield  {journal}
  {\bibinfo  {journal} {J. Cosmology Astropart. Phys.}\ }\textbf {\bibinfo
  {volume} {03}},\ \bibinfo {eid} {060} (\bibinfo {year} {2016})},\ \Eprint
  {https://arxiv.org/abs/1509.05135} {arXiv:1509.05135 [astro-ph.CO]}
  \BibitemShut {NoStop}%
\bibitem [{GP()}]{GP}%
  \BibitemOpen
  \href@noop {} {}\bibinfo {howpublished}
  {\url{http://galprop.stanford.edu/}}\BibitemShut {NoStop}%
\bibitem [{\citenamefont {{Moskalenko}}\ \emph {et~al.}(2019)\citenamefont
  {{Moskalenko}}, \citenamefont {{Johannesson}},\ and\ \citenamefont
  {{Porter}}}]{2019ICRC...36..111M}%
  \BibitemOpen
  \bibfield  {author} {\bibinfo {author} {\bibfnamefont {I.}~\bibnamefont
  {{Moskalenko}}}, \bibinfo {author} {\bibfnamefont {G.}~\bibnamefont
  {{Johannesson}}},\ and\ \bibinfo {author} {\bibfnamefont {T.}~\bibnamefont
  {{Porter}}},\ }\bibfield  {title} {\bibinfo {title} {{GALPROP Code for
  Galactic Cosmic Ray Propagation and Associated Photon Emissions}},\ }in\
  \href@noop {} {\emph {\bibinfo {booktitle} {36th International Cosmic Ray
  Conference (ICRC2019)}}},\ \bibinfo {series} {International Cosmic Ray
  Conference}, Vol.~\bibinfo {volume} {36}\ (\bibinfo {year} {2019})\ p.\
  \bibinfo {pages} {111}\BibitemShut {NoStop}%
\bibitem [{git()}]{github}%
  \BibitemOpen
  \href@noop {} {}\bibinfo {howpublished}
  {\url{https://github.com/a-e-egorov/GALPROP_DM}}\BibitemShut {NoStop}%
\bibitem [{PPP()}]{PPPC}%
  \BibitemOpen
  \href@noop {} {}\bibinfo {howpublished}
  {\url{http://www.marcocirelli.net/PPPC4DMID.html}}\BibitemShut {NoStop}%
\bibitem [{\citenamefont {{Cirelli}}\ \emph {et~al.}(2011)\citenamefont
  {{Cirelli}}, \citenamefont {{Corcella}}, \citenamefont {{Hektor}},
  \citenamefont {{H{\"u}tsi}}, \citenamefont {{Kadastik}}, \citenamefont
  {{Panci}}, \citenamefont {{Raidal}}, \citenamefont {{Sala}},\ and\
  \citenamefont {{Strumia}}}]{2011JCAP...03..051C}%
  \BibitemOpen
  \bibfield  {author} {\bibinfo {author} {\bibfnamefont {M.}~\bibnamefont
  {{Cirelli}}}, \bibinfo {author} {\bibfnamefont {G.}~\bibnamefont
  {{Corcella}}}, \bibinfo {author} {\bibfnamefont {A.}~\bibnamefont
  {{Hektor}}}, \bibinfo {author} {\bibfnamefont {G.}~\bibnamefont
  {{H{\"u}tsi}}}, \bibinfo {author} {\bibfnamefont {M.}~\bibnamefont
  {{Kadastik}}}, \bibinfo {author} {\bibfnamefont {P.}~\bibnamefont {{Panci}}},
  \bibinfo {author} {\bibfnamefont {M.}~\bibnamefont {{Raidal}}}, \bibinfo
  {author} {\bibfnamefont {F.}~\bibnamefont {{Sala}}},\ and\ \bibinfo {author}
  {\bibfnamefont {A.}~\bibnamefont {{Strumia}}},\ }\bibfield  {title} {\bibinfo
  {title} {{PPPC 4 DM ID: a poor particle physicist cookbook for dark matter
  indirect detection}},\ }\href {https://doi.org/10.1088/1475-7516/2011/03/051}
  {\bibfield  {journal} {\bibinfo  {journal} {J. Cosmology Astropart. Phys.}\
  }\textbf {\bibinfo {volume} {3}},\ \bibinfo {eid} {051} (\bibinfo {year}
  {2011})},\ \Eprint {https://arxiv.org/abs/1012.4515} {arXiv:1012.4515
  [hep-ph]} \BibitemShut {NoStop}%
\bibitem [{\citenamefont {{G{\'e}nolini}}\ \emph {et~al.}(2021)\citenamefont
  {{G{\'e}nolini}}, \citenamefont {{Boudaud}}, \citenamefont {{Cirelli}},
  \citenamefont {{Derome}}, \citenamefont {{Lavalle}}, \citenamefont
  {{Maurin}}, \citenamefont {{Salati}},\ and\ \citenamefont
  {{Weinrich}}}]{2021PhRvD.104h3005G}%
  \BibitemOpen
  \bibfield  {author} {\bibinfo {author} {\bibfnamefont {Y.}~\bibnamefont
  {{G{\'e}nolini}}}, \bibinfo {author} {\bibfnamefont {M.}~\bibnamefont
  {{Boudaud}}}, \bibinfo {author} {\bibfnamefont {M.}~\bibnamefont
  {{Cirelli}}}, \bibinfo {author} {\bibfnamefont {L.}~\bibnamefont {{Derome}}},
  \bibinfo {author} {\bibfnamefont {J.}~\bibnamefont {{Lavalle}}}, \bibinfo
  {author} {\bibfnamefont {D.}~\bibnamefont {{Maurin}}}, \bibinfo {author}
  {\bibfnamefont {P.}~\bibnamefont {{Salati}}},\ and\ \bibinfo {author}
  {\bibfnamefont {N.}~\bibnamefont {{Weinrich}}},\ }\bibfield  {title}
  {\bibinfo {title} {{New minimal, median, and maximal propagation models for
  dark matter searches with Galactic cosmic rays}},\ }\href
  {https://doi.org/10.1103/PhysRevD.104.083005} {\bibfield  {journal} {\bibinfo
   {journal} {\prd}\ }\textbf {\bibinfo {volume} {104}},\ \bibinfo {eid}
  {083005} (\bibinfo {year} {2021})},\ \Eprint
  {https://arxiv.org/abs/2103.04108} {arXiv:2103.04108 [astro-ph.HE]}
  \BibitemShut {NoStop}%
\bibitem [{\citenamefont {{Cirelli}}\ and\ \citenamefont
  {{Taoso}}(2016)}]{2016JCAP...07..041C}%
  \BibitemOpen
  \bibfield  {author} {\bibinfo {author} {\bibfnamefont {M.}~\bibnamefont
  {{Cirelli}}}\ and\ \bibinfo {author} {\bibfnamefont {M.}~\bibnamefont
  {{Taoso}}},\ }\bibfield  {title} {\bibinfo {title} {{Updated galactic radio
  constraints on Dark Matter}},\ }\href
  {https://doi.org/10.1088/1475-7516/2016/07/041} {\bibfield  {journal}
  {\bibinfo  {journal} {JCAP}\ }\textbf {\bibinfo {volume} {2016}}\bibfield
  {number} {\bibinfo  {number} { (7)},\ \bibinfo {eid} {041}},\ }\Eprint
  {https://arxiv.org/abs/1604.06267} {arXiv:1604.06267 [hep-ph]} \BibitemShut
  {NoStop}%
\bibitem [{\citenamefont {{Saikawa}}\ and\ \citenamefont
  {{Shirai}}(2020)}]{2020JCAP...08..011S}%
  \BibitemOpen
  \bibfield  {author} {\bibinfo {author} {\bibfnamefont {K.}~\bibnamefont
  {{Saikawa}}}\ and\ \bibinfo {author} {\bibfnamefont {S.}~\bibnamefont
  {{Shirai}}},\ }\bibfield  {title} {\bibinfo {title} {{Precise WIMP dark
  matter abundance and Standard Model thermodynamics}},\ }\href
  {https://doi.org/10.1088/1475-7516/2020/08/011} {\bibfield  {journal}
  {\bibinfo  {journal} {JCAP}\ }\textbf {\bibinfo {volume} {2020}}\bibfield
  {number} {\bibinfo  {number} { (8)},\ \bibinfo {eid} {011}},\ }\Eprint
  {https://arxiv.org/abs/2005.03544} {arXiv:2005.03544 [hep-ph]} \BibitemShut
  {NoStop}%
\bibitem [{sv()}]{sv}%
  \BibitemOpen
  \href@noop {} {}\bibinfo {howpublished}
  {\url{https://member.ipmu.jp/satoshi.shirai/DM2020/}}\BibitemShut {NoStop}%
\bibitem [{\citenamefont {{Banerjee}}\ and\ \citenamefont
  {{Jog}}(2008)}]{2008ApJ...685..254B}%
  \BibitemOpen
  \bibfield  {author} {\bibinfo {author} {\bibfnamefont {A.}~\bibnamefont
  {{Banerjee}}}\ and\ \bibinfo {author} {\bibfnamefont {C.~J.}\ \bibnamefont
  {{Jog}}},\ }\bibfield  {title} {\bibinfo {title} {{The Flattened Dark Matter
  Halo of M31 as Deduced from the Observed H I Scale Heights}},\ }\href
  {https://doi.org/10.1086/591223} {\bibfield  {journal} {\bibinfo  {journal}
  {ApJ}\ }\textbf {\bibinfo {volume} {685}},\ \bibinfo {pages} {254} (\bibinfo
  {year} {2008})},\ \Eprint {https://arxiv.org/abs/0806.3610} {arXiv:0806.3610
  [astro-ph]} \BibitemShut {NoStop}%
\bibitem [{\citenamefont {{Hayashi}}\ and\ \citenamefont
  {{Chiba}}(2014)}]{2014ApJ...789...62H}%
  \BibitemOpen
  \bibfield  {author} {\bibinfo {author} {\bibfnamefont {K.}~\bibnamefont
  {{Hayashi}}}\ and\ \bibinfo {author} {\bibfnamefont {M.}~\bibnamefont
  {{Chiba}}},\ }\bibfield  {title} {\bibinfo {title} {{The Prolate Dark Matter
  Halo of the Andromeda Galaxy}},\ }\href
  {https://doi.org/10.1088/0004-637X/789/1/62} {\bibfield  {journal} {\bibinfo
  {journal} {ApJ}\ }\textbf {\bibinfo {volume} {789}},\ \bibinfo {eid} {62}
  (\bibinfo {year} {2014})},\ \Eprint {https://arxiv.org/abs/1405.4606}
  {arXiv:1405.4606 [astro-ph.GA]} \BibitemShut {NoStop}%
\bibitem [{\citenamefont {{Geehan}}\ \emph {et~al.}(2006)\citenamefont
  {{Geehan}}, \citenamefont {{Fardal}}, \citenamefont {{Babul}},\ and\
  \citenamefont {{Guhathakurta}}}]{2006MNRAS.366..996G}%
  \BibitemOpen
  \bibfield  {author} {\bibinfo {author} {\bibfnamefont {J.~J.}\ \bibnamefont
  {{Geehan}}}, \bibinfo {author} {\bibfnamefont {M.~A.}\ \bibnamefont
  {{Fardal}}}, \bibinfo {author} {\bibfnamefont {A.}~\bibnamefont {{Babul}}},\
  and\ \bibinfo {author} {\bibfnamefont {P.}~\bibnamefont {{Guhathakurta}}},\
  }\bibfield  {title} {\bibinfo {title} {{Investigating the Andromeda stream -
  I. Simple analytic bulge-disc-halo model for M31}},\ }\href
  {https://doi.org/10.1111/j.1365-2966.2005.09863.x} {\bibfield  {journal}
  {\bibinfo  {journal} {MNRAS}\ }\textbf {\bibinfo {volume} {366}},\ \bibinfo
  {pages} {996} (\bibinfo {year} {2006})},\ \Eprint
  {https://arxiv.org/abs/astro-ph/0501240} {arXiv:astro-ph/0501240 [astro-ph]}
  \BibitemShut {NoStop}%
\bibitem [{\citenamefont {{Seigar}}\ \emph {et~al.}(2008)\citenamefont
  {{Seigar}}, \citenamefont {{Barth}},\ and\ \citenamefont
  {{Bullock}}}]{2008MNRAS.389.1911S}%
  \BibitemOpen
  \bibfield  {author} {\bibinfo {author} {\bibfnamefont {M.~S.}\ \bibnamefont
  {{Seigar}}}, \bibinfo {author} {\bibfnamefont {A.~J.}\ \bibnamefont
  {{Barth}}},\ and\ \bibinfo {author} {\bibfnamefont {J.~S.}\ \bibnamefont
  {{Bullock}}},\ }\bibfield  {title} {\bibinfo {title} {{A revised
  {\ensuremath{\Lambda}} CDM mass model for the Andromeda Galaxy}},\ }\href
  {https://doi.org/10.1111/j.1365-2966.2008.13732.x} {\bibfield  {journal}
  {\bibinfo  {journal} {MNRAS}\ }\textbf {\bibinfo {volume} {389}},\ \bibinfo
  {pages} {1911} (\bibinfo {year} {2008})},\ \Eprint
  {https://arxiv.org/abs/astro-ph/0612228} {arXiv:astro-ph/0612228 [astro-ph]}
  \BibitemShut {NoStop}%
\bibitem [{\citenamefont {{Chemin}}\ \emph {et~al.}(2009)\citenamefont
  {{Chemin}}, \citenamefont {{Carignan}},\ and\ \citenamefont
  {{Foster}}}]{2009ApJ...705.1395C}%
  \BibitemOpen
  \bibfield  {author} {\bibinfo {author} {\bibfnamefont {L.}~\bibnamefont
  {{Chemin}}}, \bibinfo {author} {\bibfnamefont {C.}~\bibnamefont
  {{Carignan}}},\ and\ \bibinfo {author} {\bibfnamefont {T.}~\bibnamefont
  {{Foster}}},\ }\bibfield  {title} {\bibinfo {title} {{H I Kinematics and
  Dynamics of Messier 31}},\ }\href
  {https://doi.org/10.1088/0004-637X/705/2/1395} {\bibfield  {journal}
  {\bibinfo  {journal} {ApJ}\ }\textbf {\bibinfo {volume} {705}},\ \bibinfo
  {pages} {1395} (\bibinfo {year} {2009})},\ \Eprint
  {https://arxiv.org/abs/0909.3846} {arXiv:0909.3846 [astro-ph.CO]}
  \BibitemShut {NoStop}%
\bibitem [{\citenamefont {{Corbelli}}\ \emph {et~al.}(2010)\citenamefont
  {{Corbelli}}, \citenamefont {{Lorenzoni}}, \citenamefont {{Walterbos}},
  \citenamefont {{Braun}},\ and\ \citenamefont
  {{Thilker}}}]{2010A&A...511A..89C}%
  \BibitemOpen
  \bibfield  {author} {\bibinfo {author} {\bibfnamefont {E.}~\bibnamefont
  {{Corbelli}}}, \bibinfo {author} {\bibfnamefont {S.}~\bibnamefont
  {{Lorenzoni}}}, \bibinfo {author} {\bibfnamefont {R.}~\bibnamefont
  {{Walterbos}}}, \bibinfo {author} {\bibfnamefont {R.}~\bibnamefont
  {{Braun}}},\ and\ \bibinfo {author} {\bibfnamefont {D.}~\bibnamefont
  {{Thilker}}},\ }\bibfield  {title} {\bibinfo {title} {{A wide-field H I
  mosaic of Messier 31. II. The disk warp, rotation, and the dark matter
  halo}},\ }\href {https://doi.org/10.1051/0004-6361/200913297} {\bibfield
  {journal} {\bibinfo  {journal} {A\&A}\ }\textbf {\bibinfo {volume} {511}},\
  \bibinfo {eid} {A89} (\bibinfo {year} {2010})},\ \Eprint
  {https://arxiv.org/abs/0912.4133} {arXiv:0912.4133 [astro-ph.CO]}
  \BibitemShut {NoStop}%
\bibitem [{\citenamefont {{Tamm}}\ \emph {et~al.}(2012)\citenamefont {{Tamm}},
  \citenamefont {{Tempel}}, \citenamefont {{Tenjes}}, \citenamefont
  {{Tihhonova}},\ and\ \citenamefont {{Tuvikene}}}]{2012A&A...546A...4T}%
  \BibitemOpen
  \bibfield  {author} {\bibinfo {author} {\bibfnamefont {A.}~\bibnamefont
  {{Tamm}}}, \bibinfo {author} {\bibfnamefont {E.}~\bibnamefont {{Tempel}}},
  \bibinfo {author} {\bibfnamefont {P.}~\bibnamefont {{Tenjes}}}, \bibinfo
  {author} {\bibfnamefont {O.}~\bibnamefont {{Tihhonova}}},\ and\ \bibinfo
  {author} {\bibfnamefont {T.}~\bibnamefont {{Tuvikene}}},\ }\bibfield  {title}
  {\bibinfo {title} {{Stellar mass map and dark matter distribution in M 31}},\
  }\href {https://doi.org/10.1051/0004-6361/201220065} {\bibfield  {journal}
  {\bibinfo  {journal} {A\&A}\ }\textbf {\bibinfo {volume} {546}},\ \bibinfo
  {eid} {A4} (\bibinfo {year} {2012})},\ \Eprint
  {https://arxiv.org/abs/1208.5712} {arXiv:1208.5712 [astro-ph.CO]}
  \BibitemShut {NoStop}%
\bibitem [{\citenamefont {{Bla{\~n}a D{\'\i}az}}\ \emph
  {et~al.}(2018)\citenamefont {{Bla{\~n}a D{\'\i}az}}, \citenamefont
  {{Gerhard}}, \citenamefont {{Wegg}}, \citenamefont {{Portail}}, \citenamefont
  {{Opitsch}}, \citenamefont {{Saglia}}, \citenamefont {{Fabricius}},
  \citenamefont {{Erwin}},\ and\ \citenamefont
  {{Bender}}}]{2018MNRAS.481.3210B}%
  \BibitemOpen
  \bibfield  {author} {\bibinfo {author} {\bibfnamefont {M.}~\bibnamefont
  {{Bla{\~n}a D{\'\i}az}}}, \bibinfo {author} {\bibfnamefont {O.}~\bibnamefont
  {{Gerhard}}}, \bibinfo {author} {\bibfnamefont {C.}~\bibnamefont {{Wegg}}},
  \bibinfo {author} {\bibfnamefont {M.}~\bibnamefont {{Portail}}}, \bibinfo
  {author} {\bibfnamefont {M.}~\bibnamefont {{Opitsch}}}, \bibinfo {author}
  {\bibfnamefont {R.}~\bibnamefont {{Saglia}}}, \bibinfo {author}
  {\bibfnamefont {M.}~\bibnamefont {{Fabricius}}}, \bibinfo {author}
  {\bibfnamefont {P.}~\bibnamefont {{Erwin}}},\ and\ \bibinfo {author}
  {\bibfnamefont {R.}~\bibnamefont {{Bender}}},\ }\bibfield  {title} {\bibinfo
  {title} {{Sculpting Andromeda - made-to-measure models for M31's bar and
  composite bulge: dynamics, stellar and dark matter mass}},\ }\href
  {https://doi.org/10.1093/mnras/sty2311} {\bibfield  {journal} {\bibinfo
  {journal} {MNRAS}\ }\textbf {\bibinfo {volume} {481}},\ \bibinfo {pages}
  {3210} (\bibinfo {year} {2018})},\ \Eprint {https://arxiv.org/abs/1808.07494}
  {arXiv:1808.07494 [astro-ph.GA]} \BibitemShut {NoStop}%
\bibitem [{\citenamefont {{Boldrini}}\ \emph {et~al.}(2021)\citenamefont
  {{Boldrini}}, \citenamefont {{Mohayaee}},\ and\ \citenamefont
  {{Silk}}}]{2021ApJ...919...86B}%
  \BibitemOpen
  \bibfield  {author} {\bibinfo {author} {\bibfnamefont {P.}~\bibnamefont
  {{Boldrini}}}, \bibinfo {author} {\bibfnamefont {R.}~\bibnamefont
  {{Mohayaee}}},\ and\ \bibinfo {author} {\bibfnamefont {J.}~\bibnamefont
  {{Silk}}},\ }\bibfield  {title} {\bibinfo {title} {{Flattening of Dark Matter
  Cusps during Mergers: Model of M31}},\ }\href
  {https://doi.org/10.3847/1538-4357/ac12d3} {\bibfield  {journal} {\bibinfo
  {journal} {ApJ}\ }\textbf {\bibinfo {volume} {919}},\ \bibinfo {eid} {86}
  (\bibinfo {year} {2021})},\ \Eprint {https://arxiv.org/abs/2002.12192}
  {arXiv:2002.12192 [astro-ph.GA]} \BibitemShut {NoStop}%
\bibitem [{\citenamefont {{Kamionkowski}}\ \emph {et~al.}(2010)\citenamefont
  {{Kamionkowski}}, \citenamefont {{Koushiappas}},\ and\ \citenamefont
  {{Kuhlen}}}]{2010PhRvD..81d3532K}%
  \BibitemOpen
  \bibfield  {author} {\bibinfo {author} {\bibfnamefont {M.}~\bibnamefont
  {{Kamionkowski}}}, \bibinfo {author} {\bibfnamefont {S.~M.}\ \bibnamefont
  {{Koushiappas}}},\ and\ \bibinfo {author} {\bibfnamefont {M.}~\bibnamefont
  {{Kuhlen}}},\ }\bibfield  {title} {\bibinfo {title} {{Galactic substructure
  and dark-matter annihilation in the Milky Way halo}},\ }\href
  {https://doi.org/10.1103/PhysRevD.81.043532} {\bibfield  {journal} {\bibinfo
  {journal} {Phys. Rev. D}\ }\textbf {\bibinfo {volume} {81}},\ \bibinfo {eid}
  {043532} (\bibinfo {year} {2010})},\ \Eprint
  {https://arxiv.org/abs/1001.3144} {arXiv:1001.3144 [astro-ph.GA]}
  \BibitemShut {NoStop}%
\bibitem [{\citenamefont {{Navarro}}\ \emph {et~al.}(1997)\citenamefont
  {{Navarro}}, \citenamefont {{Frenk}},\ and\ \citenamefont
  {{White}}}]{1997ApJ...490..493N}%
  \BibitemOpen
  \bibfield  {author} {\bibinfo {author} {\bibfnamefont {J.~F.}\ \bibnamefont
  {{Navarro}}}, \bibinfo {author} {\bibfnamefont {C.~S.}\ \bibnamefont
  {{Frenk}}},\ and\ \bibinfo {author} {\bibfnamefont {S.~D.~M.}\ \bibnamefont
  {{White}}},\ }\bibfield  {title} {\bibinfo {title} {{A Universal Density
  Profile from Hierarchical Clustering}},\ }\href@noop {} {\bibfield  {journal}
  {\bibinfo  {journal} {Astrophys. J.}\ }\textbf {\bibinfo {volume} {490}},\
  \bibinfo {pages} {493} (\bibinfo {year} {1997})},\ \Eprint
  {https://arxiv.org/abs/astro-ph/9611107} {astro-ph/9611107} \BibitemShut
  {NoStop}%
\bibitem [{\citenamefont {{Hoernes}}\ \emph {et~al.}(1998)\citenamefont
  {{Hoernes}}, \citenamefont {{Beck}},\ and\ \citenamefont
  {{Berkhuijsen}}}]{1998IAUS..184..351H}%
  \BibitemOpen
  \bibfield  {author} {\bibinfo {author} {\bibfnamefont {P.}~\bibnamefont
  {{Hoernes}}}, \bibinfo {author} {\bibfnamefont {R.}~\bibnamefont {{Beck}}},\
  and\ \bibinfo {author} {\bibfnamefont {E.~M.}\ \bibnamefont
  {{Berkhuijsen}}},\ }\bibfield  {title} {\bibinfo {title} {{Properties of
  synchrotron emission and magnetic fields in the central region of M31}},\
  }in\ \href@noop {} {\emph {\bibinfo {booktitle} {The Central Regions of the
  Galaxy and Galaxies}}},\ Vol.\ \bibinfo {volume} {184},\ \bibinfo {editor}
  {edited by\ \bibinfo {editor} {\bibfnamefont {Y.}~\bibnamefont {{Sofue}}}}\
  (\bibinfo {year} {1998})\ p.\ \bibinfo {pages} {351}\BibitemShut {NoStop}%
\bibitem [{\citenamefont {{Fletcher}}\ \emph {et~al.}(2004)\citenamefont
  {{Fletcher}}, \citenamefont {{Berkhuijsen}}, \citenamefont {{Beck}},\ and\
  \citenamefont {{Shukurov}}}]{2004A&A...414...53F}%
  \BibitemOpen
  \bibfield  {author} {\bibinfo {author} {\bibfnamefont {A.}~\bibnamefont
  {{Fletcher}}}, \bibinfo {author} {\bibfnamefont {E.~M.}\ \bibnamefont
  {{Berkhuijsen}}}, \bibinfo {author} {\bibfnamefont {R.}~\bibnamefont
  {{Beck}}},\ and\ \bibinfo {author} {\bibfnamefont {A.}~\bibnamefont
  {{Shukurov}}},\ }\bibfield  {title} {\bibinfo {title} {{The magnetic field of
  M31 from multi-wavelength radio polarization observations}},\ }\href
  {https://doi.org/10.1051/0004-6361:20034133} {\bibfield  {journal} {\bibinfo
  {journal} {A\&A}\ }\textbf {\bibinfo {volume} {414}},\ \bibinfo {pages} {53}
  (\bibinfo {year} {2004})},\ \Eprint {https://arxiv.org/abs/astro-ph/0310258}
  {arXiv:astro-ph/0310258 [astro-ph]} \BibitemShut {NoStop}%
\bibitem [{\citenamefont {{Han}}\ \emph {et~al.}(1998)\citenamefont {{Han}},
  \citenamefont {{Beck}},\ and\ \citenamefont
  {{Berkhuijsen}}}]{1998A&A...335.1117H}%
  \BibitemOpen
  \bibfield  {author} {\bibinfo {author} {\bibfnamefont {J.~L.}\ \bibnamefont
  {{Han}}}, \bibinfo {author} {\bibfnamefont {R.}~\bibnamefont {{Beck}}},\ and\
  \bibinfo {author} {\bibfnamefont {E.~M.}\ \bibnamefont {{Berkhuijsen}}},\
  }\bibfield  {title} {\bibinfo {title} {{New clues to the magnetic field
  structure of M 31}},\ }\href@noop {} {\bibfield  {journal} {\bibinfo
  {journal} {A\&A}\ }\textbf {\bibinfo {volume} {335}},\ \bibinfo {pages}
  {1117} (\bibinfo {year} {1998})},\ \Eprint
  {https://arxiv.org/abs/astro-ph/9805023} {arXiv:astro-ph/9805023 [astro-ph]}
  \BibitemShut {NoStop}%
\bibitem [{\citenamefont {{Beck}}(2012)}]{2012SSRv..166..215B}%
  \BibitemOpen
  \bibfield  {author} {\bibinfo {author} {\bibfnamefont {R.}~\bibnamefont
  {{Beck}}},\ }\bibfield  {title} {\bibinfo {title} {{Magnetic Fields in
  Galaxies}},\ }\href {https://doi.org/10.1007/s11214-011-9782-z} {\bibfield
  {journal} {\bibinfo  {journal} {SSRv}\ }\textbf {\bibinfo {volume} {166}},\
  \bibinfo {pages} {215} (\bibinfo {year} {2012})}\BibitemShut {NoStop}%
\bibitem [{\citenamefont {{Crocker}}\ \emph {et~al.}(2010)\citenamefont
  {{Crocker}}, \citenamefont {{Jones}}, \citenamefont {{Melia}}, \citenamefont
  {{Ott}},\ and\ \citenamefont {{Protheroe}}}]{2010Natur.463...65C}%
  \BibitemOpen
  \bibfield  {author} {\bibinfo {author} {\bibfnamefont {R.~M.}\ \bibnamefont
  {{Crocker}}}, \bibinfo {author} {\bibfnamefont {D.~I.}\ \bibnamefont
  {{Jones}}}, \bibinfo {author} {\bibfnamefont {F.}~\bibnamefont {{Melia}}},
  \bibinfo {author} {\bibfnamefont {J.}~\bibnamefont {{Ott}}},\ and\ \bibinfo
  {author} {\bibfnamefont {R.~J.}\ \bibnamefont {{Protheroe}}},\ }\bibfield
  {title} {\bibinfo {title} {{A lower limit of 50 microgauss for the magnetic
  field near the Galactic Centre}},\ }\href
  {https://doi.org/10.1038/nature08635} {\bibfield  {journal} {\bibinfo
  {journal} {Nature (London)}\ }\textbf {\bibinfo {volume} {463}},\ \bibinfo
  {pages} {65} (\bibinfo {year} {2010})},\ \Eprint
  {https://arxiv.org/abs/1001.1275} {arXiv:1001.1275 [astro-ph.GA]}
  \BibitemShut {NoStop}%
\bibitem [{\citenamefont {{Crocker}}\ \emph {et~al.}(2011)\citenamefont
  {{Crocker}}, \citenamefont {{Jones}}, \citenamefont {{Aharonian}},
  \citenamefont {{Law}}, \citenamefont {{Melia}},\ and\ \citenamefont
  {{Ott}}}]{2011MNRAS.411L..11C}%
  \BibitemOpen
  \bibfield  {author} {\bibinfo {author} {\bibfnamefont {R.~M.}\ \bibnamefont
  {{Crocker}}}, \bibinfo {author} {\bibfnamefont {D.~I.}\ \bibnamefont
  {{Jones}}}, \bibinfo {author} {\bibfnamefont {F.}~\bibnamefont
  {{Aharonian}}}, \bibinfo {author} {\bibfnamefont {C.~J.}\ \bibnamefont
  {{Law}}}, \bibinfo {author} {\bibfnamefont {F.}~\bibnamefont {{Melia}}},\
  and\ \bibinfo {author} {\bibfnamefont {J.}~\bibnamefont {{Ott}}},\ }\bibfield
   {title} {\bibinfo {title} {{{\ensuremath{\gamma}}-rays and the
  far-infrared-radio continuum correlation reveal a powerful Galactic Centre
  wind}},\ }\href {https://doi.org/10.1111/j.1745-3933.2010.00983.x} {\bibfield
   {journal} {\bibinfo  {journal} {MNRAS}\ }\textbf {\bibinfo {volume} {411}},\
  \bibinfo {pages} {L11} (\bibinfo {year} {2011})},\ \Eprint
  {https://arxiv.org/abs/1009.4340} {arXiv:1009.4340 [astro-ph.GA]}
  \BibitemShut {NoStop}%
\bibitem [{\citenamefont {{Stepanov}}\ \emph {et~al.}(2008)\citenamefont
  {{Stepanov}}, \citenamefont {{Arshakian}}, \citenamefont {{Beck}},
  \citenamefont {{Frick}},\ and\ \citenamefont
  {{Krause}}}]{2008A&A...480...45S}%
  \BibitemOpen
  \bibfield  {author} {\bibinfo {author} {\bibfnamefont {R.}~\bibnamefont
  {{Stepanov}}}, \bibinfo {author} {\bibfnamefont {T.~G.}\ \bibnamefont
  {{Arshakian}}}, \bibinfo {author} {\bibfnamefont {R.}~\bibnamefont {{Beck}}},
  \bibinfo {author} {\bibfnamefont {P.}~\bibnamefont {{Frick}}},\ and\ \bibinfo
  {author} {\bibfnamefont {M.}~\bibnamefont {{Krause}}},\ }\bibfield  {title}
  {\bibinfo {title} {{Magnetic field structures of galaxies derived from
  analysis of Faraday rotation measures, and perspectives for the SKA}},\
  }\href {https://doi.org/10.1051/0004-6361:20078678} {\bibfield  {journal}
  {\bibinfo  {journal} {A\&A}\ }\textbf {\bibinfo {volume} {480}},\ \bibinfo
  {pages} {45} (\bibinfo {year} {2008})},\ \Eprint
  {https://arxiv.org/abs/0711.1267} {arXiv:0711.1267 [astro-ph]} \BibitemShut
  {NoStop}%
\bibitem [{\citenamefont {{Haverkorn}}\ and\ \citenamefont
  {{Heesen}}(2012)}]{2012SSRv..166..133H}%
  \BibitemOpen
  \bibfield  {author} {\bibinfo {author} {\bibfnamefont {M.}~\bibnamefont
  {{Haverkorn}}}\ and\ \bibinfo {author} {\bibfnamefont {V.}~\bibnamefont
  {{Heesen}}},\ }\bibfield  {title} {\bibinfo {title} {{Magnetic Fields in
  Galactic Haloes}},\ }\href {https://doi.org/10.1007/s11214-011-9757-0}
  {\bibfield  {journal} {\bibinfo  {journal} {Space Sci. Rev.}\ }\textbf
  {\bibinfo {volume} {166}},\ \bibinfo {pages} {133} (\bibinfo {year}
  {2012})},\ \Eprint {https://arxiv.org/abs/1102.3701} {arXiv:1102.3701}
  \BibitemShut {NoStop}%
\bibitem [{\citenamefont {{Beck}}\ \emph {et~al.}(2019)\citenamefont {{Beck}},
  \citenamefont {{Chamandy}}, \citenamefont {{Elson}},\ and\ \citenamefont
  {{Blackman}}}]{2019Galax...8....4B}%
  \BibitemOpen
  \bibfield  {author} {\bibinfo {author} {\bibfnamefont {R.}~\bibnamefont
  {{Beck}}}, \bibinfo {author} {\bibfnamefont {L.}~\bibnamefont {{Chamandy}}},
  \bibinfo {author} {\bibfnamefont {E.}~\bibnamefont {{Elson}}},\ and\ \bibinfo
  {author} {\bibfnamefont {E.~G.}\ \bibnamefont {{Blackman}}},\ }\bibfield
  {title} {\bibinfo {title} {{Synthesizing Observations and Theory to
  Understand Galactic Magnetic Fields: Progress and Challenges}},\ }\href
  {https://doi.org/10.3390/galaxies8010004} {\bibfield  {journal} {\bibinfo
  {journal} {Galax}\ }\textbf {\bibinfo {volume} {8}},\ \bibinfo {pages} {4}
  (\bibinfo {year} {2019})},\ \Eprint {https://arxiv.org/abs/1912.08962}
  {arXiv:1912.08962 [astro-ph.GA]} \BibitemShut {NoStop}%
\bibitem [{\citenamefont {{Jansson}}\ and\ \citenamefont
  {{Farrar}}(2012)}]{2012ApJ...757...14J}%
  \BibitemOpen
  \bibfield  {author} {\bibinfo {author} {\bibfnamefont {R.}~\bibnamefont
  {{Jansson}}}\ and\ \bibinfo {author} {\bibfnamefont {G.~R.}\ \bibnamefont
  {{Farrar}}},\ }\bibfield  {title} {\bibinfo {title} {{A New Model of the
  Galactic Magnetic Field}},\ }\href
  {https://doi.org/10.1088/0004-637X/757/1/14} {\bibfield  {journal} {\bibinfo
  {journal} {ApJ}\ }\textbf {\bibinfo {volume} {757}},\ \bibinfo {eid} {14}
  (\bibinfo {year} {2012})},\ \Eprint {https://arxiv.org/abs/1204.3662}
  {arXiv:1204.3662 [astro-ph.GA]} \BibitemShut {NoStop}%
\bibitem [{\citenamefont {{Fornengo}}\ \emph {et~al.}(2012)\citenamefont
  {{Fornengo}}, \citenamefont {{Lineros}}, \citenamefont {{Regis}},\ and\
  \citenamefont {{Taoso}}}]{2012JCAP...01..005F}%
  \BibitemOpen
  \bibfield  {author} {\bibinfo {author} {\bibfnamefont {N.}~\bibnamefont
  {{Fornengo}}}, \bibinfo {author} {\bibfnamefont {R.~A.}\ \bibnamefont
  {{Lineros}}}, \bibinfo {author} {\bibfnamefont {M.}~\bibnamefont {{Regis}}},\
  and\ \bibinfo {author} {\bibfnamefont {M.}~\bibnamefont {{Taoso}}},\
  }\bibfield  {title} {\bibinfo {title} {{Galactic synchrotron emission from
  WIMPs at radio frequencies}},\ }\href
  {https://doi.org/10.1088/1475-7516/2012/01/005} {\bibfield  {journal}
  {\bibinfo  {journal} {J. Cosmology Astropart. Phys.}\ }\textbf {\bibinfo
  {volume} {1}},\ \bibinfo {eid} {005} (\bibinfo {year} {2012})},\ \Eprint
  {https://arxiv.org/abs/1110.4337} {arXiv:1110.4337 [astro-ph.GA]}
  \BibitemShut {NoStop}%
\bibitem [{\citenamefont {{Berkhuijsen}}\ \emph {et~al.}(2013)\citenamefont
  {{Berkhuijsen}}, \citenamefont {{Beck}},\ and\ \citenamefont
  {{Tabatabaei}}}]{2013MNRAS.435.1598B}%
  \BibitemOpen
  \bibfield  {author} {\bibinfo {author} {\bibfnamefont {E.~M.}\ \bibnamefont
  {{Berkhuijsen}}}, \bibinfo {author} {\bibfnamefont {R.}~\bibnamefont
  {{Beck}}},\ and\ \bibinfo {author} {\bibfnamefont {F.~S.}\ \bibnamefont
  {{Tabatabaei}}},\ }\bibfield  {title} {\bibinfo {title} {{How cosmic ray
  electron propagation affects radio-far-infrared correlations in M 31 and M
  33}},\ }\href {https://doi.org/10.1093/mnras/stt1400} {\bibfield  {journal}
  {\bibinfo  {journal} {MNRAS}\ }\textbf {\bibinfo {volume} {435}},\ \bibinfo
  {pages} {1598} (\bibinfo {year} {2013})},\ \Eprint
  {https://arxiv.org/abs/1307.7991} {arXiv:1307.7991 [astro-ph.GA]}
  \BibitemShut {NoStop}%
\bibitem [{\citenamefont {{Drury}}\ and\ \citenamefont
  {{Strong}}(2017)}]{2017A&A...597A.117D}%
  \BibitemOpen
  \bibfield  {author} {\bibinfo {author} {\bibfnamefont {L.~O.~C.}\
  \bibnamefont {{Drury}}}\ and\ \bibinfo {author} {\bibfnamefont {A.~W.}\
  \bibnamefont {{Strong}}},\ }\bibfield  {title} {\bibinfo {title} {{Power
  requirements for cosmic ray propagation models involving diffusive
  reacceleration; estimates and implications for the damping of interstellar
  turbulence}},\ }\href {https://doi.org/10.1051/0004-6361/201629526}
  {\bibfield  {journal} {\bibinfo  {journal} {Astron. Astrophys.}\ }\textbf
  {\bibinfo {volume} {597}},\ \bibinfo {eid} {A117} (\bibinfo {year} {2017})},\
  \Eprint {https://arxiv.org/abs/1608.04227} {arXiv:1608.04227 [astro-ph.HE]}
  \BibitemShut {NoStop}%
\bibitem [{\citenamefont {Ginzburg}(1979)}]{Ginzburg}%
  \BibitemOpen
  \bibfield  {author} {\bibinfo {author} {\bibfnamefont {V.}~\bibnamefont
  {Ginzburg}},\ }\href@noop {} {\emph {\bibinfo {title} {Theoretical physics
  and astrophysics}}}\ (\bibinfo  {publisher} {Pergamon press},\ \bibinfo
  {year} {1979})\BibitemShut {NoStop}%
\bibitem [{\citenamefont {{Groves}}\ and\ \citenamefont
  {{Krause}}(2012)}]{2012IAUS..284..112G}%
  \BibitemOpen
  \bibfield  {author} {\bibinfo {author} {\bibfnamefont {B.}~\bibnamefont
  {{Groves}}}\ and\ \bibinfo {author} {\bibfnamefont {O.}~\bibnamefont
  {{Krause}}},\ }\bibfield  {title} {\bibinfo {title} {{Hot \& cold dust in
  M31: the resolved SED of Andromeda}},\ }in\ \href
  {https://doi.org/10.1017/S174392131200885X} {\emph {\bibinfo {booktitle} {The
  Spectral Energy Distribution of Galaxies - SED 2011}}},\ Vol.\ \bibinfo
  {volume} {284},\ \bibinfo {editor} {edited by\ \bibinfo {editor}
  {\bibfnamefont {R.~J.}\ \bibnamefont {{Tuffs}}}\ and\ \bibinfo {editor}
  {\bibfnamefont {C.~C.}\ \bibnamefont {{Popescu}}}}\ (\bibinfo {year} {2012})\
  pp.\ \bibinfo {pages} {112--116}\BibitemShut {NoStop}%
\bibitem [{\citenamefont {{Planck Collaboration}}\ \emph
  {et~al.}(2015)\citenamefont {{Planck Collaboration}}, \citenamefont {{Ade}},
  \citenamefont {{Aghanim}}, \citenamefont {{Arnaud}}, \citenamefont
  {{Ashdown}}, \citenamefont {{Aumont}}, \citenamefont {{Baccigalupi}},
  \citenamefont {{Banday}}, \citenamefont {{Barreiro}}, \citenamefont
  {{Bartolo}},\ and\ \citenamefont {et~al.}}]{2015A&A...582A..28P}%
  \BibitemOpen
  \bibfield  {author} {\bibinfo {author} {\bibnamefont {{Planck
  Collaboration}}}, \bibinfo {author} {\bibfnamefont {P.~A.~R.}\ \bibnamefont
  {{Ade}}}, \bibinfo {author} {\bibfnamefont {N.}~\bibnamefont {{Aghanim}}},
  \bibinfo {author} {\bibfnamefont {M.}~\bibnamefont {{Arnaud}}}, \bibinfo
  {author} {\bibfnamefont {M.}~\bibnamefont {{Ashdown}}}, \bibinfo {author}
  {\bibfnamefont {J.}~\bibnamefont {{Aumont}}}, \bibinfo {author}
  {\bibfnamefont {C.}~\bibnamefont {{Baccigalupi}}}, \bibinfo {author}
  {\bibfnamefont {A.~J.}\ \bibnamefont {{Banday}}}, \bibinfo {author}
  {\bibfnamefont {R.~B.}\ \bibnamefont {{Barreiro}}}, \bibinfo {author}
  {\bibfnamefont {N.}~\bibnamefont {{Bartolo}}},\ and\ \bibinfo {author}
  {\bibnamefont {et~al.}},\ }\bibfield  {title} {\bibinfo {title} {{Planck
  intermediate results. XXV. The Andromeda galaxy as seen by Planck}},\ }\href
  {https://doi.org/10.1051/0004-6361/201424643} {\bibfield  {journal} {\bibinfo
   {journal} {Astron. Astrophys.}\ }\textbf {\bibinfo {volume} {582}},\
  \bibinfo {eid} {A28} (\bibinfo {year} {2015})},\ \Eprint
  {https://arxiv.org/abs/1407.5452} {arXiv:1407.5452 [astro-ph.GA]}
  \BibitemShut {NoStop}%
\bibitem [{\citenamefont {{Kent}}(1989)}]{1989AJ.....97.1614K}%
  \BibitemOpen
  \bibfield  {author} {\bibinfo {author} {\bibfnamefont {S.~M.}\ \bibnamefont
  {{Kent}}},\ }\bibfield  {title} {\bibinfo {title} {{An Improved Bulge Model
  for M31}},\ }\href {https://doi.org/10.1086/115103} {\bibfield  {journal}
  {\bibinfo  {journal} {AJ}\ }\textbf {\bibinfo {volume} {97}},\ \bibinfo
  {pages} {1614} (\bibinfo {year} {1989})}\BibitemShut {NoStop}%
\bibitem [{\citenamefont {{van den Bergh}}(1999)}]{1999A&ARv...9..273V}%
  \BibitemOpen
  \bibfield  {author} {\bibinfo {author} {\bibfnamefont {S.}~\bibnamefont {{van
  den Bergh}}},\ }\bibfield  {title} {\bibinfo {title} {{The local group of
  galaxies}},\ }\href {https://doi.org/10.1007/s001590050019} {\bibfield
  {journal} {\bibinfo  {journal} {A\&ARv}\ }\textbf {\bibinfo {volume} {9}},\
  \bibinfo {pages} {273} (\bibinfo {year} {1999})}\BibitemShut {NoStop}%
\bibitem [{\citenamefont {{Yin}}\ \emph {et~al.}(2009)\citenamefont {{Yin}},
  \citenamefont {{Hou}}, \citenamefont {{Prantzos}}, \citenamefont
  {{Boissier}}, \citenamefont {{Chang}}, \citenamefont {{Shen}},\ and\
  \citenamefont {{Zhang}}}]{2009A&A...505..497Y}%
  \BibitemOpen
  \bibfield  {author} {\bibinfo {author} {\bibfnamefont {J.}~\bibnamefont
  {{Yin}}}, \bibinfo {author} {\bibfnamefont {J.~L.}\ \bibnamefont {{Hou}}},
  \bibinfo {author} {\bibfnamefont {N.}~\bibnamefont {{Prantzos}}}, \bibinfo
  {author} {\bibfnamefont {S.}~\bibnamefont {{Boissier}}}, \bibinfo {author}
  {\bibfnamefont {R.~X.}\ \bibnamefont {{Chang}}}, \bibinfo {author}
  {\bibfnamefont {S.~Y.}\ \bibnamefont {{Shen}}},\ and\ \bibinfo {author}
  {\bibfnamefont {B.}~\bibnamefont {{Zhang}}},\ }\bibfield  {title} {\bibinfo
  {title} {{Milky Way versus Andromeda: a tale of two disks}},\ }\href
  {https://doi.org/10.1051/0004-6361/200912316} {\bibfield  {journal} {\bibinfo
   {journal} {A\&A}\ }\textbf {\bibinfo {volume} {505}},\ \bibinfo {pages}
  {497} (\bibinfo {year} {2009})},\ \Eprint {https://arxiv.org/abs/0906.4821}
  {arXiv:0906.4821 [astro-ph.GA]} \BibitemShut {NoStop}%
\bibitem [{\citenamefont {{Licquia}}\ \emph {et~al.}(2015)\citenamefont
  {{Licquia}}, \citenamefont {{Newman}},\ and\ \citenamefont
  {{Brinchmann}}}]{2015ApJ...809...96L}%
  \BibitemOpen
  \bibfield  {author} {\bibinfo {author} {\bibfnamefont {T.~C.}\ \bibnamefont
  {{Licquia}}}, \bibinfo {author} {\bibfnamefont {J.~A.}\ \bibnamefont
  {{Newman}}},\ and\ \bibinfo {author} {\bibfnamefont {J.}~\bibnamefont
  {{Brinchmann}}},\ }\bibfield  {title} {\bibinfo {title} {{Unveiling the Milky
  Way: A New Technique for Determining the Optical Color and Luminosity of Our
  Galaxy}},\ }\href {https://doi.org/10.1088/0004-637X/809/1/96} {\bibfield
  {journal} {\bibinfo  {journal} {ApJ}\ }\textbf {\bibinfo {volume} {809}},\
  \bibinfo {eid} {96} (\bibinfo {year} {2015})},\ \Eprint
  {https://arxiv.org/abs/1508.04446} {arXiv:1508.04446 [astro-ph.GA]}
  \BibitemShut {NoStop}%
\bibitem [{\citenamefont {{Tabatabaei}}\ \emph {et~al.}(2013)\citenamefont
  {{Tabatabaei}}, \citenamefont {{Berkhuijsen}}, \citenamefont {{Frick}},
  \citenamefont {{Beck}},\ and\ \citenamefont
  {{Schinnerer}}}]{2013A&A...557A.129T}%
  \BibitemOpen
  \bibfield  {author} {\bibinfo {author} {\bibfnamefont {F.~S.}\ \bibnamefont
  {{Tabatabaei}}}, \bibinfo {author} {\bibfnamefont {E.~M.}\ \bibnamefont
  {{Berkhuijsen}}}, \bibinfo {author} {\bibfnamefont {P.}~\bibnamefont
  {{Frick}}}, \bibinfo {author} {\bibfnamefont {R.}~\bibnamefont {{Beck}}},\
  and\ \bibinfo {author} {\bibfnamefont {E.}~\bibnamefont {{Schinnerer}}},\
  }\bibfield  {title} {\bibinfo {title} {{Multi-scale radio-infrared
  correlations in M 31 and M 33: The role of magnetic fields and star
  formation}},\ }\href {https://doi.org/10.1051/0004-6361/201218909} {\bibfield
   {journal} {\bibinfo  {journal} {Astron. Astrophys.}\ }\textbf {\bibinfo
  {volume} {557}},\ \bibinfo {eid} {A129} (\bibinfo {year} {2013})},\ \Eprint
  {https://arxiv.org/abs/1307.6253} {arXiv:1307.6253 [astro-ph.GA]}
  \BibitemShut {NoStop}%
\bibitem [{\citenamefont {{Battistelli}}\ \emph {et~al.}(2019)\citenamefont
  {{Battistelli}}, \citenamefont {{Fatigoni}}, \citenamefont {{Murgia}},
  \citenamefont {{Buzzelli}}, \citenamefont {{Carretti}}, \citenamefont
  {{Castangia}}, \citenamefont {{Concu}}, \citenamefont {{Cruciani}},
  \citenamefont {{de Bernardis}}, \citenamefont {{Genova-Santos}},
  \citenamefont {{Govoni}}, \citenamefont {{Guidi}}, \citenamefont {{Lamagna}},
  \citenamefont {{Luzzi}}, \citenamefont {{Masi}}, \citenamefont {{Melis}},
  \citenamefont {{Paladini}}, \citenamefont {{Piacentini}}, \citenamefont
  {{Poppi}}, \citenamefont {{Radiconi}}, \citenamefont {{Rebolo}},
  \citenamefont {{Rubino-Martin}}, \citenamefont {{Tarchi}},\ and\
  \citenamefont {{Vacca}}}]{2019ApJ...877L..31B}%
  \BibitemOpen
  \bibfield  {author} {\bibinfo {author} {\bibfnamefont {E.~S.}\ \bibnamefont
  {{Battistelli}}}, \bibinfo {author} {\bibfnamefont {S.}~\bibnamefont
  {{Fatigoni}}}, \bibinfo {author} {\bibfnamefont {M.}~\bibnamefont
  {{Murgia}}}, \bibinfo {author} {\bibfnamefont {A.}~\bibnamefont
  {{Buzzelli}}}, \bibinfo {author} {\bibfnamefont {E.}~\bibnamefont
  {{Carretti}}}, \bibinfo {author} {\bibfnamefont {P.}~\bibnamefont
  {{Castangia}}}, \bibinfo {author} {\bibfnamefont {R.}~\bibnamefont
  {{Concu}}}, \bibinfo {author} {\bibfnamefont {A.}~\bibnamefont {{Cruciani}}},
  \bibinfo {author} {\bibfnamefont {P.}~\bibnamefont {{de Bernardis}}},
  \bibinfo {author} {\bibfnamefont {R.}~\bibnamefont {{Genova-Santos}}},
  \bibinfo {author} {\bibfnamefont {F.}~\bibnamefont {{Govoni}}}, \bibinfo
  {author} {\bibfnamefont {F.}~\bibnamefont {{Guidi}}}, \bibinfo {author}
  {\bibfnamefont {L.}~\bibnamefont {{Lamagna}}}, \bibinfo {author}
  {\bibfnamefont {G.}~\bibnamefont {{Luzzi}}}, \bibinfo {author} {\bibfnamefont
  {S.}~\bibnamefont {{Masi}}}, \bibinfo {author} {\bibfnamefont
  {A.}~\bibnamefont {{Melis}}}, \bibinfo {author} {\bibfnamefont
  {R.}~\bibnamefont {{Paladini}}}, \bibinfo {author} {\bibfnamefont
  {F.}~\bibnamefont {{Piacentini}}}, \bibinfo {author} {\bibfnamefont
  {S.}~\bibnamefont {{Poppi}}}, \bibinfo {author} {\bibfnamefont
  {F.}~\bibnamefont {{Radiconi}}}, \bibinfo {author} {\bibfnamefont
  {R.}~\bibnamefont {{Rebolo}}}, \bibinfo {author} {\bibfnamefont {J.~A.}\
  \bibnamefont {{Rubino-Martin}}}, \bibinfo {author} {\bibfnamefont
  {A.}~\bibnamefont {{Tarchi}}},\ and\ \bibinfo {author} {\bibfnamefont
  {V.}~\bibnamefont {{Vacca}}},\ }\bibfield  {title} {\bibinfo {title} {{Strong
  Evidence of Anomalous Microwave Emission from the Flux Density Spectrum of
  M31}},\ }\href {https://doi.org/10.3847/2041-8213/ab21de} {\bibfield
  {journal} {\bibinfo  {journal} {Astrophys. J. Lett.}\ }\textbf {\bibinfo
  {volume} {877}},\ \bibinfo {eid} {L31} (\bibinfo {year} {2019})},\ \Eprint
  {https://arxiv.org/abs/1905.12276} {arXiv:1905.12276 [astro-ph.GA]}
  \BibitemShut {NoStop}%
\bibitem [{GPs()}]{GPs}%
  \BibitemOpen
  \href@noop {} {}\bibinfo {howpublished} {GALPROP explanatory supplement at
  \url{http://galprop.stanford.edu/code.php?option=manual}}\BibitemShut
  {NoStop}%
\bibitem [{\citenamefont {{Shimwell}}\ \emph {et~al.}(2022)\citenamefont
  {{Shimwell}}, \citenamefont {{Hardcastle}}, \citenamefont {{Tasse}},
  \citenamefont {{Best}}, \citenamefont {{R{\"o}ttgering}}, \citenamefont
  {{Williams}}, \citenamefont {{Botteon}}, \citenamefont {{Drabent}},
  \citenamefont {{Mechev}}, \citenamefont {{Shulevski}}, \citenamefont {{van
  Weeren}}, \citenamefont {{Bester}}, \citenamefont {{Br{\"u}ggen}},
  \citenamefont {{Brunetti}}, \citenamefont {{Callingham}}, \citenamefont
  {{Chy{\.z}y}}, \citenamefont {{Conway}}, \citenamefont {{Dijkema}},
  \citenamefont {{Duncan}}, \citenamefont {{de Gasperin}}, \citenamefont
  {{Hale}}, \citenamefont {{Haverkorn}}, \citenamefont {{Hugo}}, \citenamefont
  {{Jackson}}, \citenamefont {{Mevius}}, \citenamefont {{Miley}}, \citenamefont
  {{Morabito}}, \citenamefont {{Morganti}}, \citenamefont {{Offringa}},
  \citenamefont {{Oonk}}, \citenamefont {{Rafferty}}, \citenamefont
  {{Sabater}}, \citenamefont {{Smith}}, \citenamefont {{Schwarz}},
  \citenamefont {{Smirnov}}, \citenamefont {{O'Sullivan}}, \citenamefont
  {{Vedantham}}, \citenamefont {{White}}, \citenamefont {{Albert}},
  \citenamefont {{Alegre}}, \citenamefont {{Asabere}}, \citenamefont {{Bacon}},
  \citenamefont {{Bonafede}}, \citenamefont {{Bonnassieux}}, \citenamefont
  {{Brienza}}, \citenamefont {{Bilicki}}, \citenamefont {{Bonato}},
  \citenamefont {{Calistro Rivera}}, \citenamefont {{Cassano}}, \citenamefont
  {{Cochrane}}, \citenamefont {{Croston}}, \citenamefont {{Cuciti}},
  \citenamefont {{Dallacasa}}, \citenamefont {{Danezi}}, \citenamefont
  {{Dettmar}}, \citenamefont {{Di Gennaro}}, \citenamefont {{Edler}},
  \citenamefont {{En{\ss}lin}}, \citenamefont {{Emig}}, \citenamefont
  {{Franzen}}, \citenamefont {{Garc{\'\i}a-Vergara}}, \citenamefont {{Grange}},
  \citenamefont {{G{\"u}rkan}}, \citenamefont {{Hajduk}}, \citenamefont
  {{Heald}}, \citenamefont {{Heesen}}, \citenamefont {{Hoang}}, \citenamefont
  {{Hoeft}}, \citenamefont {{Horellou}}, \citenamefont {{Iacobelli}},
  \citenamefont {{Jamrozy}}, \citenamefont {{Jeli{\'c}}}, \citenamefont
  {{Kondapally}}, \citenamefont {{Kukreti}}, \citenamefont
  {{Kunert-Bajraszewska}}, \citenamefont {{Magliocchetti}}, \citenamefont
  {{Mahatma}}, \citenamefont {{Ma{\l}ek}}, \citenamefont {{Mandal}},
  \citenamefont {{Massaro}}, \citenamefont {{Meyer-Zhao}}, \citenamefont
  {{Mingo}}, \citenamefont {{Mostert}}, \citenamefont {{Nair}}, \citenamefont
  {{Nakoneczny}}, \citenamefont {{Nikiel-Wroczy{\'n}ski}}, \citenamefont
  {{Orr{\'u}}}, \citenamefont {{Pajdosz-{\'S}mierciak}}, \citenamefont
  {{Pasini}}, \citenamefont {{Prandoni}}, \citenamefont {{van Piggelen}},
  \citenamefont {{Rajpurohit}}, \citenamefont {{Retana-Montenegro}},
  \citenamefont {{Riseley}}, \citenamefont {{Rowlinson}}, \citenamefont
  {{Saxena}}, \citenamefont {{Schrijvers}}, \citenamefont {{Sweijen}},
  \citenamefont {{Siewert}}, \citenamefont {{Timmerman}}, \citenamefont
  {{Vaccari}}, \citenamefont {{Vink}}, \citenamefont {{West}}, \citenamefont
  {{Wo{\l}owska}}, \citenamefont {{Zhang}},\ and\ \citenamefont
  {{Zheng}}}]{2022A&A...659A...1S}%
  \BibitemOpen
  \bibfield  {author} {\bibinfo {author} {\bibfnamefont {T.~W.}\ \bibnamefont
  {{Shimwell}}}, \bibinfo {author} {\bibfnamefont {M.~J.}\ \bibnamefont
  {{Hardcastle}}}, \bibinfo {author} {\bibfnamefont {C.}~\bibnamefont
  {{Tasse}}}, \bibinfo {author} {\bibfnamefont {P.~N.}\ \bibnamefont {{Best}}},
  \bibinfo {author} {\bibfnamefont {H.~J.~A.}\ \bibnamefont
  {{R{\"o}ttgering}}}, \bibinfo {author} {\bibfnamefont {W.~L.}\ \bibnamefont
  {{Williams}}}, \bibinfo {author} {\bibfnamefont {A.}~\bibnamefont
  {{Botteon}}}, \bibinfo {author} {\bibfnamefont {A.}~\bibnamefont
  {{Drabent}}}, \bibinfo {author} {\bibfnamefont {A.}~\bibnamefont {{Mechev}}},
  \bibinfo {author} {\bibfnamefont {A.}~\bibnamefont {{Shulevski}}}, \bibinfo
  {author} {\bibfnamefont {R.~J.}\ \bibnamefont {{van Weeren}}}, \bibinfo
  {author} {\bibfnamefont {L.}~\bibnamefont {{Bester}}}, \bibinfo {author}
  {\bibfnamefont {M.}~\bibnamefont {{Br{\"u}ggen}}}, \bibinfo {author}
  {\bibfnamefont {G.}~\bibnamefont {{Brunetti}}}, \bibinfo {author}
  {\bibfnamefont {J.~R.}\ \bibnamefont {{Callingham}}}, \bibinfo {author}
  {\bibfnamefont {K.~T.}\ \bibnamefont {{Chy{\.z}y}}}, \bibinfo {author}
  {\bibfnamefont {J.~E.}\ \bibnamefont {{Conway}}}, \bibinfo {author}
  {\bibfnamefont {T.~J.}\ \bibnamefont {{Dijkema}}}, \bibinfo {author}
  {\bibfnamefont {K.}~\bibnamefont {{Duncan}}}, \bibinfo {author}
  {\bibfnamefont {F.}~\bibnamefont {{de Gasperin}}}, \bibinfo {author}
  {\bibfnamefont {C.~L.}\ \bibnamefont {{Hale}}}, \bibinfo {author}
  {\bibfnamefont {M.}~\bibnamefont {{Haverkorn}}}, \bibinfo {author}
  {\bibfnamefont {B.}~\bibnamefont {{Hugo}}}, \bibinfo {author} {\bibfnamefont
  {N.}~\bibnamefont {{Jackson}}}, \bibinfo {author} {\bibfnamefont
  {M.}~\bibnamefont {{Mevius}}}, \bibinfo {author} {\bibfnamefont {G.~K.}\
  \bibnamefont {{Miley}}}, \bibinfo {author} {\bibfnamefont {L.~K.}\
  \bibnamefont {{Morabito}}}, \bibinfo {author} {\bibfnamefont
  {R.}~\bibnamefont {{Morganti}}}, \bibinfo {author} {\bibfnamefont
  {A.}~\bibnamefont {{Offringa}}}, \bibinfo {author} {\bibfnamefont {J.~B.~R.}\
  \bibnamefont {{Oonk}}}, \bibinfo {author} {\bibfnamefont {D.}~\bibnamefont
  {{Rafferty}}}, \bibinfo {author} {\bibfnamefont {J.}~\bibnamefont
  {{Sabater}}}, \bibinfo {author} {\bibfnamefont {D.~J.~B.}\ \bibnamefont
  {{Smith}}}, \bibinfo {author} {\bibfnamefont {D.~J.}\ \bibnamefont
  {{Schwarz}}}, \bibinfo {author} {\bibfnamefont {O.}~\bibnamefont
  {{Smirnov}}}, \bibinfo {author} {\bibfnamefont {S.~P.}\ \bibnamefont
  {{O'Sullivan}}}, \bibinfo {author} {\bibfnamefont {H.}~\bibnamefont
  {{Vedantham}}}, \bibinfo {author} {\bibfnamefont {G.~J.}\ \bibnamefont
  {{White}}}, \bibinfo {author} {\bibfnamefont {J.~G.}\ \bibnamefont
  {{Albert}}}, \bibinfo {author} {\bibfnamefont {L.}~\bibnamefont {{Alegre}}},
  \bibinfo {author} {\bibfnamefont {B.}~\bibnamefont {{Asabere}}}, \bibinfo
  {author} {\bibfnamefont {D.~J.}\ \bibnamefont {{Bacon}}}, \bibinfo {author}
  {\bibfnamefont {A.}~\bibnamefont {{Bonafede}}}, \bibinfo {author}
  {\bibfnamefont {E.}~\bibnamefont {{Bonnassieux}}}, \bibinfo {author}
  {\bibfnamefont {M.}~\bibnamefont {{Brienza}}}, \bibinfo {author}
  {\bibfnamefont {M.}~\bibnamefont {{Bilicki}}}, \bibinfo {author}
  {\bibfnamefont {M.}~\bibnamefont {{Bonato}}}, \bibinfo {author}
  {\bibfnamefont {G.}~\bibnamefont {{Calistro Rivera}}}, \bibinfo {author}
  {\bibfnamefont {R.}~\bibnamefont {{Cassano}}}, \bibinfo {author}
  {\bibfnamefont {R.}~\bibnamefont {{Cochrane}}}, \bibinfo {author}
  {\bibfnamefont {J.~H.}\ \bibnamefont {{Croston}}}, \bibinfo {author}
  {\bibfnamefont {V.}~\bibnamefont {{Cuciti}}}, \bibinfo {author}
  {\bibfnamefont {D.}~\bibnamefont {{Dallacasa}}}, \bibinfo {author}
  {\bibfnamefont {A.}~\bibnamefont {{Danezi}}}, \bibinfo {author}
  {\bibfnamefont {R.~J.}\ \bibnamefont {{Dettmar}}}, \bibinfo {author}
  {\bibfnamefont {G.}~\bibnamefont {{Di Gennaro}}}, \bibinfo {author}
  {\bibfnamefont {H.~W.}\ \bibnamefont {{Edler}}}, \bibinfo {author}
  {\bibfnamefont {T.~A.}\ \bibnamefont {{En{\ss}lin}}}, \bibinfo {author}
  {\bibfnamefont {K.~L.}\ \bibnamefont {{Emig}}}, \bibinfo {author}
  {\bibfnamefont {T.~M.~O.}\ \bibnamefont {{Franzen}}}, \bibinfo {author}
  {\bibfnamefont {C.}~\bibnamefont {{Garc{\'\i}a-Vergara}}}, \bibinfo {author}
  {\bibfnamefont {Y.~G.}\ \bibnamefont {{Grange}}}, \bibinfo {author}
  {\bibfnamefont {G.}~\bibnamefont {{G{\"u}rkan}}}, \bibinfo {author}
  {\bibfnamefont {M.}~\bibnamefont {{Hajduk}}}, \bibinfo {author}
  {\bibfnamefont {G.}~\bibnamefont {{Heald}}}, \bibinfo {author} {\bibfnamefont
  {V.}~\bibnamefont {{Heesen}}}, \bibinfo {author} {\bibfnamefont {D.~N.}\
  \bibnamefont {{Hoang}}}, \bibinfo {author} {\bibfnamefont {M.}~\bibnamefont
  {{Hoeft}}}, \bibinfo {author} {\bibfnamefont {C.}~\bibnamefont {{Horellou}}},
  \bibinfo {author} {\bibfnamefont {M.}~\bibnamefont {{Iacobelli}}}, \bibinfo
  {author} {\bibfnamefont {M.}~\bibnamefont {{Jamrozy}}}, \bibinfo {author}
  {\bibfnamefont {V.}~\bibnamefont {{Jeli{\'c}}}}, \bibinfo {author}
  {\bibfnamefont {R.}~\bibnamefont {{Kondapally}}}, \bibinfo {author}
  {\bibfnamefont {P.}~\bibnamefont {{Kukreti}}}, \bibinfo {author}
  {\bibfnamefont {M.}~\bibnamefont {{Kunert-Bajraszewska}}}, \bibinfo {author}
  {\bibfnamefont {M.}~\bibnamefont {{Magliocchetti}}}, \bibinfo {author}
  {\bibfnamefont {V.}~\bibnamefont {{Mahatma}}}, \bibinfo {author}
  {\bibfnamefont {K.}~\bibnamefont {{Ma{\l}ek}}}, \bibinfo {author}
  {\bibfnamefont {S.}~\bibnamefont {{Mandal}}}, \bibinfo {author}
  {\bibfnamefont {F.}~\bibnamefont {{Massaro}}}, \bibinfo {author}
  {\bibfnamefont {Z.}~\bibnamefont {{Meyer-Zhao}}}, \bibinfo {author}
  {\bibfnamefont {B.}~\bibnamefont {{Mingo}}}, \bibinfo {author} {\bibfnamefont
  {R.~I.~J.}\ \bibnamefont {{Mostert}}}, \bibinfo {author} {\bibfnamefont
  {D.~G.}\ \bibnamefont {{Nair}}}, \bibinfo {author} {\bibfnamefont {S.~J.}\
  \bibnamefont {{Nakoneczny}}}, \bibinfo {author} {\bibfnamefont
  {B.}~\bibnamefont {{Nikiel-Wroczy{\'n}ski}}}, \bibinfo {author}
  {\bibfnamefont {E.}~\bibnamefont {{Orr{\'u}}}}, \bibinfo {author}
  {\bibfnamefont {U.}~\bibnamefont {{Pajdosz-{\'S}mierciak}}}, \bibinfo
  {author} {\bibfnamefont {T.}~\bibnamefont {{Pasini}}}, \bibinfo {author}
  {\bibfnamefont {I.}~\bibnamefont {{Prandoni}}}, \bibinfo {author}
  {\bibfnamefont {H.~E.}\ \bibnamefont {{van Piggelen}}}, \bibinfo {author}
  {\bibfnamefont {K.}~\bibnamefont {{Rajpurohit}}}, \bibinfo {author}
  {\bibfnamefont {E.}~\bibnamefont {{Retana-Montenegro}}}, \bibinfo {author}
  {\bibfnamefont {C.~J.}\ \bibnamefont {{Riseley}}}, \bibinfo {author}
  {\bibfnamefont {A.}~\bibnamefont {{Rowlinson}}}, \bibinfo {author}
  {\bibfnamefont {A.}~\bibnamefont {{Saxena}}}, \bibinfo {author}
  {\bibfnamefont {C.}~\bibnamefont {{Schrijvers}}}, \bibinfo {author}
  {\bibfnamefont {F.}~\bibnamefont {{Sweijen}}}, \bibinfo {author}
  {\bibfnamefont {T.~M.}\ \bibnamefont {{Siewert}}}, \bibinfo {author}
  {\bibfnamefont {R.}~\bibnamefont {{Timmerman}}}, \bibinfo {author}
  {\bibfnamefont {M.}~\bibnamefont {{Vaccari}}}, \bibinfo {author}
  {\bibfnamefont {J.}~\bibnamefont {{Vink}}}, \bibinfo {author} {\bibfnamefont
  {J.~L.}\ \bibnamefont {{West}}}, \bibinfo {author} {\bibfnamefont
  {A.}~\bibnamefont {{Wo{\l}owska}}}, \bibinfo {author} {\bibfnamefont
  {X.}~\bibnamefont {{Zhang}}},\ and\ \bibinfo {author} {\bibfnamefont
  {J.}~\bibnamefont {{Zheng}}},\ }\bibfield  {title} {\bibinfo {title} {{The
  LOFAR Two-metre Sky Survey. V. Second data release}},\ }\href
  {https://doi.org/10.1051/0004-6361/202142484} {\bibfield  {journal} {\bibinfo
   {journal} {A\&A}\ }\textbf {\bibinfo {volume} {659}},\ \bibinfo {eid} {A1}
  (\bibinfo {year} {2022})},\ \Eprint {https://arxiv.org/abs/2202.11733}
  {arXiv:2202.11733 [astro-ph.GA]} \BibitemShut {NoStop}%
\bibitem [{\citenamefont {{Beck}}\ \emph {et~al.}(1998)\citenamefont {{Beck}},
  \citenamefont {{Berkhuijsen}},\ and\ \citenamefont
  {{Hoernes}}}]{1998A&AS..129..329B}%
  \BibitemOpen
  \bibfield  {author} {\bibinfo {author} {\bibfnamefont {R.}~\bibnamefont
  {{Beck}}}, \bibinfo {author} {\bibfnamefont {E.~M.}\ \bibnamefont
  {{Berkhuijsen}}},\ and\ \bibinfo {author} {\bibfnamefont {P.}~\bibnamefont
  {{Hoernes}}},\ }\bibfield  {title} {\bibinfo {title} {{A deep lambda 20 CM
  radio continuum survey of M 31}},\ }\href
  {https://doi.org/10.1051/aas:1998187} {\bibfield  {journal} {\bibinfo
  {journal} {A\&AS}\ }\textbf {\bibinfo {volume} {129}},\ \bibinfo {pages}
  {329} (\bibinfo {year} {1998})}\BibitemShut {NoStop}%
\bibitem [{\citenamefont {{Lane}}\ \emph {et~al.}(2014)\citenamefont {{Lane}},
  \citenamefont {{Cotton}}, \citenamefont {{van Velzen}}, \citenamefont
  {{Clarke}}, \citenamefont {{Kassim}}, \citenamefont {{Helmboldt}},
  \citenamefont {{Lazio}},\ and\ \citenamefont
  {{Cohen}}}]{2014MNRAS.440..327L}%
  \BibitemOpen
  \bibfield  {author} {\bibinfo {author} {\bibfnamefont {W.~M.}\ \bibnamefont
  {{Lane}}}, \bibinfo {author} {\bibfnamefont {W.~D.}\ \bibnamefont
  {{Cotton}}}, \bibinfo {author} {\bibfnamefont {S.}~\bibnamefont {{van
  Velzen}}}, \bibinfo {author} {\bibfnamefont {T.~E.}\ \bibnamefont
  {{Clarke}}}, \bibinfo {author} {\bibfnamefont {N.~E.}\ \bibnamefont
  {{Kassim}}}, \bibinfo {author} {\bibfnamefont {J.~F.}\ \bibnamefont
  {{Helmboldt}}}, \bibinfo {author} {\bibfnamefont {T.~J.~W.}\ \bibnamefont
  {{Lazio}}},\ and\ \bibinfo {author} {\bibfnamefont {A.~S.}\ \bibnamefont
  {{Cohen}}},\ }\bibfield  {title} {\bibinfo {title} {{The Very Large Array
  Low-frequency Sky Survey Redux (VLSSr)}},\ }\href
  {https://doi.org/10.1093/mnras/stu256} {\bibfield  {journal} {\bibinfo
  {journal} {Mon. Not. R. Astron. Soc.}\ }\textbf {\bibinfo {volume} {440}},\
  \bibinfo {pages} {327} (\bibinfo {year} {2014})},\ \Eprint
  {https://arxiv.org/abs/1404.0694} {arXiv:1404.0694 [astro-ph.IM]}
  \BibitemShut {NoStop}%
\bibitem [{\citenamefont {{Beck}}\ \emph {et~al.}(2020)\citenamefont {{Beck}},
  \citenamefont {{Berkhuijsen}}, \citenamefont {{Gie{\ss}{\"u}bel}},\ and\
  \citenamefont {{Mulcahy}}}]{2020A&A...633A...5B}%
  \BibitemOpen
  \bibfield  {author} {\bibinfo {author} {\bibfnamefont {R.}~\bibnamefont
  {{Beck}}}, \bibinfo {author} {\bibfnamefont {E.~M.}\ \bibnamefont
  {{Berkhuijsen}}}, \bibinfo {author} {\bibfnamefont {R.}~\bibnamefont
  {{Gie{\ss}{\"u}bel}}},\ and\ \bibinfo {author} {\bibfnamefont {D.~D.}\
  \bibnamefont {{Mulcahy}}},\ }\bibfield  {title} {\bibinfo {title} {{Magnetic
  fields and cosmic rays in M 31. I. Spectral indices, scale lengths, Faraday
  rotation, and magnetic field pattern}},\ }\href
  {https://doi.org/10.1051/0004-6361/201936481} {\bibfield  {journal} {\bibinfo
   {journal} {A\&A}\ }\textbf {\bibinfo {volume} {633}},\ \bibinfo {eid} {A5}
  (\bibinfo {year} {2020})},\ \Eprint {https://arxiv.org/abs/1910.09634}
  {arXiv:1910.09634 [astro-ph.GA]} \BibitemShut {NoStop}%
\bibitem [{\citenamefont {{Fatigoni}}\ \emph {et~al.}(2021)\citenamefont
  {{Fatigoni}}, \citenamefont {{Radiconi}}, \citenamefont {{Battistelli}},
  \citenamefont {{Murgia}}, \citenamefont {{Carretti}}, \citenamefont
  {{Castangia}}, \citenamefont {{Concu}}, \citenamefont {{de Bernardis}},
  \citenamefont {{Fritz}}, \citenamefont {{Genova-Santos}}, \citenamefont
  {{Govoni}}, \citenamefont {{Guidi}}, \citenamefont {{Lamagna}}, \citenamefont
  {{Masi}}, \citenamefont {{Melis}}, \citenamefont {{Paladini}}, \citenamefont
  {{Perez-Toledo}}, \citenamefont {{Piacentini}}, \citenamefont {{Poppi}},
  \citenamefont {{Rebolo}}, \citenamefont {{Rubino-Martin}}, \citenamefont
  {{Surcis}}, \citenamefont {{Tarchi}},\ and\ \citenamefont
  {{Vacca}}}]{2021A&A...651A..98F}%
  \BibitemOpen
  \bibfield  {author} {\bibinfo {author} {\bibfnamefont {S.}~\bibnamefont
  {{Fatigoni}}}, \bibinfo {author} {\bibfnamefont {F.}~\bibnamefont
  {{Radiconi}}}, \bibinfo {author} {\bibfnamefont {E.~S.}\ \bibnamefont
  {{Battistelli}}}, \bibinfo {author} {\bibfnamefont {M.}~\bibnamefont
  {{Murgia}}}, \bibinfo {author} {\bibfnamefont {E.}~\bibnamefont
  {{Carretti}}}, \bibinfo {author} {\bibfnamefont {P.}~\bibnamefont
  {{Castangia}}}, \bibinfo {author} {\bibfnamefont {R.}~\bibnamefont
  {{Concu}}}, \bibinfo {author} {\bibfnamefont {P.}~\bibnamefont {{de
  Bernardis}}}, \bibinfo {author} {\bibfnamefont {J.}~\bibnamefont {{Fritz}}},
  \bibinfo {author} {\bibfnamefont {R.}~\bibnamefont {{Genova-Santos}}},
  \bibinfo {author} {\bibfnamefont {F.}~\bibnamefont {{Govoni}}}, \bibinfo
  {author} {\bibfnamefont {F.}~\bibnamefont {{Guidi}}}, \bibinfo {author}
  {\bibfnamefont {L.}~\bibnamefont {{Lamagna}}}, \bibinfo {author}
  {\bibfnamefont {S.}~\bibnamefont {{Masi}}}, \bibinfo {author} {\bibfnamefont
  {A.}~\bibnamefont {{Melis}}}, \bibinfo {author} {\bibfnamefont
  {R.}~\bibnamefont {{Paladini}}}, \bibinfo {author} {\bibfnamefont {F.~M.}\
  \bibnamefont {{Perez-Toledo}}}, \bibinfo {author} {\bibfnamefont
  {F.}~\bibnamefont {{Piacentini}}}, \bibinfo {author} {\bibfnamefont
  {S.}~\bibnamefont {{Poppi}}}, \bibinfo {author} {\bibfnamefont
  {R.}~\bibnamefont {{Rebolo}}}, \bibinfo {author} {\bibfnamefont {J.~A.}\
  \bibnamefont {{Rubino-Martin}}}, \bibinfo {author} {\bibfnamefont
  {G.}~\bibnamefont {{Surcis}}}, \bibinfo {author} {\bibfnamefont
  {A.}~\bibnamefont {{Tarchi}}},\ and\ \bibinfo {author} {\bibfnamefont
  {V.}~\bibnamefont {{Vacca}}},\ }\bibfield  {title} {\bibinfo {title} {{Study
  of the thermal and nonthermal emission components in M 31: the Sardinia Radio
  Telescope view at 6.6 GHz}},\ }\href
  {https://doi.org/10.1051/0004-6361/202040011} {\bibfield  {journal} {\bibinfo
   {journal} {A\&A}\ }\textbf {\bibinfo {volume} {651}},\ \bibinfo {eid} {A98}
  (\bibinfo {year} {2021})},\ \Eprint {https://arxiv.org/abs/2105.10453}
  {arXiv:2105.10453 [astro-ph.GA]} \BibitemShut {NoStop}%
\bibitem [{\citenamefont {{Cohen}}\ \emph {et~al.}(2007)\citenamefont
  {{Cohen}}, \citenamefont {{Lane}}, \citenamefont {{Cotton}}, \citenamefont
  {{Kassim}}, \citenamefont {{Lazio}}, \citenamefont {{Perley}}, \citenamefont
  {{Condon}},\ and\ \citenamefont {{Erickson}}}]{2007AJ....134.1245C}%
  \BibitemOpen
  \bibfield  {author} {\bibinfo {author} {\bibfnamefont {A.~S.}\ \bibnamefont
  {{Cohen}}}, \bibinfo {author} {\bibfnamefont {W.~M.}\ \bibnamefont {{Lane}}},
  \bibinfo {author} {\bibfnamefont {W.~D.}\ \bibnamefont {{Cotton}}}, \bibinfo
  {author} {\bibfnamefont {N.~E.}\ \bibnamefont {{Kassim}}}, \bibinfo {author}
  {\bibfnamefont {T.~J.~W.}\ \bibnamefont {{Lazio}}}, \bibinfo {author}
  {\bibfnamefont {R.~A.}\ \bibnamefont {{Perley}}}, \bibinfo {author}
  {\bibfnamefont {J.~J.}\ \bibnamefont {{Condon}}},\ and\ \bibinfo {author}
  {\bibfnamefont {W.~C.}\ \bibnamefont {{Erickson}}},\ }\bibfield  {title}
  {\bibinfo {title} {{The VLA Low-Frequency Sky Survey}},\ }\href
  {https://doi.org/10.1086/520719} {\bibfield  {journal} {\bibinfo  {journal}
  {AJ}\ }\textbf {\bibinfo {volume} {134}},\ \bibinfo {pages} {1245} (\bibinfo
  {year} {2007})},\ \Eprint {https://arxiv.org/abs/0706.1191} {arXiv:0706.1191
  [astro-ph]} \BibitemShut {NoStop}%
\bibitem [{\citenamefont {{Gie{\ss}{\"u}bel}}\ \emph
  {et~al.}(2013)\citenamefont {{Gie{\ss}{\"u}bel}}, \citenamefont {{Heald}},
  \citenamefont {{Beck}},\ and\ \citenamefont
  {{Arshakian}}}]{2013A&A...559A..27G}%
  \BibitemOpen
  \bibfield  {author} {\bibinfo {author} {\bibfnamefont {R.}~\bibnamefont
  {{Gie{\ss}{\"u}bel}}}, \bibinfo {author} {\bibfnamefont {G.}~\bibnamefont
  {{Heald}}}, \bibinfo {author} {\bibfnamefont {R.}~\bibnamefont {{Beck}}},\
  and\ \bibinfo {author} {\bibfnamefont {T.~G.}\ \bibnamefont {{Arshakian}}},\
  }\bibfield  {title} {\bibinfo {title} {{Polarized synchrotron radiation from
  the Andromeda galaxy M 31 and background sources at 350 MHz}},\ }\href
  {https://doi.org/10.1051/0004-6361/201321765} {\bibfield  {journal} {\bibinfo
   {journal} {Astron. Astrophys.}\ }\textbf {\bibinfo {volume} {559}},\
  \bibinfo {eid} {A27} (\bibinfo {year} {2013})},\ \Eprint
  {https://arxiv.org/abs/1309.2539} {arXiv:1309.2539 [astro-ph.CO]}
  \BibitemShut {NoStop}%
\bibitem [{\citenamefont {{Karwin}}\ \emph {et~al.}(2019)\citenamefont
  {{Karwin}}, \citenamefont {{Murgia}}, \citenamefont {{Campbell}},\ and\
  \citenamefont {{Moskalenko}}}]{2019ApJ...880...95K}%
  \BibitemOpen
  \bibfield  {author} {\bibinfo {author} {\bibfnamefont {C.~M.}\ \bibnamefont
  {{Karwin}}}, \bibinfo {author} {\bibfnamefont {S.}~\bibnamefont {{Murgia}}},
  \bibinfo {author} {\bibfnamefont {S.}~\bibnamefont {{Campbell}}},\ and\
  \bibinfo {author} {\bibfnamefont {I.~V.}\ \bibnamefont {{Moskalenko}}},\
  }\bibfield  {title} {\bibinfo {title} {{Fermi-LAT Observations of
  {\ensuremath{\gamma}}-Ray Emission toward the Outer Halo of M31}},\ }\href
  {https://doi.org/10.3847/1538-4357/ab2880} {\bibfield  {journal} {\bibinfo
  {journal} {Astrophys. J.}\ }\textbf {\bibinfo {volume} {880}},\ \bibinfo
  {eid} {95} (\bibinfo {year} {2019})},\ \Eprint
  {https://arxiv.org/abs/1812.02958} {arXiv:1812.02958 [astro-ph.HE]}
  \BibitemShut {NoStop}%
\bibitem [{\citenamefont {{Karwin}}\ \emph {et~al.}(2021)\citenamefont
  {{Karwin}}, \citenamefont {{Murgia}}, \citenamefont {{Moskalenko}},
  \citenamefont {{Fillingham}}, \citenamefont {{Burns}},\ and\ \citenamefont
  {{Fieg}}}]{2021PhRvD.103b3027K}%
  \BibitemOpen
  \bibfield  {author} {\bibinfo {author} {\bibfnamefont {C.~M.}\ \bibnamefont
  {{Karwin}}}, \bibinfo {author} {\bibfnamefont {S.}~\bibnamefont {{Murgia}}},
  \bibinfo {author} {\bibfnamefont {I.~V.}\ \bibnamefont {{Moskalenko}}},
  \bibinfo {author} {\bibfnamefont {S.~P.}\ \bibnamefont {{Fillingham}}},
  \bibinfo {author} {\bibfnamefont {A.-K.}\ \bibnamefont {{Burns}}},\ and\
  \bibinfo {author} {\bibfnamefont {M.}~\bibnamefont {{Fieg}}},\ }\bibfield
  {title} {\bibinfo {title} {{Dark matter interpretation of the Fermi-LAT
  observations toward the outer halo of M31}},\ }\href
  {https://doi.org/10.1103/PhysRevD.103.023027} {\bibfield  {journal} {\bibinfo
   {journal} {PhRvD}\ }\textbf {\bibinfo {volume} {103}},\ \bibinfo {eid}
  {023027} (\bibinfo {year} {2021})},\ \Eprint
  {https://arxiv.org/abs/2010.08563} {arXiv:2010.08563 [astro-ph.HE]}
  \BibitemShut {NoStop}%
\bibitem [{\citenamefont {{Cholis}}\ \emph {et~al.}(2021)\citenamefont
  {{Cholis}}, \citenamefont {{Zhong}}, \citenamefont {{McDermott}},\ and\
  \citenamefont {{Surdutovich}}}]{2021arXiv211209706C}%
  \BibitemOpen
  \bibfield  {author} {\bibinfo {author} {\bibfnamefont {I.}~\bibnamefont
  {{Cholis}}}, \bibinfo {author} {\bibfnamefont {Y.-M.}\ \bibnamefont
  {{Zhong}}}, \bibinfo {author} {\bibfnamefont {S.~D.}\ \bibnamefont
  {{McDermott}}},\ and\ \bibinfo {author} {\bibfnamefont {J.~P.}\ \bibnamefont
  {{Surdutovich}}},\ }\bibfield  {title} {\bibinfo {title} {{The Return of the
  Templates: Revisiting the Galactic Center Excess with Multi-Messenger
  Observations}},\ }\href@noop {} {\bibfield  {journal} {\bibinfo  {journal}
  {arXiv}\ ,\ \bibinfo {eid} {arXiv:2112.09706}} (\bibinfo {year} {2021})},\
  \Eprint {https://arxiv.org/abs/2112.09706} {arXiv:2112.09706 [astro-ph.HE]}
  \BibitemShut {NoStop}%
\bibitem [{\citenamefont {{Egorov}}\ \emph {et~al.}(2020)\citenamefont
  {{Egorov}}, \citenamefont {{Topchiev}}, \citenamefont {{Galper}},
  \citenamefont {{Dalkarov}}, \citenamefont {{Leonov}}, \citenamefont
  {{Suchkov}},\ and\ \citenamefont {{Yurkin}}}]{2020JCAP...11..049E}%
  \BibitemOpen
  \bibfield  {author} {\bibinfo {author} {\bibfnamefont {A.~E.}\ \bibnamefont
  {{Egorov}}}, \bibinfo {author} {\bibfnamefont {N.~P.}\ \bibnamefont
  {{Topchiev}}}, \bibinfo {author} {\bibfnamefont {A.~M.}\ \bibnamefont
  {{Galper}}}, \bibinfo {author} {\bibfnamefont {O.~D.}\ \bibnamefont
  {{Dalkarov}}}, \bibinfo {author} {\bibfnamefont {A.~A.}\ \bibnamefont
  {{Leonov}}}, \bibinfo {author} {\bibfnamefont {S.~I.}\ \bibnamefont
  {{Suchkov}}},\ and\ \bibinfo {author} {\bibfnamefont {Y.~T.}\ \bibnamefont
  {{Yurkin}}},\ }\bibfield  {title} {\bibinfo {title} {{Dark matter searches by
  the planned gamma-ray telescope GAMMA-400}},\ }\href
  {https://doi.org/10.1088/1475-7516/2020/11/049} {\bibfield  {journal}
  {\bibinfo  {journal} {JCAP}\ }\textbf {\bibinfo {volume} {2020}}\bibfield
  {number} {\bibinfo  {number} { (11)},\ \bibinfo {eid} {049}},\ }\Eprint
  {https://arxiv.org/abs/2005.09032} {arXiv:2005.09032 [astro-ph.HE]}
  \BibitemShut {NoStop}%
\bibitem [{\citenamefont {{Abdallah}}\ \emph {et~al.}(2016)\citenamefont
  {{Abdallah}}, \citenamefont {{Abramowski}}, \citenamefont {{Aharonian}},
  \citenamefont {{Ait Benkhali}}, \citenamefont {{Akhperjanian}}, \citenamefont
  {{Ang{\"u}ner}}, \citenamefont {{Arrieta}}, \citenamefont {{Aubert}},
  \citenamefont {{Backes}}, \citenamefont {{Balzer}},\ and\ \citenamefont
  {et~al.}}]{2016PhRvL.117k1301A}%
  \BibitemOpen
  \bibfield  {author} {\bibinfo {author} {\bibfnamefont {H.}~\bibnamefont
  {{Abdallah}}}, \bibinfo {author} {\bibfnamefont {A.}~\bibnamefont
  {{Abramowski}}}, \bibinfo {author} {\bibfnamefont {F.}~\bibnamefont
  {{Aharonian}}}, \bibinfo {author} {\bibfnamefont {F.}~\bibnamefont {{Ait
  Benkhali}}}, \bibinfo {author} {\bibfnamefont {A.~G.}\ \bibnamefont
  {{Akhperjanian}}}, \bibinfo {author} {\bibfnamefont {E.}~\bibnamefont
  {{Ang{\"u}ner}}}, \bibinfo {author} {\bibfnamefont {M.}~\bibnamefont
  {{Arrieta}}}, \bibinfo {author} {\bibfnamefont {P.}~\bibnamefont {{Aubert}}},
  \bibinfo {author} {\bibfnamefont {M.}~\bibnamefont {{Backes}}}, \bibinfo
  {author} {\bibfnamefont {A.}~\bibnamefont {{Balzer}}},\ and\ \bibinfo
  {author} {\bibnamefont {et~al.}},\ }\bibfield  {title} {\bibinfo {title}
  {{Search for Dark Matter Annihilations towards the Inner Galactic Halo from
  10 Years of Observations with H.E.S.S.}},\ }\href
  {https://doi.org/10.1103/PhysRevLett.117.111301} {\bibfield  {journal}
  {\bibinfo  {journal} {PhRvL}\ }\textbf {\bibinfo {volume} {117}},\ \bibinfo
  {eid} {111301} (\bibinfo {year} {2016})},\ \Eprint
  {https://arxiv.org/abs/1607.08142} {arXiv:1607.08142 [astro-ph.HE]}
  \BibitemShut {NoStop}%
\bibitem [{\citenamefont {{Acciari}}\ \emph {et~al.}(2022)\citenamefont
  {{Acciari}}, \citenamefont {{Ansoldi}}, \citenamefont {{Antonelli}},
  \citenamefont {{Arbet Engels}}, \citenamefont {{Artero}}, \citenamefont
  {{Asano}}, \citenamefont {{Baack}}, \citenamefont {{Babi{\'c}}},
  \citenamefont {{Baquero}}, \citenamefont {{Barres de Almeida}}, \citenamefont
  {{Barrio}}, \citenamefont {{Batkovi{\'c}}}, \citenamefont {{Becerra
  Gonz{\'a}lez}}, \citenamefont {{Bednarek}}, \citenamefont {{Bellizzi}},
  \citenamefont {{Bernardini}}, \citenamefont {{Bernardos}}, \citenamefont
  {{Berti}}, \citenamefont {{Besenrieder}}, \citenamefont {{Bhattacharyya}},
  \citenamefont {{Bigongiari}}, \citenamefont {{Biland}}, \citenamefont
  {{Blanch}}, \citenamefont {{B{\"o}kenkamp}}, \citenamefont {{Bonnoli}},
  \citenamefont {{Bo{\v{s}}njak}}, \citenamefont {{Busetto}}, \citenamefont
  {{Carosi}}, \citenamefont {{Ceribella}}, \citenamefont {{Cerruti}},
  \citenamefont {{Chai}}, \citenamefont {{Chilingarian}}, \citenamefont
  {{Cikota}}, \citenamefont {{Colak}}, \citenamefont {{Colombo}}, \citenamefont
  {{Contreras}}, \citenamefont {{Cortina}}, \citenamefont {{Covino}},
  \citenamefont {{D'Amico}}, \citenamefont {{D'Elia}}, \citenamefont {{da
  Vela}}, \citenamefont {{Dazzi}}, \citenamefont {{de Angelis}}, \citenamefont
  {{de Lotto}}, \citenamefont {{Delfino}}, \citenamefont {{Delgado}},
  \citenamefont {{Delgado Mendez}}, \citenamefont {{Depaoli}}, \citenamefont
  {{di Pierro}}, \citenamefont {{di Venere}}, \citenamefont {{Do Souto
  Espi{\~n}eira}}, \citenamefont {{Dominis Prester}}, \citenamefont {{Donini}},
  \citenamefont {{Dorner}}, \citenamefont {{Doro}}, \citenamefont
  {{Elsaesser}}, \citenamefont {{Fallah Ramazani}}, \citenamefont
  {{Fattorini}}, \citenamefont {{Fonseca}}, \citenamefont {{Font}},
  \citenamefont {{Fruck}}, \citenamefont {{Fukami}}, \citenamefont
  {{Garc{\'\i}a L{\'o}pez}}, \citenamefont {{Garczarczyk}}, \citenamefont
  {{Gasparyan}}, \citenamefont {{Gaug}}, \citenamefont {{Giglietto}},
  \citenamefont {{Giordano}}, \citenamefont {{Gliwny}}, \citenamefont
  {{Godinovi{\'c}}}, \citenamefont {{Green}}, \citenamefont {{Green}},
  \citenamefont {{Hadasch}}, \citenamefont {{Hahn}}, \citenamefont
  {{Heckmann}}, \citenamefont {{Herrera}}, \citenamefont {{Hoang}},
  \citenamefont {{Hrupec}}, \citenamefont {{H{\"u}tten}}, \citenamefont
  {{Inada}}, \citenamefont {{Ishio}}, \citenamefont {{Iwamura}}, \citenamefont
  {{Jim{\'e}nez}}, \citenamefont {{Jormanainen}}, \citenamefont {{Jouvin}},
  \citenamefont {{Karjalainen}}, \citenamefont {{Kerszberg}}, \citenamefont
  {{Kobayashi}}, \citenamefont {{Kubo}}, \citenamefont {{Kushida}},
  \citenamefont {{Lamastra}}, \citenamefont {{Lelas}}, \citenamefont {{Leone}},
  \citenamefont {{Lindfors}}, \citenamefont {{Linhoff}}, \citenamefont
  {{Lombardi}}, \citenamefont {{Longo}}, \citenamefont {{L{\'o}pez-Coto}},
  \citenamefont {{L{\'o}pez-Moya}}, \citenamefont {{L{\'o}pez-Oramas}},
  \citenamefont {{Loporchio}}, \citenamefont {{Machado de Oliveira Fraga}},
  \citenamefont {{Maggio}}, \citenamefont {{Majumdar}}, \citenamefont
  {{Makariev}}, \citenamefont {{Mallamaci}}, \citenamefont {{Maneva}},
  \citenamefont {{Manganaro}}, \citenamefont {{Mannheim}}, \citenamefont
  {{Maraschi}}, \citenamefont {{Mariotti}}, \citenamefont {{Mart{\'\i}nez}},
  \citenamefont {{Mazin}}, \citenamefont {{Menchiari}}, \citenamefont
  {{Mender}}, \citenamefont {{Mi{\'c}anovi{\'c}}}, \citenamefont {{Miceli}},
  \citenamefont {{Miener}}, \citenamefont {{Miranda}}, \citenamefont
  {{Mirzoyan}}, \citenamefont {{Molina}}, \citenamefont {{Moralejo}},
  \citenamefont {{Morcuende}}, \citenamefont {{Moreno}}, \citenamefont
  {{Moretti}}, \citenamefont {{Neustroev}}, \citenamefont {{Nigro}},
  \citenamefont {{Nilsson}}, \citenamefont {{Ninci}}, \citenamefont
  {{Nishijima}}, \citenamefont {{Noda}}, \citenamefont {{Nozaki}},
  \citenamefont {{Ohtani}}, \citenamefont {{Oka}}, \citenamefont
  {{Otero-Santos}}, \citenamefont {{Paiano}}, \citenamefont {{Palatiello}},
  \citenamefont {{Paneque}}, \citenamefont {{Paoletti}}, \citenamefont
  {{Paredes}}, \citenamefont {{Pavleti{\'c}}}, \citenamefont {{Pe{\~n}il}},
  \citenamefont {{Persic}}, \citenamefont {{Pihet}}, \citenamefont {{Prada
  Moroni}}, \citenamefont {{Prandini}}, \citenamefont {{Priyadarshi}},
  \citenamefont {{Puljak}}, \citenamefont {{Rhode}}, \citenamefont
  {{Rib{\'o}}}, \citenamefont {{Rico}}, \citenamefont {{Righi}}, \citenamefont
  {{Rugliancich}}, \citenamefont {{Saha}}, \citenamefont {{Sahakyan}},
  \citenamefont {{Saito}}, \citenamefont {{Sakurai}}, \citenamefont
  {{Satalecka}}, \citenamefont {{Saturni}}, \citenamefont {{Schleicher}},
  \citenamefont {{Schmidt}}, \citenamefont {{Schweizer}}, \citenamefont
  {{Sitarek}}, \citenamefont {{{\v{S}}nidari{\'c}}}, \citenamefont
  {{Sobczynska}}, \citenamefont {{Spolon}}, \citenamefont {{Stamerra}},
  \citenamefont {{Stri{\v{s}}kovi{\'c}}}, \citenamefont {{Strom}},
  \citenamefont {{Strzys}}, \citenamefont {{Suda}}, \citenamefont
  {{Suri{\'c}}}, \citenamefont {{Takahashi}}, \citenamefont {{Takeishi}},
  \citenamefont {{Tavecchio}}, \citenamefont {{Temnikov}}, \citenamefont
  {{Terzi{\'c}}}, \citenamefont {{Teshima}}, \citenamefont {{Tosti}},
  \citenamefont {{Truzzi}}, \citenamefont {{Tutone}}, \citenamefont {{Ubach}},
  \citenamefont {{van Scherpenberg}}, \citenamefont {{Vanzo}}, \citenamefont
  {{Vazquez Acosta}}, \citenamefont {{Ventura}}, \citenamefont {{Verguilov}},
  \citenamefont {{Vigorito}}, \citenamefont {{Vitale}}, \citenamefont {{Vovk}},
  \citenamefont {{Will}}, \citenamefont {{Wunderlich}}, \citenamefont
  {{Zari{\'c}}},\ and\ \citenamefont {{MAGIC
  Collaboration}}}]{2022PDU....3500912A}%
  \BibitemOpen
  \bibfield  {author} {\bibinfo {author} {\bibfnamefont {V.~A.}\ \bibnamefont
  {{Acciari}}}, \bibinfo {author} {\bibfnamefont {S.}~\bibnamefont
  {{Ansoldi}}}, \bibinfo {author} {\bibfnamefont {L.~A.}\ \bibnamefont
  {{Antonelli}}}, \bibinfo {author} {\bibfnamefont {A.}~\bibnamefont {{Arbet
  Engels}}}, \bibinfo {author} {\bibfnamefont {M.}~\bibnamefont {{Artero}}},
  \bibinfo {author} {\bibfnamefont {K.}~\bibnamefont {{Asano}}}, \bibinfo
  {author} {\bibfnamefont {D.}~\bibnamefont {{Baack}}}, \bibinfo {author}
  {\bibfnamefont {A.}~\bibnamefont {{Babi{\'c}}}}, \bibinfo {author}
  {\bibfnamefont {A.}~\bibnamefont {{Baquero}}}, \bibinfo {author}
  {\bibfnamefont {U.}~\bibnamefont {{Barres de Almeida}}}, \bibinfo {author}
  {\bibfnamefont {J.~A.}\ \bibnamefont {{Barrio}}}, \bibinfo {author}
  {\bibfnamefont {I.}~\bibnamefont {{Batkovi{\'c}}}}, \bibinfo {author}
  {\bibfnamefont {J.}~\bibnamefont {{Becerra Gonz{\'a}lez}}}, \bibinfo {author}
  {\bibfnamefont {W.}~\bibnamefont {{Bednarek}}}, \bibinfo {author}
  {\bibfnamefont {L.}~\bibnamefont {{Bellizzi}}}, \bibinfo {author}
  {\bibfnamefont {E.}~\bibnamefont {{Bernardini}}}, \bibinfo {author}
  {\bibfnamefont {M.}~\bibnamefont {{Bernardos}}}, \bibinfo {author}
  {\bibfnamefont {A.}~\bibnamefont {{Berti}}}, \bibinfo {author} {\bibfnamefont
  {J.}~\bibnamefont {{Besenrieder}}}, \bibinfo {author} {\bibfnamefont
  {W.}~\bibnamefont {{Bhattacharyya}}}, \bibinfo {author} {\bibfnamefont
  {C.}~\bibnamefont {{Bigongiari}}}, \bibinfo {author} {\bibfnamefont
  {A.}~\bibnamefont {{Biland}}}, \bibinfo {author} {\bibfnamefont
  {O.}~\bibnamefont {{Blanch}}}, \bibinfo {author} {\bibfnamefont
  {H.}~\bibnamefont {{B{\"o}kenkamp}}}, \bibinfo {author} {\bibfnamefont
  {G.}~\bibnamefont {{Bonnoli}}}, \bibinfo {author} {\bibfnamefont
  {{\v{Z}}.}~\bibnamefont {{Bo{\v{s}}njak}}}, \bibinfo {author} {\bibfnamefont
  {G.}~\bibnamefont {{Busetto}}}, \bibinfo {author} {\bibfnamefont
  {R.}~\bibnamefont {{Carosi}}}, \bibinfo {author} {\bibfnamefont
  {G.}~\bibnamefont {{Ceribella}}}, \bibinfo {author} {\bibfnamefont
  {M.}~\bibnamefont {{Cerruti}}}, \bibinfo {author} {\bibfnamefont
  {Y.}~\bibnamefont {{Chai}}}, \bibinfo {author} {\bibfnamefont
  {A.}~\bibnamefont {{Chilingarian}}}, \bibinfo {author} {\bibfnamefont
  {S.}~\bibnamefont {{Cikota}}}, \bibinfo {author} {\bibfnamefont {S.~M.}\
  \bibnamefont {{Colak}}}, \bibinfo {author} {\bibfnamefont {E.}~\bibnamefont
  {{Colombo}}}, \bibinfo {author} {\bibfnamefont {J.~L.}\ \bibnamefont
  {{Contreras}}}, \bibinfo {author} {\bibfnamefont {J.}~\bibnamefont
  {{Cortina}}}, \bibinfo {author} {\bibfnamefont {S.}~\bibnamefont {{Covino}}},
  \bibinfo {author} {\bibfnamefont {G.}~\bibnamefont {{D'Amico}}}, \bibinfo
  {author} {\bibfnamefont {V.}~\bibnamefont {{D'Elia}}}, \bibinfo {author}
  {\bibfnamefont {P.}~\bibnamefont {{da Vela}}}, \bibinfo {author}
  {\bibfnamefont {F.}~\bibnamefont {{Dazzi}}}, \bibinfo {author} {\bibfnamefont
  {A.}~\bibnamefont {{de Angelis}}}, \bibinfo {author} {\bibfnamefont
  {B.}~\bibnamefont {{de Lotto}}}, \bibinfo {author} {\bibfnamefont
  {M.}~\bibnamefont {{Delfino}}}, \bibinfo {author} {\bibfnamefont
  {J.}~\bibnamefont {{Delgado}}}, \bibinfo {author} {\bibfnamefont
  {C.}~\bibnamefont {{Delgado Mendez}}}, \bibinfo {author} {\bibfnamefont
  {D.}~\bibnamefont {{Depaoli}}}, \bibinfo {author} {\bibfnamefont
  {F.}~\bibnamefont {{di Pierro}}}, \bibinfo {author} {\bibfnamefont
  {L.}~\bibnamefont {{di Venere}}}, \bibinfo {author} {\bibfnamefont
  {E.}~\bibnamefont {{Do Souto Espi{\~n}eira}}}, \bibinfo {author}
  {\bibfnamefont {D.}~\bibnamefont {{Dominis Prester}}}, \bibinfo {author}
  {\bibfnamefont {A.}~\bibnamefont {{Donini}}}, \bibinfo {author}
  {\bibfnamefont {D.}~\bibnamefont {{Dorner}}}, \bibinfo {author}
  {\bibfnamefont {M.}~\bibnamefont {{Doro}}}, \bibinfo {author} {\bibfnamefont
  {D.}~\bibnamefont {{Elsaesser}}}, \bibinfo {author} {\bibfnamefont
  {V.}~\bibnamefont {{Fallah Ramazani}}}, \bibinfo {author} {\bibfnamefont
  {A.}~\bibnamefont {{Fattorini}}}, \bibinfo {author} {\bibfnamefont {M.~V.}\
  \bibnamefont {{Fonseca}}}, \bibinfo {author} {\bibfnamefont {L.}~\bibnamefont
  {{Font}}}, \bibinfo {author} {\bibfnamefont {C.}~\bibnamefont {{Fruck}}},
  \bibinfo {author} {\bibfnamefont {S.}~\bibnamefont {{Fukami}}}, \bibinfo
  {author} {\bibfnamefont {R.~J.}\ \bibnamefont {{Garc{\'\i}a L{\'o}pez}}},
  \bibinfo {author} {\bibfnamefont {M.}~\bibnamefont {{Garczarczyk}}}, \bibinfo
  {author} {\bibfnamefont {S.}~\bibnamefont {{Gasparyan}}}, \bibinfo {author}
  {\bibfnamefont {M.}~\bibnamefont {{Gaug}}}, \bibinfo {author} {\bibfnamefont
  {N.}~\bibnamefont {{Giglietto}}}, \bibinfo {author} {\bibfnamefont
  {F.}~\bibnamefont {{Giordano}}}, \bibinfo {author} {\bibfnamefont
  {P.}~\bibnamefont {{Gliwny}}}, \bibinfo {author} {\bibfnamefont
  {N.}~\bibnamefont {{Godinovi{\'c}}}}, \bibinfo {author} {\bibfnamefont
  {J.~G.}\ \bibnamefont {{Green}}}, \bibinfo {author} {\bibfnamefont
  {D.}~\bibnamefont {{Green}}}, \bibinfo {author} {\bibfnamefont
  {D.}~\bibnamefont {{Hadasch}}}, \bibinfo {author} {\bibfnamefont
  {A.}~\bibnamefont {{Hahn}}}, \bibinfo {author} {\bibfnamefont
  {L.}~\bibnamefont {{Heckmann}}}, \bibinfo {author} {\bibfnamefont
  {J.}~\bibnamefont {{Herrera}}}, \bibinfo {author} {\bibfnamefont
  {J.}~\bibnamefont {{Hoang}}}, \bibinfo {author} {\bibfnamefont
  {D.}~\bibnamefont {{Hrupec}}}, \bibinfo {author} {\bibfnamefont
  {M.}~\bibnamefont {{H{\"u}tten}}}, \bibinfo {author} {\bibfnamefont
  {T.}~\bibnamefont {{Inada}}}, \bibinfo {author} {\bibfnamefont
  {K.}~\bibnamefont {{Ishio}}}, \bibinfo {author} {\bibfnamefont
  {Y.}~\bibnamefont {{Iwamura}}}, \bibinfo {author} {\bibfnamefont
  {I.}~\bibnamefont {{Jim{\'e}nez}}}, \bibinfo {author} {\bibfnamefont
  {J.}~\bibnamefont {{Jormanainen}}}, \bibinfo {author} {\bibfnamefont
  {L.}~\bibnamefont {{Jouvin}}}, \bibinfo {author} {\bibfnamefont
  {M.}~\bibnamefont {{Karjalainen}}}, \bibinfo {author} {\bibfnamefont
  {D.}~\bibnamefont {{Kerszberg}}}, \bibinfo {author} {\bibfnamefont
  {Y.}~\bibnamefont {{Kobayashi}}}, \bibinfo {author} {\bibfnamefont
  {H.}~\bibnamefont {{Kubo}}}, \bibinfo {author} {\bibfnamefont
  {J.}~\bibnamefont {{Kushida}}}, \bibinfo {author} {\bibfnamefont
  {A.}~\bibnamefont {{Lamastra}}}, \bibinfo {author} {\bibfnamefont
  {D.}~\bibnamefont {{Lelas}}}, \bibinfo {author} {\bibfnamefont
  {F.}~\bibnamefont {{Leone}}}, \bibinfo {author} {\bibfnamefont
  {E.}~\bibnamefont {{Lindfors}}}, \bibinfo {author} {\bibfnamefont
  {L.}~\bibnamefont {{Linhoff}}}, \bibinfo {author} {\bibfnamefont
  {S.}~\bibnamefont {{Lombardi}}}, \bibinfo {author} {\bibfnamefont
  {F.}~\bibnamefont {{Longo}}}, \bibinfo {author} {\bibfnamefont
  {R.}~\bibnamefont {{L{\'o}pez-Coto}}}, \bibinfo {author} {\bibfnamefont
  {M.}~\bibnamefont {{L{\'o}pez-Moya}}}, \bibinfo {author} {\bibfnamefont
  {A.}~\bibnamefont {{L{\'o}pez-Oramas}}}, \bibinfo {author} {\bibfnamefont
  {S.}~\bibnamefont {{Loporchio}}}, \bibinfo {author} {\bibfnamefont
  {B.}~\bibnamefont {{Machado de Oliveira Fraga}}}, \bibinfo {author}
  {\bibfnamefont {C.}~\bibnamefont {{Maggio}}}, \bibinfo {author}
  {\bibfnamefont {P.}~\bibnamefont {{Majumdar}}}, \bibinfo {author}
  {\bibfnamefont {M.}~\bibnamefont {{Makariev}}}, \bibinfo {author}
  {\bibfnamefont {M.}~\bibnamefont {{Mallamaci}}}, \bibinfo {author}
  {\bibfnamefont {G.}~\bibnamefont {{Maneva}}}, \bibinfo {author}
  {\bibfnamefont {M.}~\bibnamefont {{Manganaro}}}, \bibinfo {author}
  {\bibfnamefont {K.}~\bibnamefont {{Mannheim}}}, \bibinfo {author}
  {\bibfnamefont {L.}~\bibnamefont {{Maraschi}}}, \bibinfo {author}
  {\bibfnamefont {M.}~\bibnamefont {{Mariotti}}}, \bibinfo {author}
  {\bibfnamefont {M.}~\bibnamefont {{Mart{\'\i}nez}}}, \bibinfo {author}
  {\bibfnamefont {D.}~\bibnamefont {{Mazin}}}, \bibinfo {author} {\bibfnamefont
  {S.}~\bibnamefont {{Menchiari}}}, \bibinfo {author} {\bibfnamefont
  {S.}~\bibnamefont {{Mender}}}, \bibinfo {author} {\bibfnamefont
  {S.}~\bibnamefont {{Mi{\'c}anovi{\'c}}}}, \bibinfo {author} {\bibfnamefont
  {D.}~\bibnamefont {{Miceli}}}, \bibinfo {author} {\bibfnamefont
  {T.}~\bibnamefont {{Miener}}}, \bibinfo {author} {\bibfnamefont {J.~M.}\
  \bibnamefont {{Miranda}}}, \bibinfo {author} {\bibfnamefont {R.}~\bibnamefont
  {{Mirzoyan}}}, \bibinfo {author} {\bibfnamefont {E.}~\bibnamefont
  {{Molina}}}, \bibinfo {author} {\bibfnamefont {A.}~\bibnamefont
  {{Moralejo}}}, \bibinfo {author} {\bibfnamefont {D.}~\bibnamefont
  {{Morcuende}}}, \bibinfo {author} {\bibfnamefont {V.}~\bibnamefont
  {{Moreno}}}, \bibinfo {author} {\bibfnamefont {E.}~\bibnamefont {{Moretti}}},
  \bibinfo {author} {\bibfnamefont {V.}~\bibnamefont {{Neustroev}}}, \bibinfo
  {author} {\bibfnamefont {C.}~\bibnamefont {{Nigro}}}, \bibinfo {author}
  {\bibfnamefont {K.}~\bibnamefont {{Nilsson}}}, \bibinfo {author}
  {\bibfnamefont {D.}~\bibnamefont {{Ninci}}}, \bibinfo {author} {\bibfnamefont
  {K.}~\bibnamefont {{Nishijima}}}, \bibinfo {author} {\bibfnamefont
  {K.}~\bibnamefont {{Noda}}}, \bibinfo {author} {\bibfnamefont
  {S.}~\bibnamefont {{Nozaki}}}, \bibinfo {author} {\bibfnamefont
  {Y.}~\bibnamefont {{Ohtani}}}, \bibinfo {author} {\bibfnamefont
  {T.}~\bibnamefont {{Oka}}}, \bibinfo {author} {\bibfnamefont
  {J.}~\bibnamefont {{Otero-Santos}}}, \bibinfo {author} {\bibfnamefont
  {S.}~\bibnamefont {{Paiano}}}, \bibinfo {author} {\bibfnamefont
  {M.}~\bibnamefont {{Palatiello}}}, \bibinfo {author} {\bibfnamefont
  {D.}~\bibnamefont {{Paneque}}}, \bibinfo {author} {\bibfnamefont
  {R.}~\bibnamefont {{Paoletti}}}, \bibinfo {author} {\bibfnamefont {J.~M.}\
  \bibnamefont {{Paredes}}}, \bibinfo {author} {\bibfnamefont {L.}~\bibnamefont
  {{Pavleti{\'c}}}}, \bibinfo {author} {\bibfnamefont {P.}~\bibnamefont
  {{Pe{\~n}il}}}, \bibinfo {author} {\bibfnamefont {M.}~\bibnamefont
  {{Persic}}}, \bibinfo {author} {\bibfnamefont {M.}~\bibnamefont {{Pihet}}},
  \bibinfo {author} {\bibfnamefont {P.~G.}\ \bibnamefont {{Prada Moroni}}},
  \bibinfo {author} {\bibfnamefont {E.}~\bibnamefont {{Prandini}}}, \bibinfo
  {author} {\bibfnamefont {C.}~\bibnamefont {{Priyadarshi}}}, \bibinfo {author}
  {\bibfnamefont {I.}~\bibnamefont {{Puljak}}}, \bibinfo {author}
  {\bibfnamefont {W.}~\bibnamefont {{Rhode}}}, \bibinfo {author} {\bibfnamefont
  {M.}~\bibnamefont {{Rib{\'o}}}}, \bibinfo {author} {\bibfnamefont
  {J.}~\bibnamefont {{Rico}}}, \bibinfo {author} {\bibfnamefont
  {C.}~\bibnamefont {{Righi}}}, \bibinfo {author} {\bibfnamefont
  {A.}~\bibnamefont {{Rugliancich}}}, \bibinfo {author} {\bibfnamefont
  {L.}~\bibnamefont {{Saha}}}, \bibinfo {author} {\bibfnamefont
  {N.}~\bibnamefont {{Sahakyan}}}, \bibinfo {author} {\bibfnamefont
  {T.}~\bibnamefont {{Saito}}}, \bibinfo {author} {\bibfnamefont
  {S.}~\bibnamefont {{Sakurai}}}, \bibinfo {author} {\bibfnamefont
  {K.}~\bibnamefont {{Satalecka}}}, \bibinfo {author} {\bibfnamefont {F.~G.}\
  \bibnamefont {{Saturni}}}, \bibinfo {author} {\bibfnamefont {B.}~\bibnamefont
  {{Schleicher}}}, \bibinfo {author} {\bibfnamefont {K.}~\bibnamefont
  {{Schmidt}}}, \bibinfo {author} {\bibfnamefont {T.}~\bibnamefont
  {{Schweizer}}}, \bibinfo {author} {\bibfnamefont {J.}~\bibnamefont
  {{Sitarek}}}, \bibinfo {author} {\bibfnamefont {I.}~\bibnamefont
  {{{\v{S}}nidari{\'c}}}}, \bibinfo {author} {\bibfnamefont {D.}~\bibnamefont
  {{Sobczynska}}}, \bibinfo {author} {\bibfnamefont {A.}~\bibnamefont
  {{Spolon}}}, \bibinfo {author} {\bibfnamefont {A.}~\bibnamefont
  {{Stamerra}}}, \bibinfo {author} {\bibfnamefont {J.}~\bibnamefont
  {{Stri{\v{s}}kovi{\'c}}}}, \bibinfo {author} {\bibfnamefont {D.}~\bibnamefont
  {{Strom}}}, \bibinfo {author} {\bibfnamefont {M.}~\bibnamefont {{Strzys}}},
  \bibinfo {author} {\bibfnamefont {Y.}~\bibnamefont {{Suda}}}, \bibinfo
  {author} {\bibfnamefont {T.}~\bibnamefont {{Suri{\'c}}}}, \bibinfo {author}
  {\bibfnamefont {M.}~\bibnamefont {{Takahashi}}}, \bibinfo {author}
  {\bibfnamefont {R.}~\bibnamefont {{Takeishi}}}, \bibinfo {author}
  {\bibfnamefont {F.}~\bibnamefont {{Tavecchio}}}, \bibinfo {author}
  {\bibfnamefont {P.}~\bibnamefont {{Temnikov}}}, \bibinfo {author}
  {\bibfnamefont {T.}~\bibnamefont {{Terzi{\'c}}}}, \bibinfo {author}
  {\bibfnamefont {M.}~\bibnamefont {{Teshima}}}, \bibinfo {author}
  {\bibfnamefont {L.}~\bibnamefont {{Tosti}}}, \bibinfo {author} {\bibfnamefont
  {S.}~\bibnamefont {{Truzzi}}}, \bibinfo {author} {\bibfnamefont
  {A.}~\bibnamefont {{Tutone}}}, \bibinfo {author} {\bibfnamefont
  {S.}~\bibnamefont {{Ubach}}}, \bibinfo {author} {\bibfnamefont
  {J.}~\bibnamefont {{van Scherpenberg}}}, \bibinfo {author} {\bibfnamefont
  {G.}~\bibnamefont {{Vanzo}}}, \bibinfo {author} {\bibfnamefont
  {M.}~\bibnamefont {{Vazquez Acosta}}}, \bibinfo {author} {\bibfnamefont
  {S.}~\bibnamefont {{Ventura}}}, \bibinfo {author} {\bibfnamefont
  {V.}~\bibnamefont {{Verguilov}}}, \bibinfo {author} {\bibfnamefont {C.~F.}\
  \bibnamefont {{Vigorito}}}, \bibinfo {author} {\bibfnamefont
  {V.}~\bibnamefont {{Vitale}}}, \bibinfo {author} {\bibfnamefont
  {I.}~\bibnamefont {{Vovk}}}, \bibinfo {author} {\bibfnamefont
  {M.}~\bibnamefont {{Will}}}, \bibinfo {author} {\bibfnamefont
  {C.}~\bibnamefont {{Wunderlich}}}, \bibinfo {author} {\bibfnamefont
  {D.}~\bibnamefont {{Zari{\'c}}}},\ and\ \bibinfo {author} {\bibnamefont
  {{MAGIC Collaboration}}},\ }\bibfield  {title} {\bibinfo {title} {{Combined
  searches for dark matter in dwarf spheroidal galaxies observed with the MAGIC
  telescopes, including new data from Coma Berenices and Draco}},\ }\href
  {https://doi.org/10.1016/j.dark.2021.100912} {\bibfield  {journal} {\bibinfo
  {journal} {PDU}\ }\textbf {\bibinfo {volume} {35}},\ \bibinfo {eid} {100912}
  (\bibinfo {year} {2022})},\ \Eprint {https://arxiv.org/abs/2111.15009}
  {arXiv:2111.15009 [astro-ph.HE]} \BibitemShut {NoStop}%
\bibitem [{\citenamefont {{Calore}}\ \emph {et~al.}(2022)\citenamefont
  {{Calore}}, \citenamefont {{Cirelli}}, \citenamefont {{Derome}},
  \citenamefont {{Genolini}}, \citenamefont {{Maurin}}, \citenamefont
  {{Salati}},\ and\ \citenamefont {{Serpico}}}]{2022arXiv220203076C}%
  \BibitemOpen
  \bibfield  {author} {\bibinfo {author} {\bibfnamefont {F.}~\bibnamefont
  {{Calore}}}, \bibinfo {author} {\bibfnamefont {M.}~\bibnamefont {{Cirelli}}},
  \bibinfo {author} {\bibfnamefont {L.}~\bibnamefont {{Derome}}}, \bibinfo
  {author} {\bibfnamefont {Y.}~\bibnamefont {{Genolini}}}, \bibinfo {author}
  {\bibfnamefont {D.}~\bibnamefont {{Maurin}}}, \bibinfo {author}
  {\bibfnamefont {P.}~\bibnamefont {{Salati}}},\ and\ \bibinfo {author}
  {\bibfnamefont {P.~D.}\ \bibnamefont {{Serpico}}},\ }\bibfield  {title}
  {\bibinfo {title} {{AMS-02 antiprotons and dark matter: Trimmed hints and
  robust bounds}},\ }\href@noop {} {\bibfield  {journal} {\bibinfo  {journal}
  {arXiv}\ ,\ \bibinfo {eid} {arXiv:2202.03076}} (\bibinfo {year} {2022})},\
  \Eprint {https://arxiv.org/abs/2202.03076} {arXiv:2202.03076 [hep-ph]}
  \BibitemShut {NoStop}%
\bibitem [{\citenamefont {{de Gasperin}}\ \emph {et~al.}(2021)\citenamefont
  {{de Gasperin}}, \citenamefont {{Williams}}, \citenamefont {{Best}},
  \citenamefont {{Br{\"u}ggen}}, \citenamefont {{Brunetti}}, \citenamefont
  {{Cuciti}}, \citenamefont {{Dijkema}}, \citenamefont {{Hardcastle}},
  \citenamefont {{Norden}}, \citenamefont {{Offringa}}, \citenamefont
  {{Shimwell}}, \citenamefont {{van Weeren}}, \citenamefont {{Bomans}},
  \citenamefont {{Bonafede}}, \citenamefont {{Botteon}}, \citenamefont
  {{Callingham}}, \citenamefont {{Cassano}}, \citenamefont {{Chy{\.z}y}},
  \citenamefont {{Emig}}, \citenamefont {{Edler}}, \citenamefont {{Haverkorn}},
  \citenamefont {{Heald}}, \citenamefont {{Heesen}}, \citenamefont
  {{Iacobelli}}, \citenamefont {{Intema}}, \citenamefont {{Kadler}},
  \citenamefont {{Ma{\l}ek}}, \citenamefont {{Mevius}}, \citenamefont
  {{Miley}}, \citenamefont {{Mingo}}, \citenamefont {{Morabito}}, \citenamefont
  {{Sabater}}, \citenamefont {{Morganti}}, \citenamefont {{Orr{\'u}}},
  \citenamefont {{Pizzo}}, \citenamefont {{Prandoni}}, \citenamefont
  {{Shulevski}}, \citenamefont {{Tasse}}, \citenamefont {{Vaccari}},
  \citenamefont {{Zarka}},\ and\ \citenamefont
  {{R{\"o}ttgering}}}]{2021A&A...648A.104D}%
  \BibitemOpen
  \bibfield  {author} {\bibinfo {author} {\bibfnamefont {F.}~\bibnamefont {{de
  Gasperin}}}, \bibinfo {author} {\bibfnamefont {W.~L.}\ \bibnamefont
  {{Williams}}}, \bibinfo {author} {\bibfnamefont {P.}~\bibnamefont {{Best}}},
  \bibinfo {author} {\bibfnamefont {M.}~\bibnamefont {{Br{\"u}ggen}}}, \bibinfo
  {author} {\bibfnamefont {G.}~\bibnamefont {{Brunetti}}}, \bibinfo {author}
  {\bibfnamefont {V.}~\bibnamefont {{Cuciti}}}, \bibinfo {author}
  {\bibfnamefont {T.~J.}\ \bibnamefont {{Dijkema}}}, \bibinfo {author}
  {\bibfnamefont {M.~J.}\ \bibnamefont {{Hardcastle}}}, \bibinfo {author}
  {\bibfnamefont {M.~J.}\ \bibnamefont {{Norden}}}, \bibinfo {author}
  {\bibfnamefont {A.}~\bibnamefont {{Offringa}}}, \bibinfo {author}
  {\bibfnamefont {T.}~\bibnamefont {{Shimwell}}}, \bibinfo {author}
  {\bibfnamefont {R.}~\bibnamefont {{van Weeren}}}, \bibinfo {author}
  {\bibfnamefont {D.}~\bibnamefont {{Bomans}}}, \bibinfo {author}
  {\bibfnamefont {A.}~\bibnamefont {{Bonafede}}}, \bibinfo {author}
  {\bibfnamefont {A.}~\bibnamefont {{Botteon}}}, \bibinfo {author}
  {\bibfnamefont {J.~R.}\ \bibnamefont {{Callingham}}}, \bibinfo {author}
  {\bibfnamefont {R.}~\bibnamefont {{Cassano}}}, \bibinfo {author}
  {\bibfnamefont {K.~T.}\ \bibnamefont {{Chy{\.z}y}}}, \bibinfo {author}
  {\bibfnamefont {K.~L.}\ \bibnamefont {{Emig}}}, \bibinfo {author}
  {\bibfnamefont {H.}~\bibnamefont {{Edler}}}, \bibinfo {author} {\bibfnamefont
  {M.}~\bibnamefont {{Haverkorn}}}, \bibinfo {author} {\bibfnamefont
  {G.}~\bibnamefont {{Heald}}}, \bibinfo {author} {\bibfnamefont
  {V.}~\bibnamefont {{Heesen}}}, \bibinfo {author} {\bibfnamefont
  {M.}~\bibnamefont {{Iacobelli}}}, \bibinfo {author} {\bibfnamefont {H.~T.}\
  \bibnamefont {{Intema}}}, \bibinfo {author} {\bibfnamefont {M.}~\bibnamefont
  {{Kadler}}}, \bibinfo {author} {\bibfnamefont {K.}~\bibnamefont
  {{Ma{\l}ek}}}, \bibinfo {author} {\bibfnamefont {M.}~\bibnamefont
  {{Mevius}}}, \bibinfo {author} {\bibfnamefont {G.}~\bibnamefont {{Miley}}},
  \bibinfo {author} {\bibfnamefont {B.}~\bibnamefont {{Mingo}}}, \bibinfo
  {author} {\bibfnamefont {L.~K.}\ \bibnamefont {{Morabito}}}, \bibinfo
  {author} {\bibfnamefont {J.}~\bibnamefont {{Sabater}}}, \bibinfo {author}
  {\bibfnamefont {R.}~\bibnamefont {{Morganti}}}, \bibinfo {author}
  {\bibfnamefont {E.}~\bibnamefont {{Orr{\'u}}}}, \bibinfo {author}
  {\bibfnamefont {R.}~\bibnamefont {{Pizzo}}}, \bibinfo {author} {\bibfnamefont
  {I.}~\bibnamefont {{Prandoni}}}, \bibinfo {author} {\bibfnamefont
  {A.}~\bibnamefont {{Shulevski}}}, \bibinfo {author} {\bibfnamefont
  {C.}~\bibnamefont {{Tasse}}}, \bibinfo {author} {\bibfnamefont
  {M.}~\bibnamefont {{Vaccari}}}, \bibinfo {author} {\bibfnamefont
  {P.}~\bibnamefont {{Zarka}}},\ and\ \bibinfo {author} {\bibfnamefont
  {H.}~\bibnamefont {{R{\"o}ttgering}}},\ }\bibfield  {title} {\bibinfo {title}
  {{The LOFAR LBA Sky Survey. I. Survey description and preliminary data
  release}},\ }\href {https://doi.org/10.1051/0004-6361/202140316} {\bibfield
  {journal} {\bibinfo  {journal} {A\&A}\ }\textbf {\bibinfo {volume} {648}},\
  \bibinfo {eid} {A104} (\bibinfo {year} {2021})},\ \Eprint
  {https://arxiv.org/abs/2102.09238} {arXiv:2102.09238 [astro-ph.IM]}
  \BibitemShut {NoStop}%
\bibitem [{\citenamefont {{Ferri{\`e}re}}\ \emph {et~al.}(2007)\citenamefont
  {{Ferri{\`e}re}}, \citenamefont {{Gillard}},\ and\ \citenamefont
  {{Jean}}}]{2007A&A...467..611F}%
  \BibitemOpen
  \bibfield  {author} {\bibinfo {author} {\bibfnamefont {K.}~\bibnamefont
  {{Ferri{\`e}re}}}, \bibinfo {author} {\bibfnamefont {W.}~\bibnamefont
  {{Gillard}}},\ and\ \bibinfo {author} {\bibfnamefont {P.}~\bibnamefont
  {{Jean}}},\ }\bibfield  {title} {\bibinfo {title} {{Spatial distribution of
  interstellar gas in the innermost 3 kpc of our galaxy}},\ }\href
  {https://doi.org/10.1051/0004-6361:20066992} {\bibfield  {journal} {\bibinfo
  {journal} {Astron. Astrophys.}\ }\textbf {\bibinfo {volume} {467}},\ \bibinfo
  {pages} {611} (\bibinfo {year} {2007})},\ \Eprint
  {https://arxiv.org/abs/astro-ph/0702532} {arXiv:astro-ph/0702532 [astro-ph]}
  \BibitemShut {NoStop}%
\bibitem [{\citenamefont {{Dickey}}\ and\ \citenamefont
  {{Lockman}}(1990)}]{1990ARA&A..28..215D}%
  \BibitemOpen
  \bibfield  {author} {\bibinfo {author} {\bibfnamefont {J.~M.}\ \bibnamefont
  {{Dickey}}}\ and\ \bibinfo {author} {\bibfnamefont {F.~J.}\ \bibnamefont
  {{Lockman}}},\ }\bibfield  {title} {\bibinfo {title} {{H I in the galaxy.}},\
  }\href {https://doi.org/10.1146/annurev.aa.28.090190.001243} {\bibfield
  {journal} {\bibinfo  {journal} {Ann. Rev. Astron. Astrophys.}\ }\textbf
  {\bibinfo {volume} {28}},\ \bibinfo {pages} {215} (\bibinfo {year}
  {1990})}\BibitemShut {NoStop}%
\bibitem [{\citenamefont {{Nieten}}\ \emph {et~al.}(2006)\citenamefont
  {{Nieten}}, \citenamefont {{Neininger}}, \citenamefont {{Gu{\'e}lin}},
  \citenamefont {{Ungerechts}}, \citenamefont {{Lucas}}, \citenamefont
  {{Berkhuijsen}}, \citenamefont {{Beck}},\ and\ \citenamefont
  {{Wielebinski}}}]{2006A&A...453..459N}%
  \BibitemOpen
  \bibfield  {author} {\bibinfo {author} {\bibfnamefont {C.}~\bibnamefont
  {{Nieten}}}, \bibinfo {author} {\bibfnamefont {N.}~\bibnamefont
  {{Neininger}}}, \bibinfo {author} {\bibfnamefont {M.}~\bibnamefont
  {{Gu{\'e}lin}}}, \bibinfo {author} {\bibfnamefont {H.}~\bibnamefont
  {{Ungerechts}}}, \bibinfo {author} {\bibfnamefont {R.}~\bibnamefont
  {{Lucas}}}, \bibinfo {author} {\bibfnamefont {E.~M.}\ \bibnamefont
  {{Berkhuijsen}}}, \bibinfo {author} {\bibfnamefont {R.}~\bibnamefont
  {{Beck}}},\ and\ \bibinfo {author} {\bibfnamefont {R.}~\bibnamefont
  {{Wielebinski}}},\ }\bibfield  {title} {\bibinfo {title} {{Molecular gas in
  the Andromeda galaxy}},\ }\href {https://doi.org/10.1051/0004-6361:20035672}
  {\bibfield  {journal} {\bibinfo  {journal} {Astron. Astrophys.}\ }\textbf
  {\bibinfo {volume} {453}},\ \bibinfo {pages} {459} (\bibinfo {year}
  {2006})},\ \Eprint {https://arxiv.org/abs/astro-ph/0512563}
  {arXiv:astro-ph/0512563 [astro-ph]} \BibitemShut {NoStop}%
\bibitem [{\citenamefont {{Bonnarel}}\ \emph {et~al.}(2000)\citenamefont
  {{Bonnarel}}, \citenamefont {{Fernique}}, \citenamefont {{Bienaym{\'e}}},
  \citenamefont {{Egret}}, \citenamefont {{Genova}}, \citenamefont {{Louys}},
  \citenamefont {{Ochsenbein}}, \citenamefont {{Wenger}},\ and\ \citenamefont
  {{Bartlett}}}]{2000A&AS..143...33B}%
  \BibitemOpen
  \bibfield  {author} {\bibinfo {author} {\bibfnamefont {F.}~\bibnamefont
  {{Bonnarel}}}, \bibinfo {author} {\bibfnamefont {P.}~\bibnamefont
  {{Fernique}}}, \bibinfo {author} {\bibfnamefont {O.}~\bibnamefont
  {{Bienaym{\'e}}}}, \bibinfo {author} {\bibfnamefont {D.}~\bibnamefont
  {{Egret}}}, \bibinfo {author} {\bibfnamefont {F.}~\bibnamefont {{Genova}}},
  \bibinfo {author} {\bibfnamefont {M.}~\bibnamefont {{Louys}}}, \bibinfo
  {author} {\bibfnamefont {F.}~\bibnamefont {{Ochsenbein}}}, \bibinfo {author}
  {\bibfnamefont {M.}~\bibnamefont {{Wenger}}},\ and\ \bibinfo {author}
  {\bibfnamefont {J.~G.}\ \bibnamefont {{Bartlett}}},\ }\bibfield  {title}
  {\bibinfo {title} {{The ALADIN interactive sky atlas. A reference tool for
  identification of astronomical sources}},\ }\href
  {https://doi.org/10.1051/aas:2000331} {\bibfield  {journal} {\bibinfo
  {journal} {Astron. Astrophys. Suppl. Ser.}\ }\textbf {\bibinfo {volume}
  {143}},\ \bibinfo {pages} {33} (\bibinfo {year} {2000})}\BibitemShut
  {NoStop}%
\bibitem [{num()}]{numpy}%
  \BibitemOpen
  \href@noop {} {}\bibinfo {howpublished}
  {\url{https://numpy.org/}}\BibitemShut {NoStop}%
\bibitem [{hea()}]{healpy}%
  \BibitemOpen
  \href@noop {} {}\bibinfo {howpublished}
  {\url{https://github.com/healpy}}\BibitemShut {NoStop}%
\bibitem [{WPD()}]{WPD}%
  \BibitemOpen
  \href@noop {} {}\bibinfo {howpublished}
  {\url{https://apps.automeris.io/wpd/}}\BibitemShut {NoStop}%
\end{thebibliography}%

\end{document}